\documentclass[fleqn,usenatbib]{mnras}

\usepackage{mathptmx}

\usepackage[T1]{fontenc}
\usepackage[normalem]{ulem}

\DeclareRobustCommand{\VAN}[3]{#2}
\let\VANthebibliography\thebibliography
\def\thebibliography{\DeclareRobustCommand{\VAN}[3]{##3}\VANthebibliography}

\usepackage{graphicx}	
\usepackage{amsmath}	

\usepackage{natbib,etoolbox,lipsum,hyperref}
\usepackage{xfrac}
\usepackage{pdflscape}
\usepackage{subcaption}
\usepackage{soul}
\usepackage{ulem}
\usepackage{comment}
\usepackage{epstopdf}
\usepackage{placeins}
\usepackage{threeparttable} 

\title[Comet C/2020 V2 (ZTF)]{Long-term monitoring of a dynamically new comet C/2020 V2 (ZTF)}

\author[Goldy Ahuja et al.]{
Goldy Ahuja$^{1,2}$\thanks{E-mail: goldy@prl.res.in | goldezz20@gmail.com},
K. Aravind$^{3,1}$,
Shashikiran Ganesh$^{1}$,
Said Hmiddouch$^{3,4}$,
Mathieu Vander Donckt$^{3}$,
\newauthor
Emmanuel Jehin$^{3}$,
Devendra Sahu$^{5}$,
and T. Sivarani$^{5}$
\\
$^{1}$Physical Research Laboratory, Ahmedabad, Gujarat-380009, India\\
$^{2}$Indian Institute of Technology Gandhinagar, Palaj, Gujarat-382355, India\\
$^{3}$Space sciences, Technologies \& Astrophysics Research (STAR) Institute, University of Liège, Liège, Belgium\\
$^{4}$Cadi Ayyad University (UCA), Oukaimeden Observatory (OUCA), Faculté des Sciences Semlalia (FSSM),\\
High Energy Physics, Astrophysics and Geoscience Laboratory (LPHEAG), Marrakech, Morocco\\
$^{5}$Indian Institute of Astrophysics, Koramangala, Bengaluru, Karnataka – 560034, India
}

\date{Accepted XXX. Received YYY; in original form ZZZ}

\pubyear{2025}

\begin{document}
\label{firstpage}
\pagerange{\pageref{firstpage}--\pageref{lastpage}}
\maketitle

\begin{abstract}
Comet C/2020 V2 (ZTF) is categorized as a dynamically new long-period comet, making its first approach to the inner Solar system. We have observed this comet for around 32 months (from January 2022 to July 2024) at heliocentric distances from 5.41 au (pre-perihelion) to 5.26 au (post-perihelion) through various telescopes, employing photometric (60 epochs) and spectroscopic techniques (5 epochs).
Using these observations, we derived the production rates of different molecules such as CN$(0-0)$, C$_2(\Delta \nu=0)$, and C$_3$ and calculated the production rate ratios. The values of the ratios closest to the perihelion are found to be $\log$ (C$_2/$CN) $ = -0.04 \pm 0.03$ and $\log$ (C$_3/$CN) $ = -0.70 \pm 0.04$, which implies a typical carbon composition.  The mean photometric broadband colours are found to be $B-V = 0.77\pm0.04$, $V-R = 0.43\pm0.04$, $R-I = 0.42\pm0.06$, and $B-R = 1.19\pm0.04$. 
The stability of the molecular production rate ratios and mean photometric broadband colours, pre- and post-perihelion, implies a homogeneous composition.
The mean reflectivity gradient for $B-V$ colour is $10.90 \pm 3.62$ $\%/1000$ \AA; $V-R$ colour is $6.15 \pm 3.51$ $\%/1000$ \AA; and for $R-I$ colour is $4.94 \pm 3.56$ $\%/1000$ \AA\ which is similar to the mean value of the dynamically new comets.
Additionally, using an asymmetric non-gravitational force model, we report the comet's nuclear radius to be $1.1 \pm 0.1$ km. Our results are expected to provide inputs to the selection of a potential dynamically new comet as a target for the Comet Interceptor mission.

\end{abstract}




\section{Introduction}



Comets are known as the pristine bodies of the solar system. They are remnants of the proto-solar nebula, which are distributed in two big reservoirs in the early solar system. These reservoirs, i.e., the Kuiper Belt and the Oort Cloud, are predicted to be formed due to the giant planetary migration towards the inner solar system \citep{Tsiganis2005, Gomes2005, Morbidelli_2007, Walsh2011, BRASSER_Oort}. 

Initially, comets are classified based on their orbital period, where Short-Period comets, hereafter, SPCs, and Long-Period Comets, hereafter, LPCs, are differentiated by whether their orbital period is less than or greater than 200 years \citep{Comet_Taxonomy}. 
The SPCs are further divided into Jupiter Family Comets and Halley-type Comets, based on the distribution of the comets in the inclination vs eccentricity plot (see Fig. 3 in \citet{Comet_Taxonomy} and Fig. 1(a) in \citet{Nesvorný_2017}). Jupiter-Family Comets, hereafter JFCs, show the clump distribution with a median inclination of 12.4$^{\circ}$ \citep{Brasser_2015_JFC}, and revolve in a prograde orbit for a period of less than 20 years. Whereas Halley-type comets, hereafter HTCs, have a median inclination of 75$^{\circ}$ \citep{Wang_2014}. Several HTCs do have a retrograde orbit \citep{levision_1992_JFC}.

Based on numerical simulation by \citet{LEVISON_1997}, JFCs have been found to be coming from the Kuiper-Belt. Also, another possibility of the origin of JFCs is from the Scattered disk \citep{jfc_sdo}. while the origin of HTCs, due to their varying inclination, is possible from the Oort Cloud \citep{Wang_2014,Nesvorný_2017} or from the scattered disk \citep{Levison_2006}.

The LPCs are thought to originate from the Oort Cloud \citep{Oort_1950}. The LPCs can be further divided based on whether the comet is coming into the inner solar system for the first time, i.e., a new comet, or it has already visited before, i.e., a returning comet. Those comets whose semi-major axis (a) is more than 10000 au are called the dynamically new comets, while if the semi-major axis (a) is less than 10000 au, they are called returning comets \citep{Comet_Taxonomy}. Recently, there has been a debate on the limit (check \citet{kroli_paper1} and \citet{Kroli_2011}). Studying Dynamically New Comets, hereafter DNCs, is essential as they are coming towards the Sun for the first time, and it is expected that they 
have never been exposed to the solar radiation after their initial formation and probable migration. 
The understanding of DNCs is important for the future Comet Interceptor mission \citep{cometInterceptor_2024}, which will be launched to study a least processed object, one that has never been to the inner solar system since its formation. Therefore, long-term monitoring of these objects is important to understand their evolution as they come to and go beyond perihelion. 

Comet C/2020 V2 (ZTF), hereafter V2, was discovered by the Zwicky Transient Facility in November 2020 at a distance of 8.7 au from the Sun. It approached perihelion on 08 May 2023 at a distance of 2.228 au. Comet V2 is a DNC with a semi-major axis of -2321 au at Epoch JD2459827.5 and an eccentricity of 1.0009 with a 1-$\sigma$ eccentricity of $2.669\times10^{-6}$ as given in \textsc{NASA JPL SBDB}\footnote{
\url{https://ssd.jpl.nasa.gov/tools/sbdb_lookup.html\#/?sstr=C/2020\%20V2}
}. It is currently in a hyperbolic orbit. However, just being in a hyperbolic orbit doesn't conclude that the comet is a DNC because its orbital parameters can be perturbed due to interaction with giant planets, galactic tidal effects, or non-gravitational forces, which result in changes in its trajectory.
Therefore, to predict the orbital parameter of DNC, we need to know its osculating elements in the past at -250 au (negative sign representing that the comet is approaching the Sun) as mentioned in the model by \citet{kroli_paper1} and \citet{kroli_paper2}. Using the CODE catalogue\footnote{\url{https://pad2.astro.amu.edu.pl/CODE/orbit.php?int=2020v2ec&orb=previous}}, which is based on \citet{kroli_paper1}, the average $1/a_{orig}$ is 14.88 $\times$ $10^{-6}$, which describe this comet as a DNC. Since it is now travelling in a non-periodic orbit, this gives us a unique opportunity to study this body before it leaves the solar system unless its orbit gets significantly perturbed to a returning trajectory.

In this work, we discuss the results obtained from photometric and spectroscopic observations of comet V2 from the TRAPPIST telescopes and multiple Indian observatories, respectively. We briefly describe the different telescope facilities used in section \ref{sec: Method} and the data reduction methods for photometry and spectroscopy in section \ref{sec: data reduction}. We mention the data analysis method to calculate the gas production rates and the dust proxy parameter, $Af\rho$ in \ref{Haser} and \ref{dust}. The major spectroscopic and photometric results are discussed in section \ref{sec: low spectrometry} and \ref{sec: Broad Photometry}. The discussions on both techniques are given in \ref{discussion}.

\section{Observations}\label{sec: Method}
We obtained low-resolution optical spectroscopic and photometric observations using the 2-m Himalayan Chandra Telescope (HCT). We also used the 1.2-m PRL Mount Abu telescope to obtain low-resolution spectra. Additionally, we observed the comet using broadband photometric observations and comet Hale-Bopp (HB) narrow-band filters \citep{Farnham2000} mounted on TRAPPIST (Transiting Planets and Planetesimals Small Telescope). 
\subsection{Himalayan Chandra Telescope (HCT)}\label{HCT}
Himalayan Chandra Telescope (HCT)\footnote{\url{https://www.iiap.res.in/?q=telescope_iao}} is a 2-m Ritchey-Chretien, Cassegrain focus, $f/9$ telescope situated at ( Lat: 32.78 deg N; Long 75.96 deg E; Alt: 4500 m) the Indian Astronomical Observatory (IAO), in Hanle, India \citep{HCT_paper}. The telescope is operated by the Indian Institute of Astrophysics (IIA). The plate scale of the HFOSC instrument is 0.296 arcsec pixel$^{-1}$. The active area of the CCD is 2048 pixel $\times$ 4096 pixel, which gives the field of view of $\sim$ 10 arcmin $\times$ 20 arcmin. We used HFOSC in both photometric and spectroscopic modes. For spectroscopy, we have taken observations in the CCD dimensions of 1500 pixel $\times$ 3500 pixel to get the spatial coverage of the 11 arcmin across the cometary coma. We used Grism 7, which provides a resolving power (R=$\Delta \lambda/ \lambda$) of 1330 and wavelength coverage of 3800-6840 \AA. We also used Grism 8 for one observation to increase the spectral range and look for any absorption/emission at a redder wavelength. Grism 8 has a medium resolving power of 2190 with a spectral range of 5800-8350 \AA. The comet is observed with a slit having a width of 1.92 arcsec and a length of 11 arcmin, while the standard stars used for flux calibration are observed with a slit of width 15.4 arcsec and a length of 11 arcmin. The slit orientation is in the East-West (E-W) direction. We used the keystone mode mentioned in \citet{Aravind_Borisov} to make sure that the telescope was tracking the comet throughout the exposure. The \textsc{NASA JPL HORIZONS}\footnote{\url{https://ssd.jpl.nasa.gov/horizons.cgi.}} was used to obtain the comet ephemerides at each epoch. We have taken a separate sky frame after moving the telescope 1 degree away from the photocenter in the direction of the comet's declination motion. We observed the standard stars BD+284211 (sdO) \citep{oke_1990} and HD74721 (A0V) \citep{HD_1918,barnes_1984} with the wider slit to avoid slit losses. We also observed HD19445, a solar analog (G2V) star, to correct for the comet's continuum. Zero exposure frames were taken to correct the CCD bias, lamp flats to correct the pixel-to-pixel response, and FeNe (iron-neon) and FeAr (iron-argon) lamp spectra for wavelength calibration of the Gr8 and Gr7 spectra, respectively.

For the photometry, we used the CCD with dimensions of 2048 pixel $\times$ 2048 pixel, which provides a field of view of 10 arcmin $\times$ 10 arcmin. We observed the comet with the Bessel B, V, R, and I filters. We also observed the Landolt standard field, RU149 \citep{Landolt_1992}, at each epoch in all the filters. The bias and twilight flats were obtained regularly. The average seeing in the photometric and spectroscopic observations from HFOSC was 2-2.5 arcsec.

\subsection{PRL Mount Abu Observatory}\label{PRL}
\begin{figure}
    \centering
    \includegraphics[width=\linewidth]{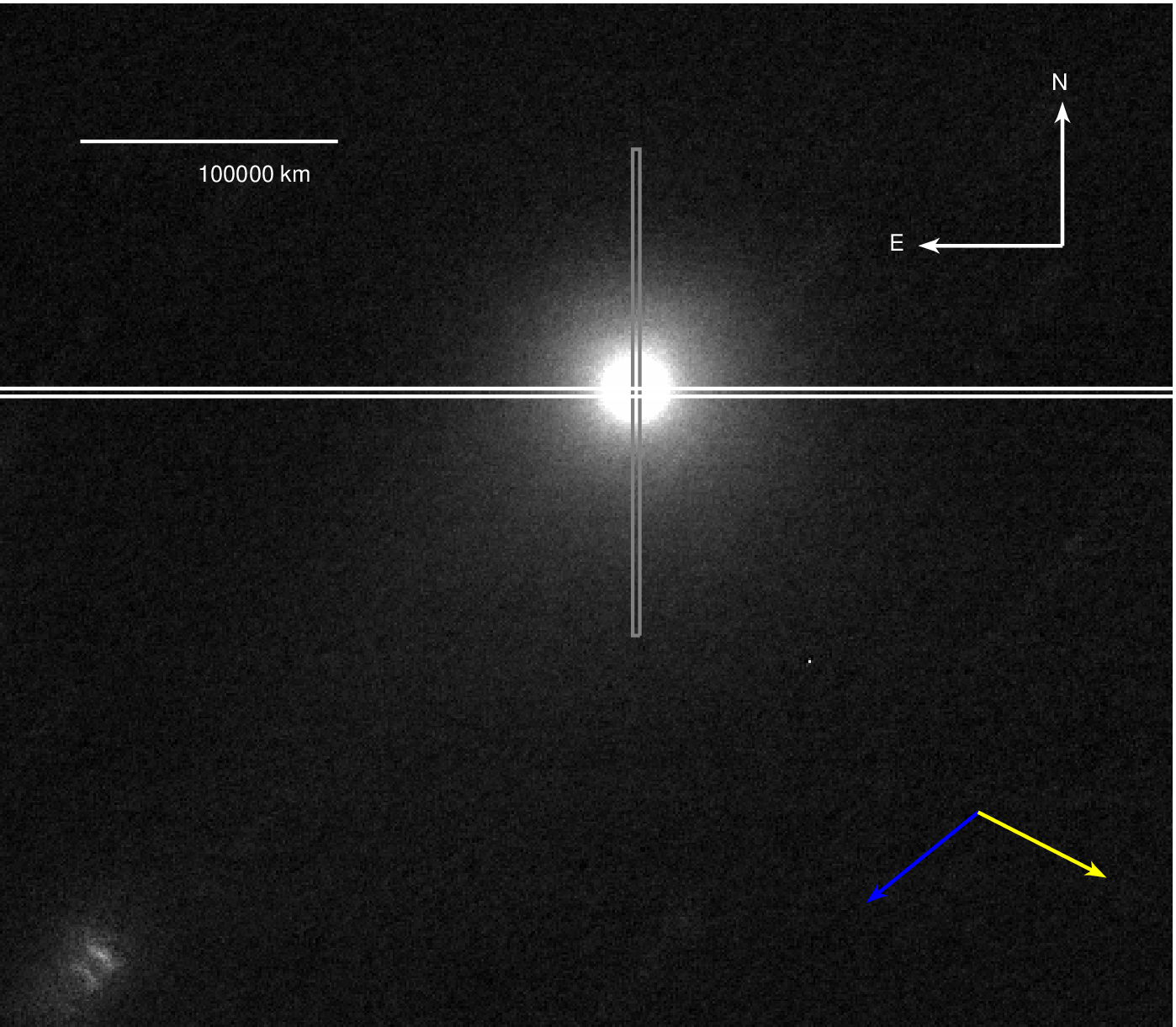}
    \caption{Slits corresponding to HFOSC (\textit{white}) and LISA (\textit{grey}) are marked on the comet V2 image observed using HFOSC instrument. The dust-tail orientation and the Sun's direction are marked in \textit{blue} and \textit{yellow}.}
    \label{fig: slit}
\end{figure}

Another low-resolution spectrograph, LISA, was used at the PRL 1.2-m f/13 Mount Abu telescope at Mount Abu (Lat.: 24.65 N; Long.: 72.78 E, Alt: 1680m)\footnote{\url{https://www.prl.res.in/~miro/telescopes.html}}. LISA provides a wavelength coverage of 3800-7000 \AA. \citet{kumar_2016} and \citet{Aravind_JAA} have provided further details of this instrument. Since LISA works best on an f/5 telescope, we include a focal reducer at the telescopic entrance of the instrument. The plate scale of the instrument is 0.326 arcsec pixel$^{-1}$. The long slit is 2 arcmin in length and 1.76 arcsec in width, having an orientation in the North-South (N-S) direction. Since we do not have the photometric observation on LISA, we have used the photometric image obtained from HFOSC and added the slits of both LISA and HFOSC instruments as shown in Fig. \ref{fig: slit}. We collected bias frames, flat frames to correct for the pixel response, and ArNe (Argon-Neon) lamps to wavelength calibrate the spectra. We observed standard stars HD74721 (A0V) for flux calibration and HD19445 (G2V) \citep{HD_1918} to remove the solar continuum from the comet spectrum. For the LISA observations, the seeing was around 1.8 arcsec. The details of the observations are given in Table \ref{sec: observation_details}.

\begin{table*}
\caption{Details of observations from Indian telescopes}
\setlength{\tabcolsep}{3.5pt}
\begin{threeparttable}
\resizebox{\textwidth}{!}{%
\begin{tabular}{lccccccccc}
\hline
\multicolumn{1}{|c|}{{Date}} &
\multicolumn{1}{|c|}{{DTP\tnote{a}}} &
  \multicolumn{1}{c|}{{r\tnote{b}}}&
  \multicolumn{1}{c|}{{$\Delta$\tnote{c}}} &
  \multicolumn{1}{c|}{{Time}}&
  \multicolumn{1}{c|}{{Telescope}}&
  \multicolumn{1}{c|}{{Instrument}}&
  \multicolumn{1}{c|}{{Observation}}&
  \multicolumn{1}{c|}{{Phase}}&
  \multicolumn{1}{c|}{{Airmass}}\\
\multicolumn{1}{|l|}{} &
\multicolumn{1}{|c|}{(Days)} &
  \multicolumn{1}{c|}{{(au)}} &
  \multicolumn{1}{c|}{{(au)}} &
  \multicolumn{1}{c|}{{(UT)}} &
  \multicolumn{1}{c|}{{Facility}} &
  \multicolumn{1}{c|}{{}} &
  \multicolumn{1}{c|}{{Type}} &
  \multicolumn{1}{c|}{{Angle ($^{\circ}$)}} &
  \multicolumn{1}{c|}{{}}\\\hline
04-10-2022 & -215 & -3.27 & 3.58 & 22:40 & HCT & HFOSC & Imaging (R), Spectroscopy (Gr7) & 15.96 & 1.96\\
22-11-2022 & -166 & -2.91 & 2.54 & 20:16 & HCT & HFOSC & Imaging (B, V, R), Spectroscopy (Gr7, Gr8) & 19.38 & 1.81\\
16-12-2022 & -142 & -2.75 & 2.17 & 14:59 & HCT & HFOSC & Imaging (B, V, R), Spectroscopy (Gr7) & 18.70 & 1.99\\
22-12-2022 & -136 & -2.72 & 2.12 & 18:08 & PRL & LISA & Spectroscopy & 18.76 & 2.10\\
21-09-2023 & 136 & 2.71 & 1.86 & 20:10 & HCT & HFOSC & Spectroscopy (Gr7) & 13.68 & 1.93\\\hline
\end{tabular}
}
\begin{tablenotes}
\footnotesize
\item[a] Days To Perihelion.
\item[b] Heliocentric Distance.
\item[c] Geocentric Distance.
\end{tablenotes}
\end{threeparttable}
\label{sec: observation_details}
\end{table*}

\subsection{TRAPPIST}\label{TRAPPIST}
TRAPPIST (Transiting Planets and Planetesimals Small Telescope) is a network of two 60-cm robotic telescopes situated in two different hemispheres \footnote{\url{https://www.trappist.uliege.be/cms/c_5006023/fr/trappist}}. TRAPPIST-North is built at the Oukaimeden Observatory in the Atlas Mountains in Morocco \citep{TRAPPIST_North_Barkaoui_2017} while TRAPPIST-South is located in the Chilean mountains
at the La Silla Observatory of ESO \citep{JEHIN_TRAP}. These telescopes are remotely controlled. The telescopes are equipped with the cometary Hale-Bopp (HB) narrow-band filter set to sample different molecular gaseous emissions such as OH, NH, CN, C$_2$, and C$_3$. The telescopes also have dust continuum filters, i.e., BC, GC, and RC for blue, green, and red continuum. The dust-continuum filters are emission-free filters that are used to obtain the proxy dust production parameter and remove the dust contamination from the gas filters. These narrow-band filters were built to observe the emissions only, unlike broadband filters, which cover a wide range of the spectrum. The details of these filters are mentioned in \citet{Farnham2000}. The telescopes also have broad-band Bessel filters, B, V, R, and I. TRAPPIST-South is equipped with a 2K $\times$ 2K FLI Proline CCD with a pixel scale of 0.65 arcsec pixel$^{-1}$, resulting in a field of view of 22 arcmin $\times$ 22 arcmin and TRAPPIST-North with an Andor IKONL BEX2 DD (0.59 arcsec pixel$^{-1}$) with a 20 arcmin $\times$ 20 arcmin field of view. The seeing varied between 1.5-3 arcsec during the observations from TRAPPIST telescopes.

\section{Data Reduction}\label{sec: data reduction}
In the following section, we discuss the different steps of data reduction performed on the spectrometric data (see section \ref{sec: data red -spec}) and the photometric data (see section \ref{sec: data red -phot}).
\subsection{Spectroscopic Data Reduction}\label{sec: data red -spec}
Having obtained the spectroscopic data from different telescopes, we now have to obtain the final calibrated frame before carrying out further calculations. The details about the data reduction are explained in the same order as mentioned in section \ref{sec: Method}. For instruments like HFOSC on HCT and LISA on Mount Abu Observatory, the data reduction codes have been self-scripted in \textsc{Python} for bias correction and flat-fielding. Since HCT is located at 4500m, the frames are heavily affected by cosmic rays. We have used the LA Cosmic package developed by \citet{LACosmic} to remove the cosmic ray contamination. 

Using self-scripted \textsc{Python} code, we have extracted a one-dimensional spectrum after removing the background from the standard star, solar analog star, and comet V2. In the case of a star, we create the background by taking the median values of the counts of at least 50 pixels away from both sides of the photocenter and subtracting it from the star spectrum. In the case of comets, we subtracted the separate sky frame from the comet frame. 

After extracting the one-dimensional spectra, different \textsc{IRAF} packages, such as \texttt{identify} and \texttt{dispcor}, have been used to create the wavelength solution by identifying the lines of the arc lamps. After that, \texttt{standard} and \texttt{sensfunc} have been employed to make the sensitivity function. Finally, the \texttt{calibrate} package has been applied to obtain the flux-calibrated data.

The final step is to remove the solar continuum spectrum from the comet's spectrum. For this, we use the method explained in \citet{continuum_removal}. The first step is to normalize the flux-calibrated comet's spectrum with the spectrum of the solar analog star. Then, to correct for the reddening, we use the continuum windows mentioned in \citet{continuum_removal} to make the comet and solar polynomials, then take the ratio of these two polynomials and multiply with the solar spectrum to get the synthetic dust spectrum. Finally, we subtract this spectrum from the comet spectrum to get the continuum-corrected, flux-calibrated comet spectrum.

This is the general method followed for reducing the two-dimensional raw data to get the final one-dimensional spectrum to be used for analysis. However, if the slit is not wide enough, then other factors also need to be considered, such as the slit correction and the atmospheric dispersion correction, which will be discussed in the following sub-subsections.

\subsubsection{Slit Correction}
As we have mentioned in section \ref{PRL}, the LISA slit size is 1.76 arcsec; therefore, in conditions of bad seeing, the complete flux of a standard star will not enter the slit, affecting the sensitivity function of the instrument. This results in the overestimation of the object's absolute flux. Hence, to correct for it, we must multiply our object's flux-calibrated spectrum by the slit-correction factor. The slit correction factor is based on calculating the FWHM by fitting a Gaussian profile along a spatial axis. It is then used to calculate the error function. This function varies with wavelength. The complete details are given by \citet{Lee_Pak}. This correction factor was not required while reducing the HCT data because the slit used to observe the standard star was wide enough (15.4 arcsec).

\subsubsection{Atmospheric Dispersion Correction}
We also need to correct LISA spectra for the atmospheric dispersion, which depends on the star's airmass, the dry pressure at the observatory elevation (in mm hg), the water vapour pressure (in mm hg), and the temperature (in Celsius). The dispersion causes the star's light at different wavelengths to get shifted by some arcseconds with respect to the central wavelength at 5000 \AA, which is responsible for the low photon counts at the bluer and redder sides of this central wavelength. The detailed methods are given in \citet{Filippenko_1982} and \citet{Stone_1996}.

\subsection{Photometric Data Reduction}\label{sec: data red -phot}
The photometric datasets come from two telescopes, HCT and TRAPPIST. We describe the data reduction of the two different telescopes below:

\subsubsection{HFOSC Photometric Data Reduction}
To reduce the images observed from HFOSC, we used self-scripted \textsc{Python} codes to create the master-bias and master-flat files. We then subtracted the master bias and divided the master flat files from all the comet and field star frames. The \textsc{Photutils} module from the \textsc{Astropy} package \citep{Astropy_2013} was used to calibrate the standard star frames. We took the extinction coefficients from \citet{Stalin_2008} for the location of HCT. The apertures, which were 2-3 $\times$ Full Width Half Maximum (FWHM), were used to measure the magnitudes of the standard field stars and calculate the zero point. The computed zero point is further used to obtain the magnitude of the comet for an aperture size of 5 arcsec.

\subsubsection{TRAPPIST Photometric Data Reduction}
Comet V2 was observed using both narrowband and broadband filters with the TRAPPIST-North (TN) and TRAPPIST-South (TS) telescopes \citep{JEHIN_TRAP}. The observation campaign was conducted from January 11 2022 (r=5.41 au, pre-perihelion) to August 13 2024 (r = 5.26 au, post-perihelion), covering the comet's evolution over 32 months around perihelion to analyse long-term variations in the gas and dust activity. For data reduction, we applied standard procedures, including bias, dark, and flat-field corrections, followed by sky contamination removal and flux calibration using standard stars. Absolute flux calibration was performed when conditions were photometric by observing standard stars defined for the Comet Hale-Bopp (HB) narrow-band filter set \citep{Farnham2000}.

\section{Data Analysis}\label{sec: Data Analysis}
In this section, we have used the continuum corrected spectrum to calculate the gas production rates of different molecules such as CN, C$_2$, and C$_3$ using Haser modelling \citep{1957BSRSL..43..740H} (details provided in \ref{Column} and \ref{Haser}) and the dust proxy parameter Af$\rho$ (details provided in \ref{dust}).
\subsection{Column Density Profile}\label{Column}
After obtaining the final reduced (wavelength calibrated, flux calibrated and continuum corrected) spectra, we calculated the column density along the slit direction (y) using the method in section 5 of the paper by \citet{Langland-Shula2011}.
\begin{equation}
    N(y)=\frac{4\pi}{g}\frac{F(y)}{\Omega}.
	\label{eq: Column Density}
\end{equation}

Here, $g$ is the fluorescence efficiency of the molecules, and $\Omega$ is the solid angle defined by the multiplication of the slit width and pixels extracted in the spatial axis (\textit{p}). To find \textit{p}, we measured the Half Width Half Maxima (HWHM) of the standard star along the spatial direction and used it to bin the pixels. $F$ is the integrated flux along the slit direction of different molecules integrated at the wavelength range mentioned in Table 5 of \citet{Langland-Shula2011}. The values of $g$ for different molecules are given in Table 6 of \citet{Langland-Shula2011}. The fluorescence efficiency factor has an inverse relationship with the heliocentric distance ($g \propto r_h^{-n}$). In the case of CN, due to the Swings effect \citep{Swings_1941}, the fluorescence efficiency depends on the comet's velocity with respect to the Sun. Based on the Swings effect, \citet{Schleicher_2010} has given the method to calculate the $g$-factor using the double interpolation of the heliocentric distance and velocity. Using the values of different parameters, we calculate the column density from equation (\ref{eq: Column Density}).

\subsection{Haser Model}\label{Haser}
The observed column density described in section \ref{Column} is then compared with the theoretical column density generated using Haser's model given by \citet{1957BSRSL..43..740H}.

Haser's model is a simple two-component model in which gas is assumed to sublimate in a spherically symmetric way at a constant velocity from the nucleus of radius R. It considers that the daughter molecules come out as the only product of the parent molecules. The number density of daughter molecules in the coma is given by: 
\begin{equation}
    n(R)=\frac{Q}{4\pi v_{flow}R^{2}}\frac{\beta_0}{\beta_0 - \beta_1}(e^{-\beta_1R} -e^{-\beta_0R})
	\label{eq: Theoretical Column Density-I}
\end{equation}
Where $\beta_0$ and $\beta_1$ are the inverse scale lengths of the parent and daughter molecules. We have used the scale lengths given by \citet{AHEARN1995223}. $v_{flow}$ is the outflow velocity of the daughter molecules, which has been considered to be 1 km s$^{-1}$ and $Q$ is the production rate of the daughter molecules. 

The density will be integrated along the line of sight distance $z$, which is related to the nucleus size R and slit aperture $y$ along the spatial axis. It is given by $R^2 = y^2 + z^2$, if the comet is placed at the centre of the slit $x = 0$. Therefore, we get:
\begin{equation}
    N(y)=\int_{-\infty}^{\infty} (n(R))\,dz
	\label{eq: Theoretical Column Density-II}
\end{equation}
On integrating it, we get:
\begin{equation}
    N(y)=\frac{Q}{2\pi v_{flow}}\frac{\beta_0}{\beta_0 - \beta_1} \int_{0}^{\infty} (e^{-\beta_1R} - e^{-\beta_0R})\,dz
	\label{eq: Theoretical Column Density-III}
\end{equation}

The chi-square minimisation between the observed column density and the theoretical column density was used to obtain the production rates of different molecules. For TRAPPIST, the emission from the gaseous species OH, CN, C$_3$, and C$_2$ were measured using HB narrowband filters, and their production rates were calculated using the Haser model. The model fitting was performed at a projected distance of 10,000 km from the nucleus. Similarly, for the long-slit spectra, the HB filter passband for CN, C$_3$, and C$_2$ were used to calculate the production rates of these molecules from 10000 km onwards.

When the molecular emission is not strong in the comet spectrum, there is difficulty in removing the continuum, which could add a significant error in the computation of the column density profile, like for this comet. Therefore, we have also used another method described by \citet{Aravind_Borisov} and \citet{Langland-Shula2011}. The method allows us to calculate the number of molecules for a given aperture using the relation given below:
\begin{equation}
    N = \frac{4\pi\Delta^2}{g}\times F
\end{equation}
Here, $g$ is the fluorescence efficiency for different molecules, $\Delta$ is the geocentric distance, and $F$ is the flux calculated for different molecules in an aperture. The number of molecules is then extrapolated to mimic the number of molecules in the cometary coma using the Haser Fraction \citep{Fink_1996} given by Schleicher\footnote{\url{https://asteroid.lowell.edu/comet/haser}}. The total number of molecules is divided by the respective molecular lifetime given in \citet{AHEARN1995223} to get the production rate of different molecules.

\subsection{\texorpdfstring{Dust Proxy Parameter - Af$\rho$}{Dust Proxy Parameter - Afrho}}\label{dust}
The next important parameter in the case of comets is to know the dust activity in the cometary coma. It also helps to calculate the dust-to-gas ratio, a parameter to decide whether the comet is dust- or gas-rich. \citet{AHearn1984} has introduced a parameter $Af\rho$, which is a proxy of the dust production rate in comets. $Af\rho$ is a product of the bond Albedo ($A$), the filling factor of grains in our field of view ($f$), and the aperture radius $\rho$. Filling factor, $f$ is given by:
\begin{equation}
    f=\frac{N(\rho)\sigma}{\pi\rho^2}
	\label{eq: filling factor}
\end{equation}
where $N(\rho)$ is the number of grains present in a circular aperture of radius $\rho$, $ \sigma$ is the area of a single grain, and $A$ is a dimensionless quantity that depends on the scattering or phase angle $\theta$. As $N(\rho)$ is a dimensionless quantity, $\sigma$ has dimension of $cm^{2}$ and $\rho$ has a dimension of $cm$, $f$ becomes a dimensionless quantity which makes $Af\rho$ depend only on the dimension of $\rho$ i.e. in $cm$.

\cite{AHearn1984} have given the details to calculate this parameter. They have defined the total luminosity of a comet as:
\begin{equation}
    L=AN(\rho)\sigma \frac{F_{\odot}}{r^2}
	\label{eq: Luminosity_comet-I}
\end{equation}
where $A$, $\sigma$ are defined earlier, $F_{\odot}$ is the solar flux at 1 au, and $r$ is the heliocentric distance, given in au. Using the definition of filling factor given by equation (\ref{eq: filling factor}), we can get:
\begin{equation}
    L= Af\pi \rho^2 \frac{F_{\odot}}{r^2}
	\label{eq: Luminosity_comet-II}
\end{equation}
Also, the luminosity of the comet can be defined as:
\begin{equation}
    L= 4 \pi \Delta^2 F_{com}
	\label{eq: Luminosity_comet-III}
\end{equation}
where $F_{com}$ is the flux of a comet within the continuum region, i.e., Blue-Continuum (BC), Green-Continuum (GC), and Red-Continuum (RC), defined by \citet{Farnham2000}.
From equation (\ref{eq: Luminosity_comet-III}) and (\ref{eq: Luminosity_comet-II}), we can get:
\begin{equation}
    Af =  \frac{(2\Delta r)^2}{\rho^2} \frac{F_{com}}{F_{\odot}}
	\label{eq: Luminosity_comet-IV}
\end{equation}
or, the dust proxy parameter $Af\rho$ is,
\begin{equation}
    Af\rho =  \frac{(2\Delta r)^2}{\rho^2} \frac{F_{com}}{F_{\odot}} \rho
	\label{eq: Luminosity_comet-V}
\end{equation} 
For narrowband photometry, the Af$\rho$ parameter was derived from the dust continuum profiles using HB cometary dust filters BC, GC, and RC. The phase angle correction to $0^{\circ}$ was applied using the details mentioned in \cite{Schleicher2007} to obtain A(0)f$\rho$. The main uncertainties in production rates and Af$\rho$ stem from absolute flux calibration and sky background subtraction. An uncertainty of 5\% was considered for extinction coefficients, which is negligible at low airmass but becomes significant at high airmass. Sky background uncertainty was estimated at the three-sigma level and incorporated into the final uncertainties on production rates. The total errors reported in the following sections are a quadratic combination of sky background and extinction coefficient uncertainties.

For long-slit spectroscopy, we have summed three to five pixels along the spatial axis based on seeing. We used the geometric correction factor given by \citet{Langland-Shula2011} to convert the long slit into a circular aperture. The full disk flux in different continuum bands, such as BC and GC, is convolved with the filter profile of these bands. Using the equation (\ref{eq: Luminosity_comet-V}), we get the value of $Af\rho$ at a particular phase angle value, $\theta$. and normalised to $0^{\circ}$. It is done by dividing the value of $Af\rho$ by the phase function $S(\theta)$ given by \citet{SCHLEICHER_Afp}. We have used the online phase function table on the Schleicher website \footnote{\url{https://asteroid.lowell.edu/comet/dustphase/table}}.

\section{Results}\label{sec: low spectrometry}
The results obtained using the spectroscopic and photometric observations are given below:
\subsection{Spectroscopic Results}
We have taken spectroscopic observations at six different epochs (ref: table \ref{sec: observation_details}), four from the HFOSC instrument on the 2 m HCT and two from the LISA instrument mounted on the PRL 1.2 m telescope. The details of the HFOSC and LISA instruments are mentioned in sections \ref{HCT} and \ref{PRL}. Out of the six spectra, five spectra are observed pre-perihelion, and one spectrum is taken post-perihelion.

\begin{figure}
    \centering
    \includegraphics[width=\linewidth]{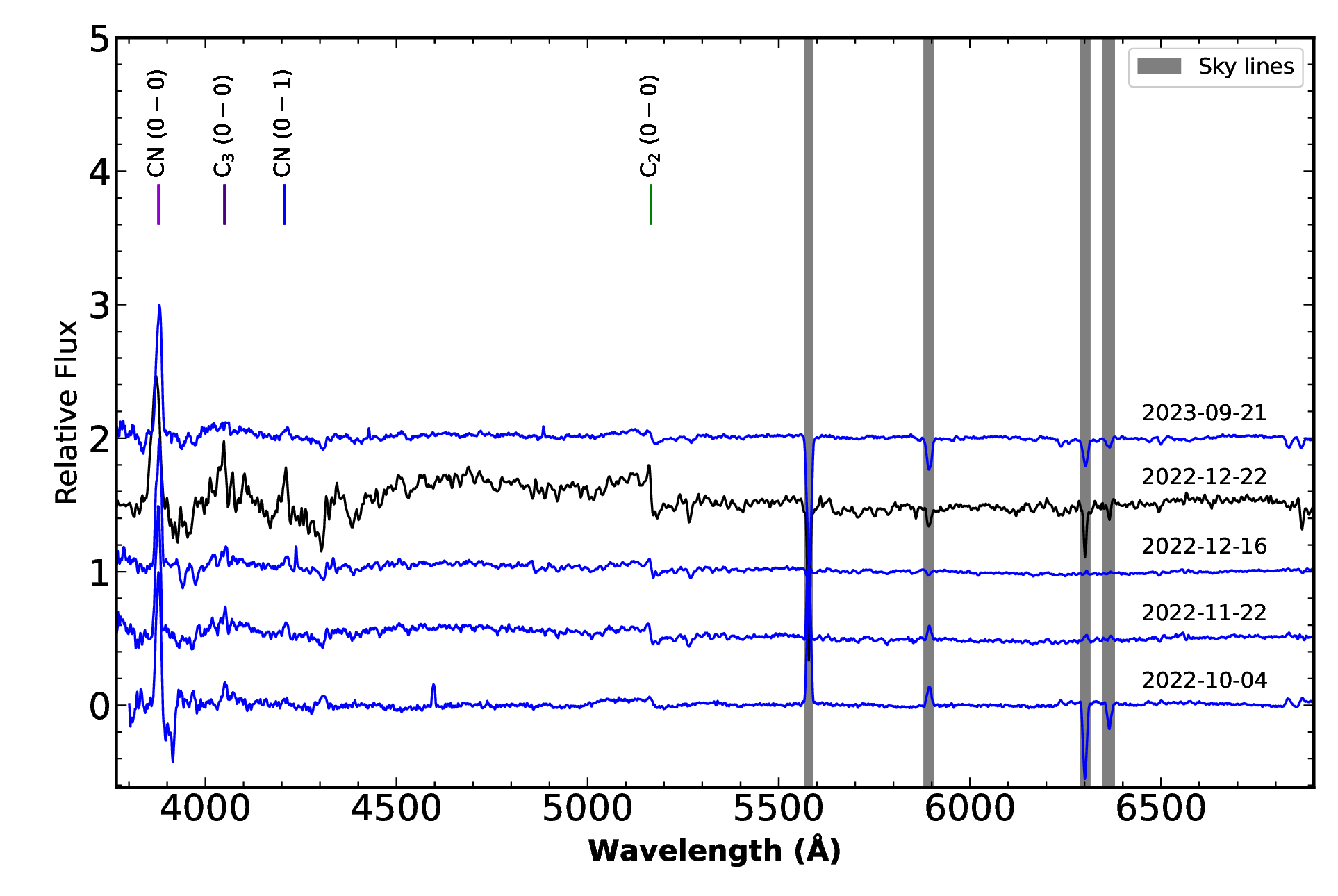}
    \caption{Spectra of comet V2 (ZTF) observed from HFOSC/HCT (in blue) and LISA/PRL (in black). Prominent emission bands are marked, and the skylines are masked. The spectrum is normalised and spaced by 0.5 relative flux units for clarity.}
    \label{spec_V2}
\end{figure}

We have reduced the low-resolution spectroscopic data sets using the methods given in section \ref{sec: data red -spec} and calculated the production rates of different molecules such as CN $(0-0)$, C$_2 (\Delta=0)$, and C$_3(0-0)$. The prominent molecular emission bands are marked in Fig. \ref{spec_V2}. The details of the production rates are given in Table \ref{sec: Production rate- II}.

\begin{table*}
\caption{Spectroscopic molecular gas production rates, Af$\rho$ in Blue-Continuum (BC) and Green-Continuum (GC) are calculated using observations from HFOSC/HCT and LISA/PRL. Broad-band photometric calculations are done using observations from HFOSC/HCT}
\setlength{\tabcolsep}{2.5pt}
        \resizebox{\textwidth}{!}{%
\begin{tabular}{lcrccccccccc}
\hline
\multicolumn{1}{|c|}{{Date}} &
  \multicolumn{1}{c|}{{DTP}} &
  \multicolumn{1}{c|}{{r}} &
  \multicolumn{1}{c|}{{$\Delta$}} &
  \multicolumn{3}{c|}{{Production Rates ($\times$ 10$^{25}$ molecules s$^{-1}$)}} &
  \multicolumn{2}{c|}{{A($\theta=0^{\circ}$)f$\rho$ ($\times$ 10$^{3}$ cm)}} &
  \multicolumn{3}{c|}{{Magnitudes (Dimensionless)}}\\
\multicolumn{1}{|l|}{} &
  \multicolumn{1}{c|}{{(Days)}} &
  \multicolumn{1}{c|}{{(au)}} &
  \multicolumn{1}{c|}{{(au)}} &
  \multicolumn{1}{c|}{{Q(CN)}} &
  \multicolumn{1}{c|}{{Q(C$_2$)}} &
  \multicolumn{1}{c|}{{Q(C$_3$)}} &
  \multicolumn{1}{c|}{{BC}} & 
  \multicolumn{1}{c|}{{GC}} &
  \multicolumn{1}{c|}{{B}} & 
  \multicolumn{1}{c|}{{V}} &
  \multicolumn{1}{c|}{{R}}\\\hline
04-10-2022 & -215 & -3.27 & 3.58 & 14.7 $\pm$ 1.0 & 6.9 $\pm$ 1.2 & 2.1 $\pm$ 1.3 & 4748 $\pm$ 2128 & 6623 $\pm$ 1703 & -- & -- & 13.56 $\pm$ 0.09\\
22-11-2022 & -166 & -2.91 & 2.54 & 19.3 $\pm$ 1.3 & 13.8 $\pm$ 3.9 & 5.1 $\pm$ 2.2 & 11274 $\pm$ 552 & 12192 $\pm$ 393 & 14.47 $\pm$ 0.10 & 13.45 $\pm$ 0.10 & 12.93 $\pm$ 0.10\\
16-12-2022 & -142 & -2.75 & 2.17 & 18.9 $\pm$ 1.0 & 13.9 $\pm$ 1.7 & 2.3 $\pm$ 1.0 & 9539 $\pm$ 364 & 11529 $\pm$ 268 & 13.89 $\pm$ 0.06 & 12.93 $\pm$ 0.10 & 12.43 $\pm$ 0.10\\
22-12-2022 & -136 & -2.71 & 2.12 & 18.0 $\pm$ 0.9 & 13.8 $\pm$ 2.9 & 2.5 $\pm$ 0.6 & 9108 $\pm$ 561 & 7277 $\pm$ 561 & -- & -- & --\\
21-09-2023 & 136  & 2.71 & 1.86 & 7.4 $\pm$ 0.4 & 4.9 $\pm$ 1.5 & 2.0 $\pm$ 0.7 & 3884 $\pm$ 230 & 4752 $\pm$ 153 & -- & -- & --\\
\hline
\end{tabular}
}
\label{sec: Production rate- II}
\end{table*}

\subsection{Photometric Results}\label{sec: Broad Photometry}
In Table \ref{sec: Production rate- I}, we mention the molecular gas production rates like CN($0-0$) at 3880 \AA, C$_2$($\Delta \nu = 0$) at 5165\AA, C$_3$(0-0) at 4050 \AA, and OH ($0-0$) at 3097 \AA. The images through the CN, C$_{2}$, C$_{3}$ gas filters are shown in the Appendix (Fig. \ref{CN_2211}, \ref{C2_2211} and \ref{C3_2211}). These molecular production rates were calculated using observations from the TRAPPIST-North. We have also derived the dust proxy parameter (Af$\rho$) defined in section \ref{dust}. The variation of production rates of different molecules are shown in the Appendix (Fig. \ref{QOH}, \ref{QCN}, \ref{QC2} and \ref{QC3}) and the variation of the dust proxy parameter are plotted in the Appendix (Fig. \ref{A_BC}, \ref{A_GC}, \ref{A_RC} and \ref{A_R}).

\begin{figure*}
    \centering
    \includegraphics[width=0.9\linewidth]{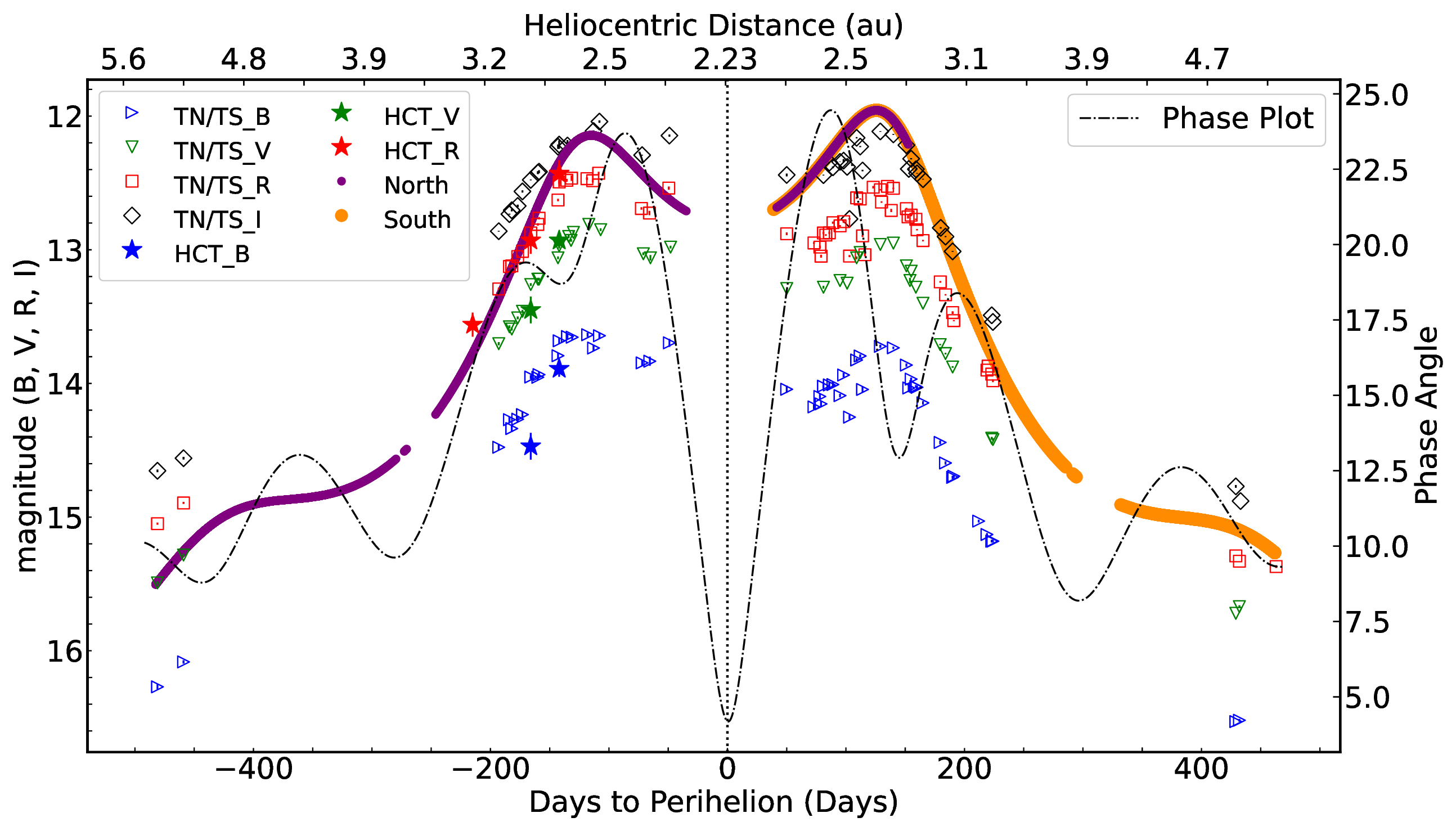}
    \caption{Light curve of comet V2 in B (in blue), V (in green), R (in red), and I (in black) filters from the HCT and TRAPPIST. The availability of the comet for observations from the HCT and TN is shown as purple dots (labelled North). Similarly, orange dots are used for the TS (labelled South). They are marked as per the expected magnitude given by \textsc{NASA JPL HORIZONS}. The vertical dotted line, marked in this figure (and all others), represents the perihelion distance, which is 2.23 au. The dash-dot line represents the variation of the phase angle (Sun-Comet-Earth angle) with heliocentric distance.
    }
    \label{fig: light curve}
\end{figure*}

Table \ref{sec: BB Magnitude} presents the apparent photometric magnitudes of comet V2 in broad-band filters, i.e., B, V, R, and I, from TRAPPIST-North and TRAPPIST-South telescopes, and Table \ref{sec: Production rate- I} mentions the magnitude in broadband B, V, and R filters observed from HCT. We have plotted the light curves in B, V, R, and I filters comprising both HCT and TRAPPIST telescopes in Fig. \ref{fig: light curve}. The dash-dot curve represents the variation of phase angle (Sun - Comet - Observer angle) with heliocentric distance. 
The low phase angles close to perihelion (Fig. \ref{fig: light curve}) show that the comet was behind the Sun, as seen from the Earth, and hence it was not conducive for observations. 

\begin{table*}
\caption{TRAPPIST photometric gas production rates and Af$\rho$ in Blue-Continuum (BC), Green-Continuum (GC), and Red-Continuum (RC), observed from TRAPPIST-North Observatory}
\setlength{\tabcolsep}{3.0pt}
        \resizebox{\textwidth}{!}{%
\begin{tabular}{lcrccccccccr}
\hline
\multicolumn{1}{|c|}{{Date}} &
  \multicolumn{1}{c|}{{DTP}} &
  \multicolumn{1}{c|}{{r}} &
  \multicolumn{1}{c|}{{$\Delta$}} &
  \multicolumn{4}{l|}{\hspace{0.5cm}{Production Rates ($\times$ 10$^{25}$ molecules s$^{-1}$)}} &
  \multicolumn{3}{c|}{{A($\theta=0^{\circ}$)f$\rho$ (cm)}}\\
\multicolumn{1}{|c|}{} &
  \multicolumn{1}{c|}{{(Days)}} &
  \multicolumn{1}{c|}{{(au)}} &
  \multicolumn{1}{c|}{{(au)}} &
  \multicolumn{1}{c|}{{Q(CN)}} &
  \multicolumn{1}{c|}{{Q(C\textsubscript{2})}} &
  \multicolumn{1}{c|}{{Q(C\textsubscript{3})}} &
  \multicolumn{1}{c|}{{Q(OH) $\times$ 10$^{3}$}} &
  \multicolumn{1}{c|}{{BC}} & 
  \multicolumn{1}{c|}{{GC}} & 
  \multicolumn{1}{c|}{{RC}}\\\hline
04-11-2022 & -184 & -3.04 & 2.91 & 11.3 $\pm$ 0.6 & 6.8 $\pm$ 0.5 & -- & 0.7 $\pm$ 0.4 & 8644 $\pm$ 164 & -- & 9561 $\pm$ 43\\
11-11-2022 & -177 & -2.99 & 2.76 & 11.2 $\pm$ 0.6 & 7.8 $\pm$ 0.7 & 1.8 $\pm$ 0.3 & 0.5 $\pm$ 0.6 & 7846 $\pm$ 201 & -- & 9533 $\pm$ 115\\
15-11-2022 & -173 & -2.96 & 2.67 & 12.3 $\pm$ 0.6 & 9.7 $\pm$ 0.8 & 2.9 $\pm$ 0.2 & 1.0 $\pm$ 0.4 & 8155 $\pm$ 166 & -- & 9525 $\pm$ 59\\
22-11-2022 & -166 & -2.91 & 2.53 & 12.8 $\pm$ 0.6 & 9.3 $\pm$ 0.8 & 2.4 $\pm$ 0.2 & 1.3 $\pm$ 0.4 & 8780 $\pm$ 158 & -- & 9783 $\pm$ 50\\
28-11-2022 & -160 & -2.87 & 2.42 & 12.5 $\pm$ 0.6 & 9.8 $\pm$ 0.8 & 2.9 $\pm$ 0.2 & 1.0 $\pm$ 0.5 & 8360 $\pm$ 153 & -- & 9940 $\pm$ 47\\
29-11-2022 & -159 & -2.86 & 2.41 & 12.9 $\pm$ 0.6 & 8.1 $\pm$ 0.6 & 2.2 $\pm$ 0.2 & 1.0 $\pm$ 0.4 & 8585 $\pm$ 158 & -- & 10674 $\pm$ 41\\
15-12-2022 & -143 & -2.76 & 2.18 & 16.0 $\pm$ 0.9 & 10.3 $\pm$ 1.0 & 3.6 $\pm$ 0.3 & -- & 7834 $\pm$ 164 & -- & 9090 $\pm$ 67\\
16-12-2022 & -142 & -2.75 & 2.17 & 15.7 $\pm$ 0.9 & -- & -- & 2.0 $\pm$ 0.7 & 9117 $\pm$ 192 & -- & --\\
23-12-2022 & -136 & -2.71 & 2.11 & 16.2 $\pm$ 0.8 & 11.8 $\pm$ 1.0 & 2.7 $\pm$ 0.3 & 2.4 $\pm$ 0.8 & 8635 $\pm$ 183 & 9224 $\pm$ 116 & 10435 $\pm$ 54\\
27-12-2022 & -132 & -2.68 & 2.08 & 17.4 $\pm$ 1.0 & 14.7 $\pm$ 0.9 & 3.9 $\pm$ 0.3 & 2.1 $\pm$ 0.7 & 8789 $\pm$ 179 & 9190 $\pm$ 108 & 10900 $\pm$ 60\\
09-01-2023 & -119 & -2.60 & 2.07 & 18.0 $\pm$ 0.8 & 17.8 $\pm$ 1.1 & 3.9 $\pm$ 0.3 & 3.4 $\pm$ 0.8 & 8713 $\pm$ 166 & 9200 $\pm$ 126 & 10543 $\pm$ 103\\
14-01-2023 & -114 & -2.58 & 2.08 & 18.4 $\pm$ 0.8 & 15.7 $\pm$ 1.1 & 4.3 $\pm$ 0.3 & 4.2 $\pm$ 1.0 & 8257 $\pm$ 180 & 8724 $\pm$ 113 & 10318 $\pm$ 100\\
19-01-2023 & -109 & -2.55 & 2.11 & 20.0 $\pm$ 0.8 & 16.1 $\pm$ 1.0 & 3.7 $\pm$ 0.3 & 4.2 $\pm$ 0.8 & 8504 $\pm$ 187 & -- & 10494 $\pm$ 153\\
03-03-2023 & -66 & -2.35 & 2.65 & 18.0 $\pm$ 0.7 & 17.0 $\pm$ 1.0 & 3.8 $\pm$ 0.3 & 6.9 $\pm$ 1.4 & 7246 $\pm$ 219 & 7475 $\pm$ 146 & 8927 $\pm$ 82\\
18-08-2023 & 103 & 2.52 & 2.08 & 8.8 $\pm$ 0.3 & 11.6 $\pm$ 0.7 & -- & -- & 3616 $\pm$ 61 & -- & 5520 $\pm $33\\
29-08-2023 & 114 & 2.58 & 1.95 & 8.6 $\pm$ 0.4 & 10.0 $\pm$ 0.6 & 1.6 $\pm$ 0.2 & -- & 4905 $\pm$ 120 & 5355 $\pm$ 71 & 6097 $\pm$ 36\\
19-09-2023 & 134 & 2.70 & 1.86 & 8.2 $\pm$ 0.5 & 6.8 $\pm$ 0.8 & -- & 0.3 $\pm$ 0.4 & 6258 $\pm$ 57 & -- & --\\
23-09-2023 & 138 & 2.72 & 1.87 & 7.3 $\pm$ 0.5 & 7.6 $\pm$ 0.8 & -- & 0.3 $\pm$ 0.5 & 5615 $\pm$ 151 & -- & 6630 $\pm$ 50\\
07-10-2023 & 152 & 2.81 & 1.99 & 6.2 $\pm$ 0.6 & 4.0 $\pm$ 0.7 & -- & -- & 5902 $\pm$ 176 & -- & 6553 $\pm$ 48\\\hline
\end{tabular}
}
\label{sec: Production rate- I}
\end{table*}

\section{Discussion}\label{discussion}

\subsection{Production Rates and Rate Ratios}\label{Spec_discussion}

\begin{figure}
    \centering
    \includegraphics[width=1.0\linewidth]{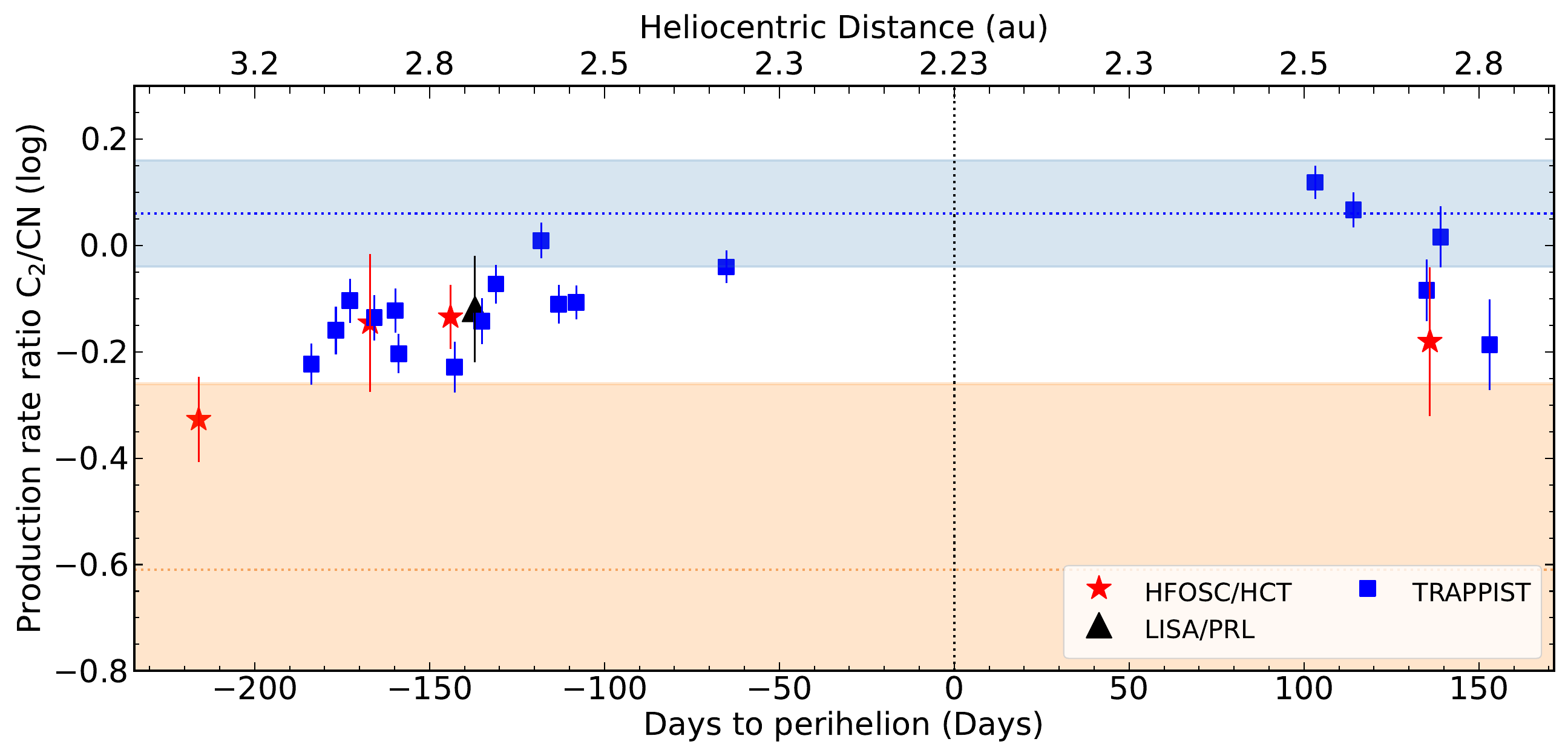}
    \caption{Variation of C$_2$/CN with the heliocentric distance. We have marked the 1-$\sigma$ range of the typical region with \textit{blue} and the 1-$\sigma$ range of the depleted region with \textit{orange}. The observation from TRAPPIST is marked as \textit{blue square}, the observation from HFOSC/HCT is marked as \textit{red star}, and the observation from LISA/PRL is marked as \textit{black triangle}.}
    \label{C2_CN}
\end{figure}
\begin{figure}
    \centering
    \includegraphics[width=1.0\linewidth]{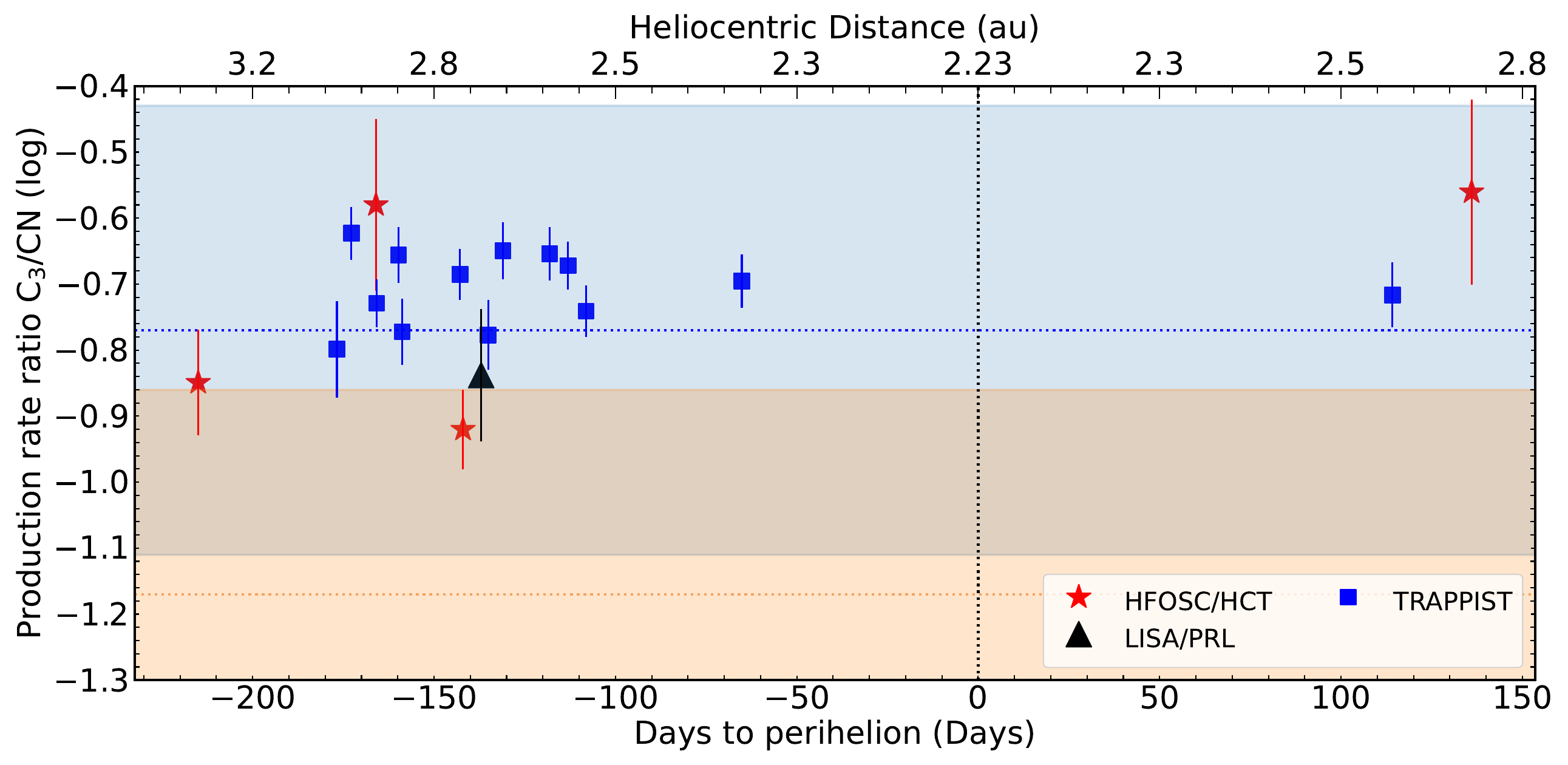}
    \caption{Variation of C$_3$/CN with the heliocentric distance. We have marked the 1-$\sigma$ range of the typical region with \textit{blue} and the 1-$\sigma$ range of the depleted region with \textit{orange}.}
    \label{C3_CN}
\end{figure}
\begin{figure}
    \centering
    \includegraphics[width=1.0\linewidth]{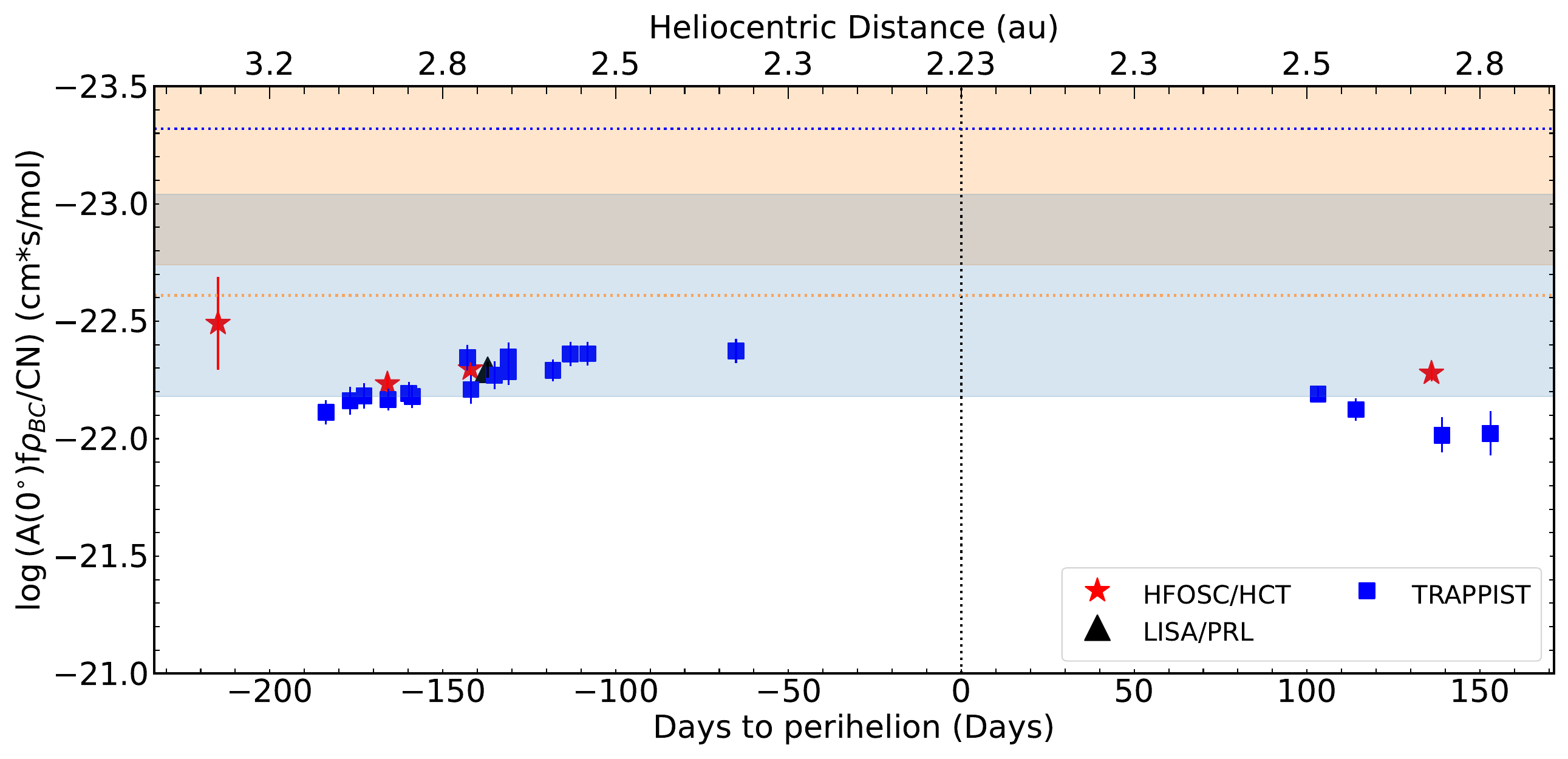}
    \caption{Variation of Af$\rho_{BC}$/CN with the heliocentric distance. We have marked the 1-$\sigma$ range of dust-rich with \textit{blue} and the 1-$\sigma$ range of dust-poor with \textit{orange}.}
    \label{Afp_CN}
\end{figure}

The molecular production rates and the rate ratios for the spectroscopic and photometric measurements are given in Tables \ref{sec: Production rate- II} and \ref{sec: Production rate- I}, respectively. The variation of the production rates of OH, CN, C$_2$, C$_3$ molecules with respect to heliocentric distance (in au) and Days to Perihelion (in days) are given in Fig. \ref{QOH}, \ref{QCN}, \ref{QC2}, and \ref{QC3}. The computed gas production rates and the rate ratios can provide hints of the chemical composition of a comet. The intensity ratios of C$_2$ to CN are directly correlated to the aromatic rings, as mentioned by \citet{Portnov_2003} and \citet{Mousavi_2015} using Laser-Induced Breakdown Spectroscopy (LIBS). \citet{AHEARN1995223} has provided the limits of the production rate ratios, such as C$_2$/CN and C$_3$/CN, to understand whether the comet has a carbon-typical composition or a carbon-depleted composition.

In Fig. \ref{C2_CN} and \ref{C3_CN}, we have plotted the production rate ratios, C$_2$/CN and C$_3$/CN, as a function of the days to perihelion (DTP) in days and the heliocentric distance in au. We have shown two regions: one in \textit{blue}, which represents the range for typical carbon composition of the comets, and the second in \textit{orange}, which represents the depleted composition of comets. The centre line of every region represents the respective median value. The observations from the LISA instrument are shown as a black triangle, from HFOSC are shown as a star, and from TRAPPIST are marked as a red square. The vertical dotted line represents the perihelion distance, which is 2.23 au. Fig. \ref{C2_CN} shows a slight variation in the production rate ratio C$_2$/CN with the heliocentric distance. We see the comet appears to become carbon-depleted at a distance larger than about 2.5 au, which could be due to the drop in the production of C$_2$ molecules.
However, in Fig. \ref{C3_CN}, the production rate ratio C$_3$/CN is consistently typical over a large range in heliocentric distance. Therefore, we believe that comet V2 has a typical carbon chemical composition. In addition, we have plotted the dust-to-gas ratio, which is defined as the ratio of the dust proxy parameter Af$\rho$ in the Blue-Continuum filter (BC) to the production rate of the CN molecule. As shown in Fig. \ref{Afp_CN}, the comet has a dust-rich composition. 

We have also fitted a power law function to the variation of the production rates of different molecules, such as OH, CN, C$_2$, C$_3$, pre- and post-perihelion. The computed power law slopes are given in Table \ref{sec: Power Law}. We could not calculate the power law slopes for OH and C$_3$ due to insufficient points post-perihelion. The slope of C$_2$ post-perihelion is seemingly higher due to the sharp drop in C$_2$ production beyond 2.5 au. This is why the production rate ratio for C$_2$/CN should be looked at under 2.5 au to determine the carbon-depleted/typical nature of the comet, as also mentioned in \citet{Opitom_2015}.
\begin{table}
\centering
\caption{Computed power-law slopes for different molecules, i.e., OH, CN, C$_2$, and C$_3$ at pre- and post-perihelion.}
\setlength{\tabcolsep}{12pt}
        \resizebox{\linewidth}{!}{%
\begin{tabular}{ccc}
\hline
\multicolumn{1}{|c|}{{Species}} &
  \multicolumn{2}{c|}{{r-dependence}}\\
\multicolumn{1}{|c|}{} &
  \multicolumn{1}{c|}{{pre-perihelion}} &
  \multicolumn{1}{c|}{{post-perihelion}}\\\hline
OH & -10.07 $\pm$ 0.75 & --\\
CN & -1.84 $\pm$ 0.49 & -3.56 $\pm$ 1.65\\
C$_2$ & -3.65 $\pm$ 0.52 & -9.13 $\pm$ 2.00\\
C$_3$ & -3.51 $\pm$ 1.50 & --\\
\hline
\end{tabular}
}
\label{sec: Power Law}
\end{table}

\subsection{Photometric Colours}\label{phot_discussion}
\begin{figure}
    \centering
    \includegraphics[width= \linewidth]{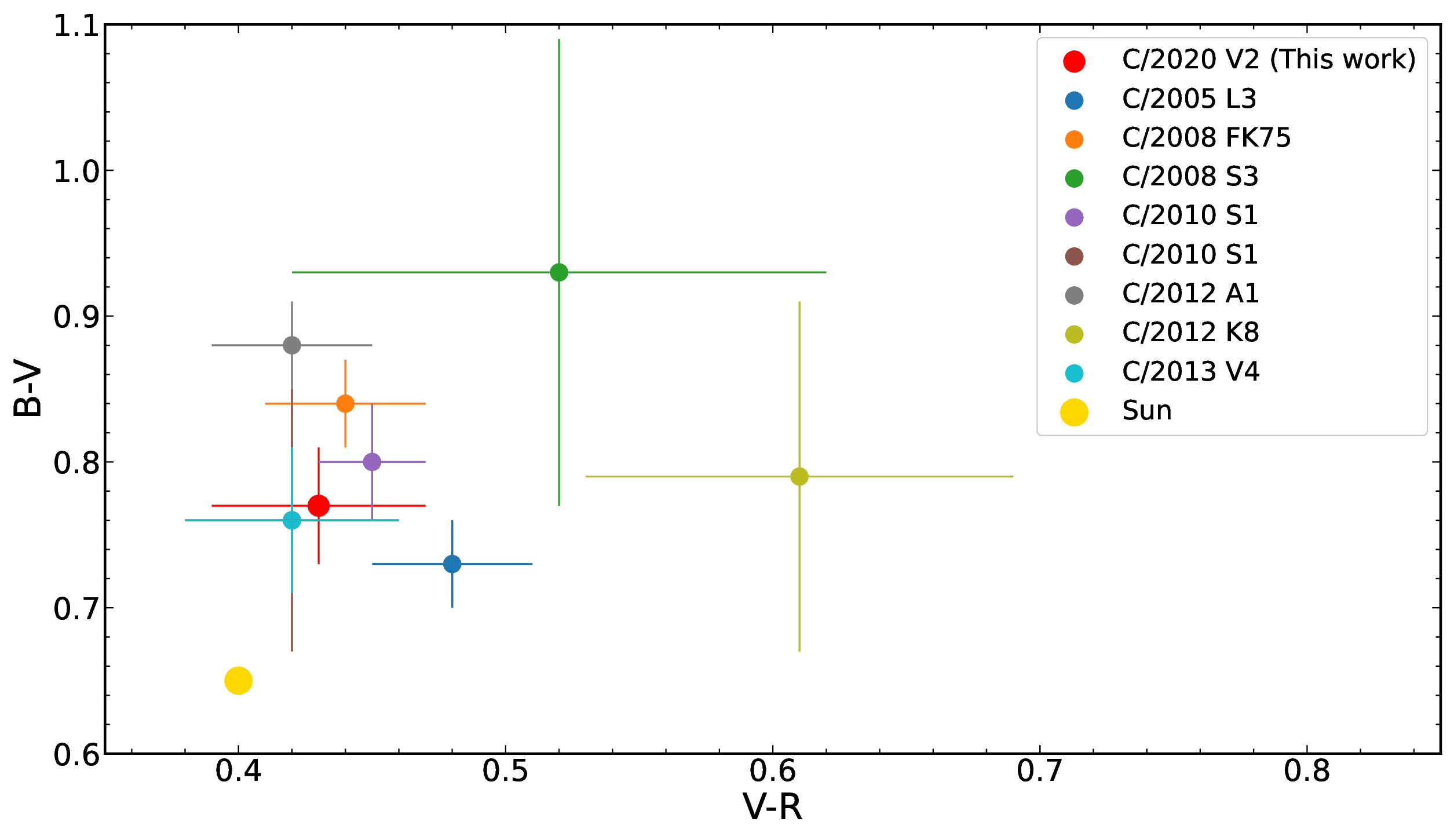}
    \caption{Colour-colour plot of $B-V$ vs $V-R$ of different DNCs (Data is taken from \citet{kulyk_2018}). The solar colour is marked as \textit{yellow} and median colour of comet V2 is marked as \textit{red}.}
    \label{DNC_V2}
\end{figure}

\begin{figure}
    \centering
    \includegraphics[width= \linewidth]{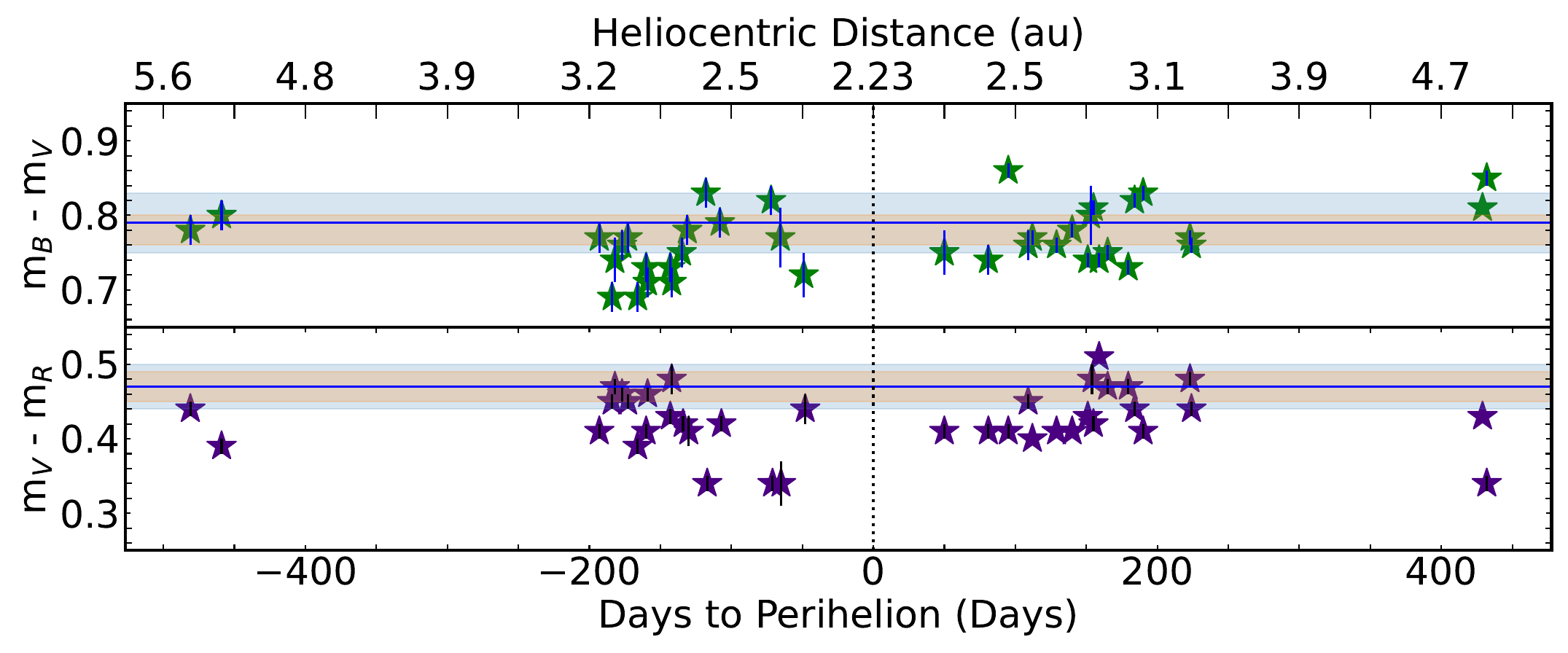}
        \caption{Variation of the colour indices $B-V$ (above) and $V-R$ (below) with days to perihelion and heliocentric distance (in au). The 1-$\sigma$ range for DNCs \citep{kulyk_2018} is marked as blue and for LPCs \citep{Jewitt_2015} as orange. The blue line is the median value of the DNCs \citep{kulyk_2018}.}
    \label{colour1}
\end{figure}

\begin{figure}
    \centering
    \includegraphics[width= \linewidth]{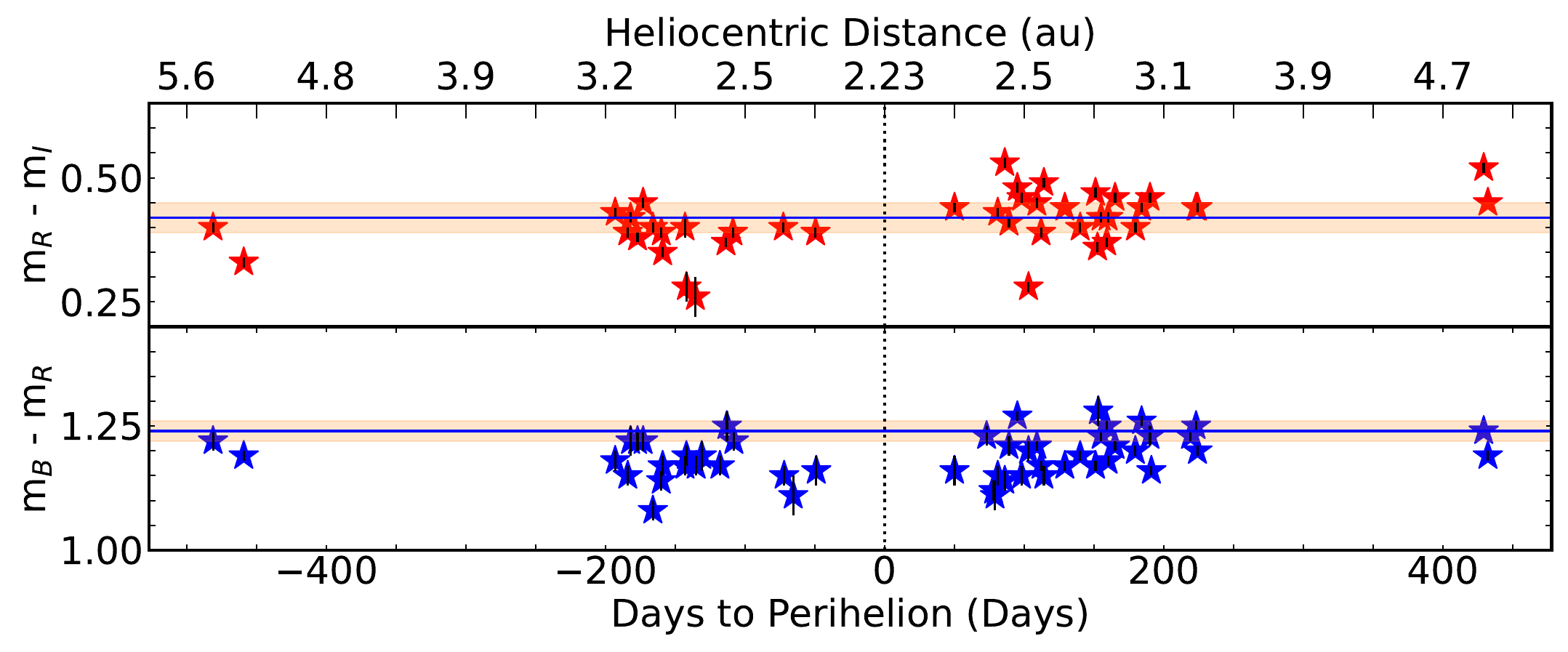}
    \caption{Variation of the colour indices $R-I$ (above) and $B-R$ (below) with days to perihelion and heliocentric distance (in au). The 1-$\sigma$ range for LPCs \citep{Jewitt_2015} is marked as orange. The blue line is the median value of the LPCs \citep{Jewitt_2015}.}
    \label{colour2}
\end{figure}

Using the values in the Table \ref{sec: BB Magnitude}, we have calculated different colours such as $B-V$, $V-R$, $R-I$, and $B-R$. The mean values of these colours are $B-V = 0.77$ $\pm$ $0.04$, $V-R = 0.43$ $\pm$ $0.04$, $R-I = 0.42$ $\pm$ $0.06$, and $B-R = 1.19$ $\pm$ $0.04$. We have compared these colours with the known median colours of active LPCs, calculated by \citet{Jewitt_2015}, which are $B-V = 0.78$ $\pm$ $0.02$, $V-R = 0.47$ $\pm$ $0.02$, $R-I = 0.42$ $\pm$ $0.03$, $B-R = 1.24$ $\pm$ $0.02$. Since this comet is a DNC, we also compared its colours with \citet{kulyk_2018}, which has shown that the median $B-V$ and $V-I$ colours for DNCs are $0.80$ $\pm$ $0.07$ and $0.44$ $\pm$ $0.07$. We have made the colour-colour plot of the $B-V$ and $V-R$ colours of different DNCs along with comet V2, as shown in Fig. \ref{DNC_V2} and found that comet V2's mean colours are in close agreement with many reported DNCs.
In Fig. \ref{colour1} and \ref{colour2}, we have marked the 1-$\sigma$ range of DNCs with \textit{blue} and the 1-$\sigma$ range of LPCs with \textit{orange}. As shown in the figures, there is a significant variation in the different colours of the comet, which could be due to the varying activity of comet V2, as also mentioned in \citet{Voitko_EPSC2024}.

The mean $B-V$ pre-perihelion colour is $0.76$ $\pm$ $0.04$ and post-perihelion is $0.78$ $\pm$ $0.04$. Similarly, the mean value of $V-R$ pre-perihelion is $0.43$ $\pm$ $0.04$ and post-perihelion is $0.43$ $\pm$ $0.04$. For $R-I$, mean pre-perihelion colour is $0.38$ $\pm$ $0.05$ and post-perihelion is $0.43$ $\pm$ $0.05$. For $B-R$, the mean value for pre-perihelion is $1.18$ $\pm$ $0.04$ and for post-perihelion is $1.21$ $\pm$ $0.04$. As can be seen, the mean colours of the comet during pre- and post-perihelion have not changed in the apparition, which hints at the homogeneous properties of the dust on the nucleus.

We also checked the contamination by gas contribution in the broadband colours using our spectroscopic datasets. We used the \textsc{Pyphot} package \citep{zenodopyphot,filter_2020} to calculate the magnitude of the flux-calibrated spectrum by first considering the spectrum, which includes the contribution of molecular gas emissions, and second, we used the continuum spectrum of the comet by tracing the dust continuum directly from the observed spectrum. We calculated the magnitude of different filters in both spectra and found that the difference between magnitudes derived from the two spectra is within 0.1, which comes under the error bar of our measurement as mentioned in Table \ref{sec: Production rate- II}. Hence, we are confident that the computed colours are not significantly affected by the gas emissions.
\subsection{Reflective Gradient}
The reddening of cometary colours gives us a qualitative idea of the size of dust grains present in the cometary comae. \citet{AHearn1984} has defined the parameter known as reflective gradient, which is defined as $(dS/d\lambda)/\Tilde{S}$, where $S$ is the reflectance of the comet, which is the ratio of the comet flux ($F(\lambda)$) to the solar flux ($F_{\odot}(\lambda)$), i.e., $S(\lambda) = \frac{F(\lambda)}{F_{\odot}(\lambda)}$. $d\lambda$ is defined as the difference between two wavelengths, which will be the central bandwidth for different filters, and $\Tilde{S}$ is the mean value of the reflectance in the bandpass \citep{Jewitt_2002}. Mathematically, it can be written as:

\begin{equation}\label{ref_grad2}
    S'(\lambda) = \frac{2}{\Delta\lambda}\frac{S(\lambda_2)-S(\lambda_1)}{S(\lambda_2)+S(\lambda_1)} 
\end{equation}

where $\Delta\lambda = \lambda_2 - \lambda_1$, is given in \AA. The unit of the reflective gradient, $S'(\lambda)$ is given in $\%$ (1000 \AA)$^{-1}$. In units of magnitude, we get \citep{kulyk_2018}:
\begin{equation}\label{ref_grad3}
    S'(\lambda) = \frac{2}{\Delta\lambda}\frac{10^{0.4(CI_{comet}-CI_{sun})}-1}{10^{0.4(CI_{comet}-CI_{sun})}+1}
\end{equation}

Here, $CI_{comet}$ and $CI_{sun}$ are the comet and solar colours. The solar colours have been taken from \citet{Holmberg_2006}. Using equation (\ref{ref_grad3}), we get the mean value of the normalized spectral gradient for $B-V$ to be $10.90 \pm 3.62$ $\%$ $\%$ (1000 \AA)$^{-1}$, $V-R$ to be $6.15 \pm 3.51$ $\%$ (1000 \AA)$^{-1}$, and $R-I$ to be $4.94 \pm 3.56$ $\%$ (1000 \AA)$^{-1}$. The trend of decreasing reflectivity with increasing wavelength matches the trend mentioned by \citet{Jewitt_1986, kulyk_2018, Shubina_2024}.

\citet{kulyk_2018} has provided colour values of 14 comets, including 10 DNCs. Out of these, we have looked at 10 DNCs and found the mean values of the spectral colour gradients for $B-V$ and $V-R$ to be $12 \pm 4$ $\%$ (1000 \AA)$^{-1}$ and $5 \pm 2$ $\%$ (1000 \AA)$^{-1}$ respectively, which are in close agreement to what we have found for comet V2.

\subsection{Coma Morphology}\label{sec: coma morphology}
The surface of a comet is not completely active. There are some active regions that are responsible for the comet's activity. This has been discovered in comets such as 1P/Halley \citep{halley_1987_jet} and Comet 43P/Wolf-Harrington \citep{43P_2000_jet}. 
We have studied the coma morphology using the spectra and images of V2 on common dates as discussed in two further sections below. 

\subsubsection{\texorpdfstring{Measurement of dust activity levels through Af$\rho$}{Measurement of dust activity levels through Afrho}}
\begin{figure}
\centering
\subcaptionbox{Af$\rho$ variation of BC filter in \textit{Blue} and of GC filter in \textit{Green} at the East-side (left) and the West-side (right) on 2022 November 22. The dotted red line shows the Half Width Half Maxima (HWHM), i.e., 1.24 arcsec.
\label{afp_2211}}%
  [1\linewidth]{\includegraphics[width = 1.05\linewidth]{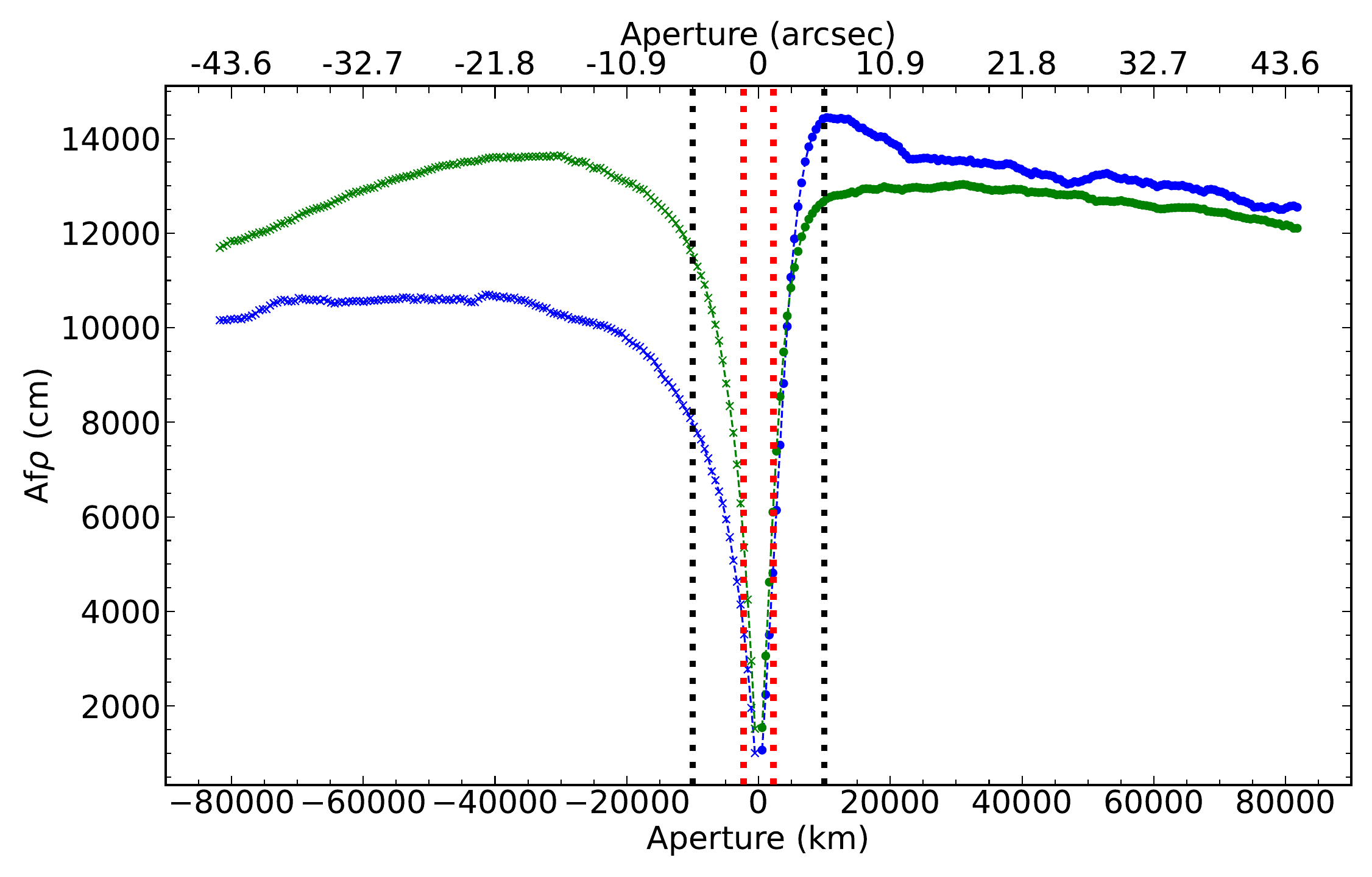}}
  \hfill
\subcaptionbox{Af$\rho$ variation of BC filter in \textit{Blue} and of GC filter in \textit{Green} at the East-side (left) and the West-side (right) on 2022 December 16. The dotted red line shows the Half Width Half Maxima (HWHM), i.e., 1.21 arcsec.
\label{afp_1612}}
  [1\linewidth]{\includegraphics[width = 1.05\linewidth]{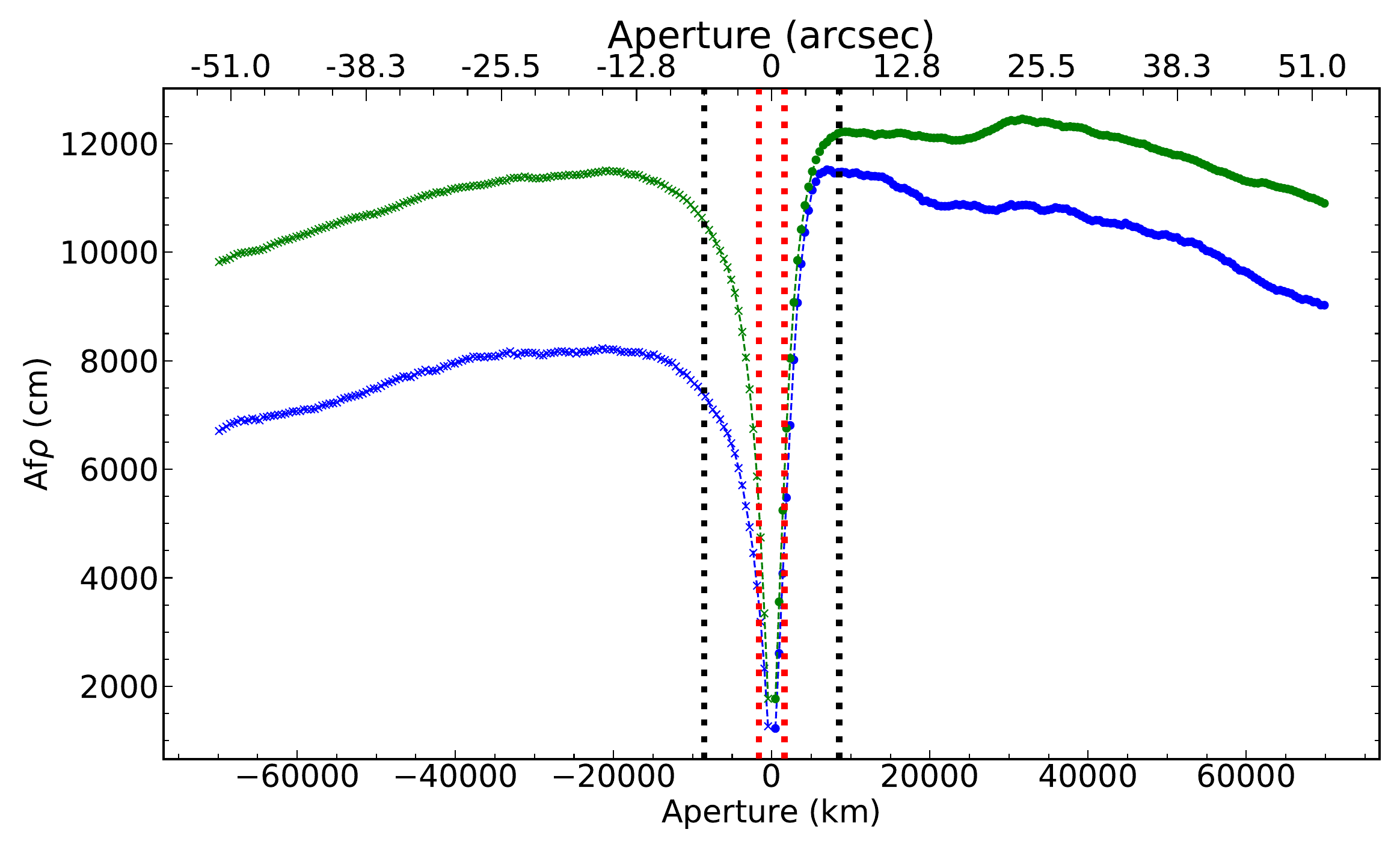}}
\caption{The variation of Af$\rho$ at both sides shows the asymmetry of the emission in the comet. The dotted black line shows the distance at 10000 km from the photocenter.}
\end{figure}

From our spectroscopic observations of 2022 November 22 and 2022 December 16, we calculated the dust proxy parameter (A(0$^{\circ}$)f$\rho$) using equation (\ref{eq: Luminosity_comet-V}). Spectra are extracted for apertures in the direction of the East and West sides of the photocenter to check the emission trend. Fig. \ref{afp_2211} and \ref{afp_1612} show the variation of Af$\rho$ of the comet in the BC and GC filter passbands on both sides. The positive numbers represent the West side of the comet, and the negative numbers represent the East side of the comet. The significant difference between the two sides is prominent. This shows the asymmetric emissions of the dust. To look for a connection between this asymmetric emission with features in the images, we carried out image processing techniques such as those described by \citet{Larson_Sekanina_1984} and \citet{ Samarsinha_2006} to reveal the presence of jets or other asymmetric features in the coma. 

\subsubsection{Investigation of coma structures and asymmetries using image enhancement techniques}
\begin{figure*}
    \centering
    \subcaptionbox{Direct image\label{B_2211}}{
        \includegraphics[width=0.4\linewidth]{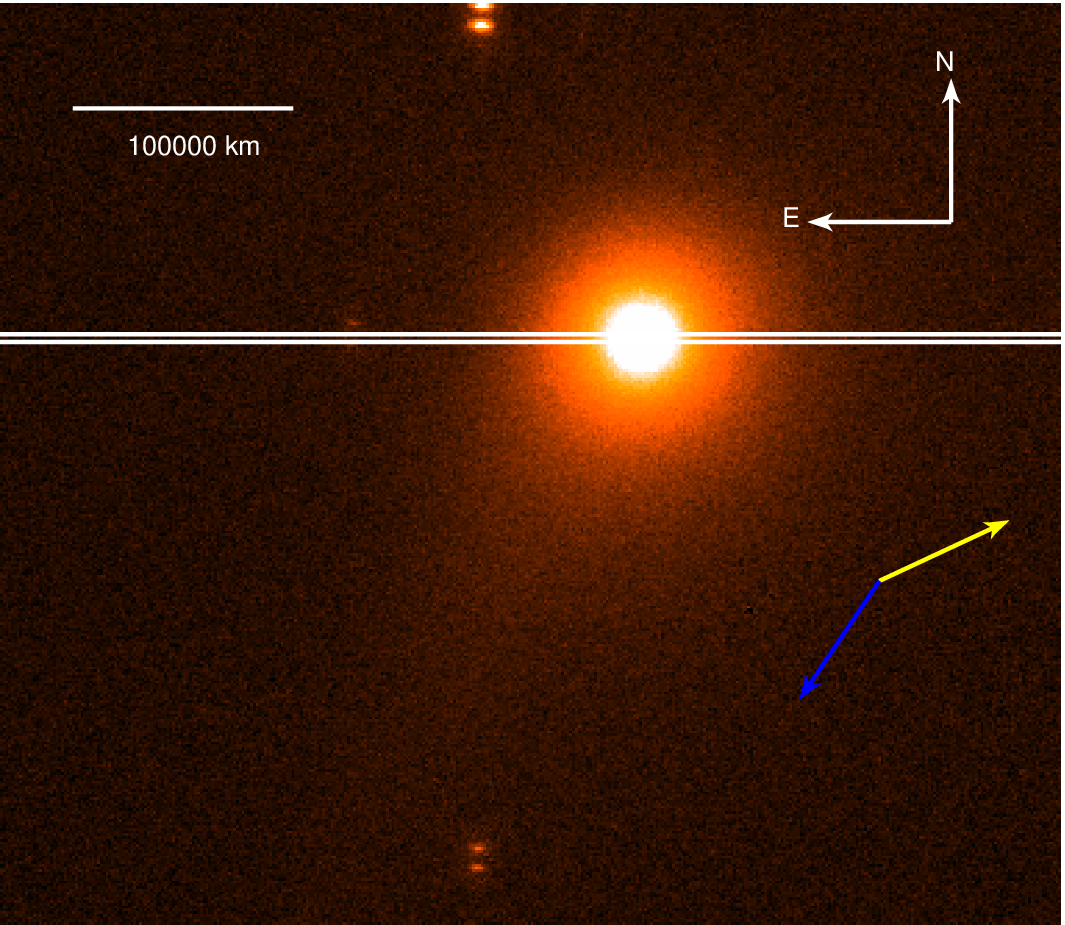}
    }
    \subcaptionbox{After LS processing at 45$^\circ$ angle\label{LS_B_2211}}{
        \includegraphics[width=0.4\linewidth]{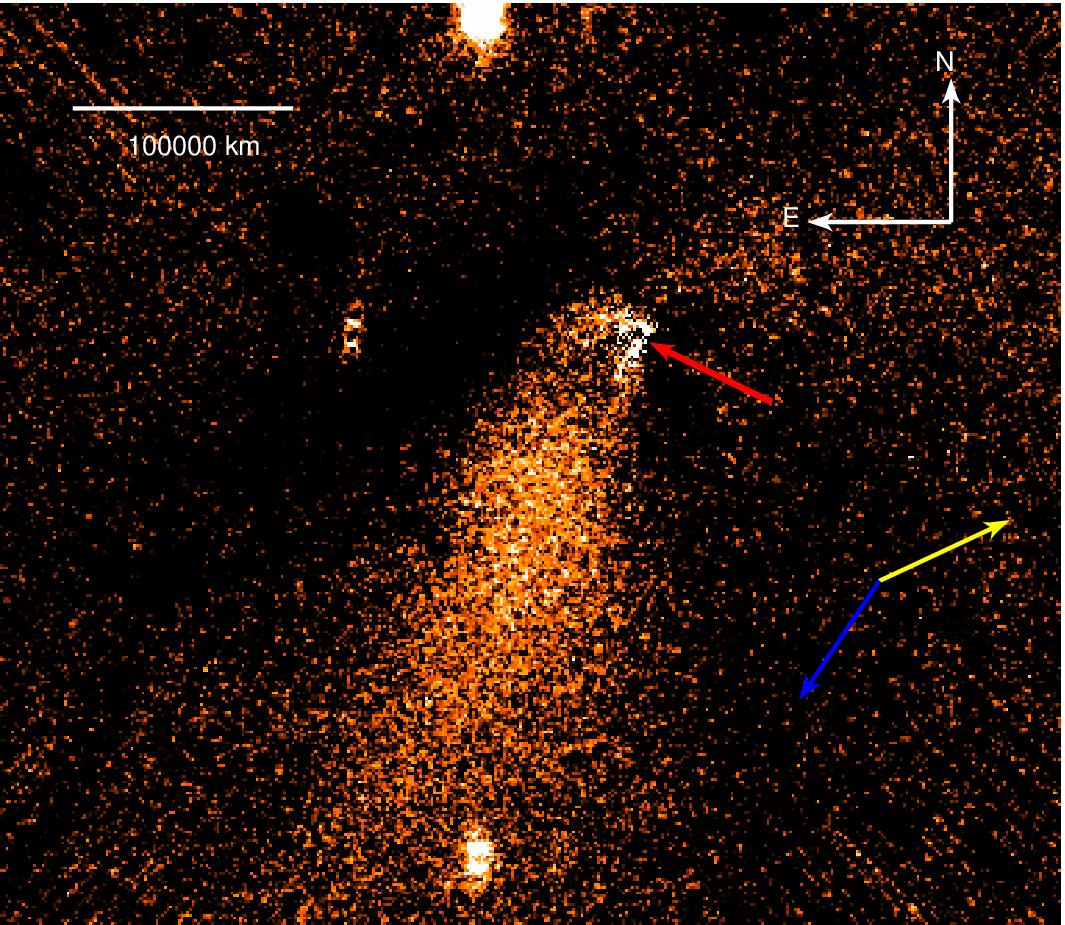}
    }
    \caption{B band image observed using HFOSC, HCT on 2022 November 22. The position of the spectrograph slit is in the East-West direction, as shown. The dust-tail orientation and the Sun's direction are marked in \textit{blue} and \textit{yellow}. The tail and the projection of the tail (anti-tail) are visible on either side of the photocenter. The \textit{red} arrow is pointing towards the photocenter.
    }
    \label{fig:LS_B_processing}
\end{figure*}

\begin{figure*}
    \centering
    \subcaptionbox{Direct image 
    \label{B_1612}}{
        \includegraphics[width=0.4\linewidth]{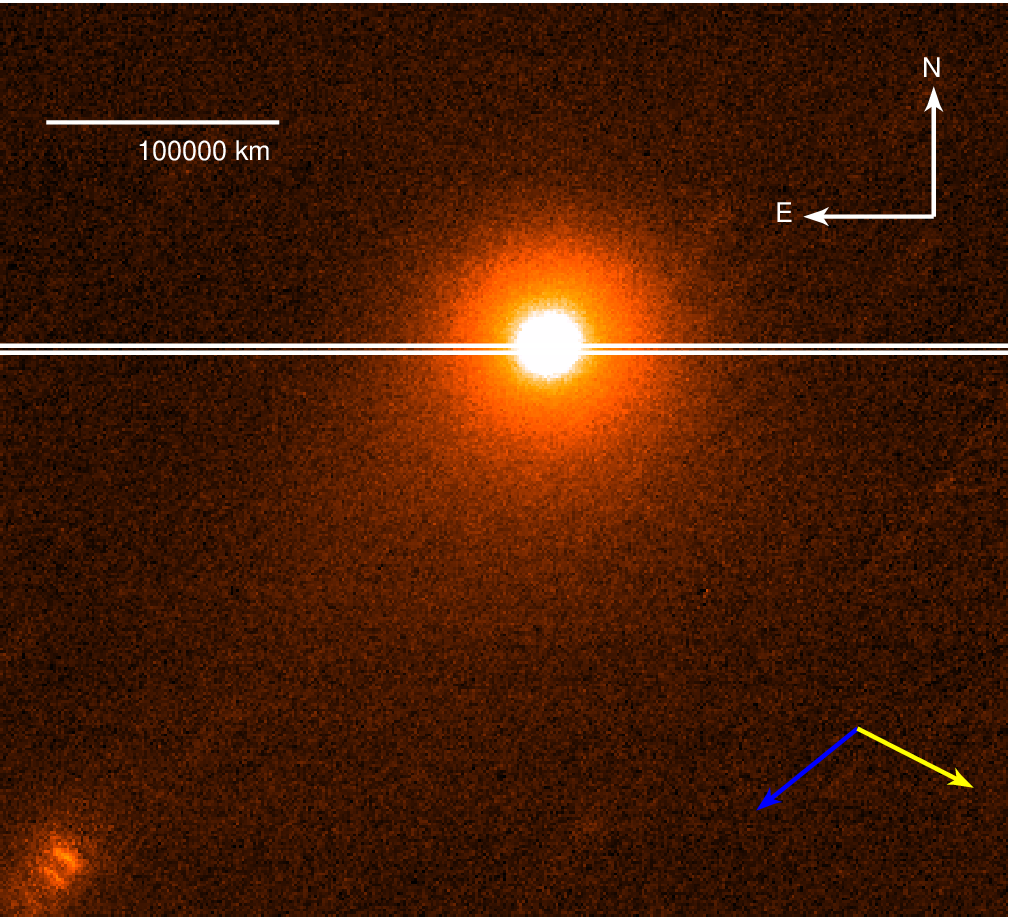}
    }
    \subcaptionbox{After LS processing at 45$^\circ$ angle
    \label{LS_B_1612}}{
        \includegraphics[width=0.4\linewidth]{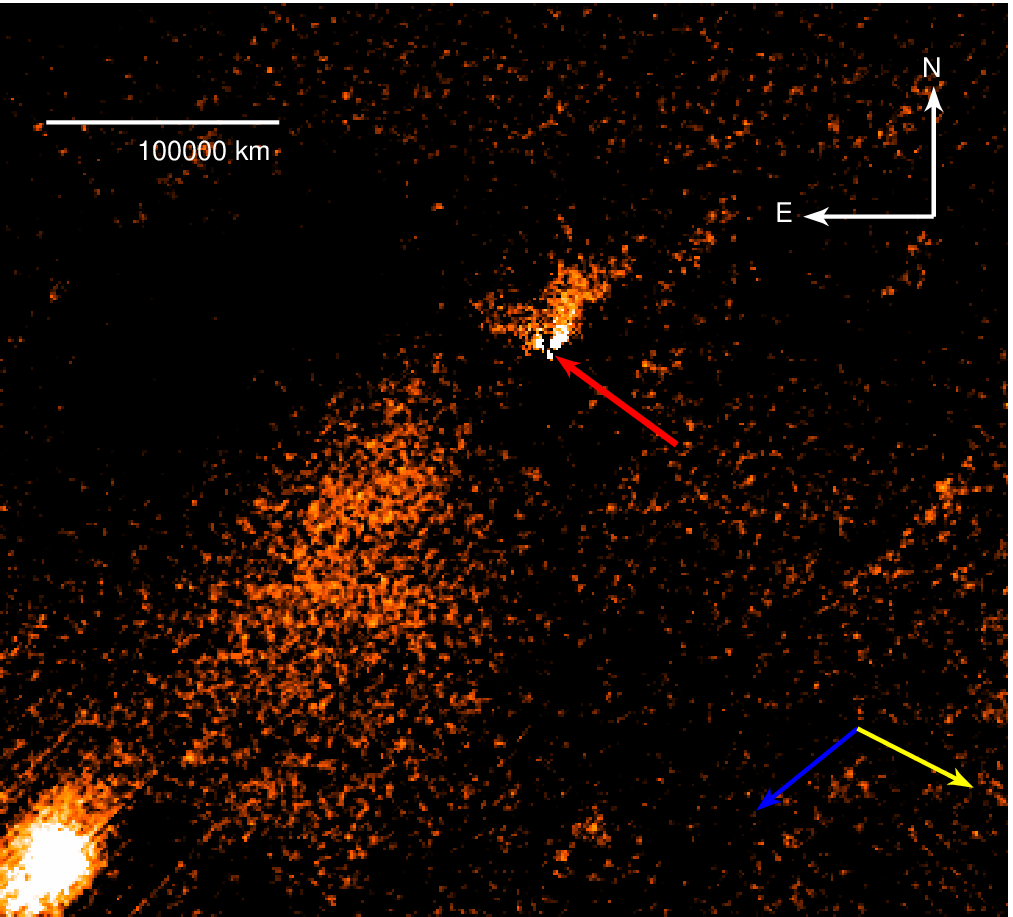}
    }
    \caption{B band image observed from HFOSC, HCT on 2022 December 16. The position of the spectrograph slit is in the East-West direction, as shown. The dust-tail orientation and the Sun's direction are marked in \textit{blue} and \textit{yellow}. The tail and the projection of the tail (anti-tail) are visible on either side of the photocenter. The \textit{red} arrow is pointing towards the photocenter.}
    \label{fig:LS_B_processing_1}
\end{figure*}

For our work, we used the modified version of the Larson-Sekanina (LS) processing technique to extract hidden features. The details are mentioned in \citet{GARCIA_2020} and \citet{156P_Aravind}. With this technique, the image is rotated clockwise and counter-clockwise at a certain angle, centred on the photocenter, and then subtracted from the original image. After that, the two images are summed to get the final image. We applied this technique to the Bessel B filter images obtained using the HFOSC instrument on 2022 November 22 and 2022 December 16. We found two regions with an excess in emission; one is strong, and one is relatively weak, as seen in the two images \ref{LS_B_2211} and \ref{LS_B_1612}. We have labelled the slit position, which is in the East (Left) - West (Right) direction. When we compare the images with the spectroscopic results, we see that the Af$\rho$ is following the same trend. 
We have also applied this technique on the different narrow-band molecular gas filter images obtained using the TRAPPIST telescopes to see similar features on 2022 November 22 (check appendix Fig. \ref{fig: CN_combined_2211}, \ref{fig: C2_combined_2211} and \ref{fig: C3_combined_2211}), and on 2022 December 16 (check appendix Fig. \ref{fig: CN_combined_1612}, \ref{fig: C2_combined_1612} and \ref{fig: C3_combined_1612}). We see similar structures in different filters by other image processing techniques, such as applying division by Azimuthal median and division by azimuthal average \citep{Samarasinha2013}.
The strong and weak dust features are consistent with the view of the comet along the orbital plane (orbital plane angle $-7.5^{\circ}$ and $-15.6^{\circ}$, respectively, for the two dates), showing them to coincide with the tail and 'apparent anti-tail' of the comet. 


\subsection{Nuclear Radius}
The nuclear radius of the comet is an important parameter for understanding the size of the minor planetesimals distributed in the solar system. There are many methods, such as using RADAR observations \citep{Harmon_2005}, or using a diffraction-limited imaging telescope. \citet{Meech_2004} and \citet{Lamy_hst_1995} have used the Hubble Space Telescope (HST) to derive the nuclear radius of the comet.

Here, to calculate the size of the comet nucleus, we have used the TRAPPIST photometric observations in the Bessel-R filter and used the relation given by \citet{Lamy_2004} for calculating the radius of the comet,
\begin{equation}
    r_N = \frac{1.496 \times 10^{11}}{\sqrt{p}} \times 10^{0.2 \times [R_{\odot} - R_N(1,1,0)]}
\end{equation}
Here, p is the geometric albedo of the nucleus, which is taken to be 0.04 \citep{Licandro_2016}. $R_{\odot}$ is the apparent magnitude of the Sun in the Bessel-R band, i.e., -27.15 \citep{Willmer_2018}. $R_N(1,1,0)$ is the absolute magnitude of the nucleus, that is, at $r=\Delta=1$ and phase angle, $\alpha = 0$ degree. It is given by
\begin{equation}
    R_N(1,1,0) = R_N - 5\log(r\Delta) + 2.5 log(\phi(\alpha))
\end{equation}
Here, $\phi(\alpha)$ is the phase function corresponding to the phase angle $\alpha$ (Sun-Comet-Observer angle). Since an active coma region can affect the computed radius, it is always better to perform these calculations for comets at large heliocentric distances in order to minimize the coma contribution.

If we start using the details above and assuming that all the light from the comet comes from reflection, we get a radius of around 60-90 km. However, the famous comet C/2014 UN271, whose nuclear radius is around 60 km, was detected around 29 au \citep{Hui_2022}. So, if comet V2 was of a similar size, it should have been discovered much earlier than its distance of 8 au at discovery.

Also, it has been known that LPCs are active at even far away distances. Hence, we cannot remove the contribution of the outer coma, which is the reason for the overestimation of the nuclear magnitude and, hence, the nuclear radius. This method by \citet{Lamy_2004} is useful when we are sure that there is a minimal coma contribution. However, for an extended body like a comet, it is very difficult to observe them without any significant contribution from the coma. Therefore, we need to estimate the nuclear radius using a different approach.

The method by \citet{Sosa_2011} uses the non-gravitational parameters to calculate the radius of the comet. The complete force equation of a comet facing non-gravitational forces is:
\begin{equation}
    \frac{d^2\textbf{r}}{dt^2} = -\frac{Gm_\odot \textbf{r}}{r^3} + \nabla \mathcal{R} + A_1 g(r) \hat{\textbf{r}} + A_2 g(r) \hat{\textbf{t}} + A_3 g(r) \hat{\textbf{n}}
\end{equation}
where $G$ is the gravitational constant, $\mathcal{R}$ is the planetary disturbing function, $M_\odot$ is the Sun's mass, $r$ is the heliocentric distance, and $g(r)$ is the empirical function of the H$_2$O or CO sublimation rate, which is given by:
\begin{equation}
    g(r) = \alpha \left(\frac{r}{r_0}\right)^{-m} \left[1 + \left(\frac{r}{r_0}\right)^n \right]^{-k}
\end{equation}
The different values of parameters such as $\alpha$, $m$, $n$, $k$, $r_0$ depend on whether the water or CO is sublimating. Using the details mentioned in \citet{krowlikoska_ngf_2017}, the above parameters depend on the value of perihelion. Since the perihelion of comet V2 is less than 4 au, the $g(r)$ function represents the water sublimation, and hence the parametric values are $\alpha = 0.1113$, $m=2.15$, $n=5.093$, $k=4.6142$, $r_0 = 2.808$ au.

In the non-gravitational models, the $A_1$, $A_2$, and $A_3$ represent the modulus of the non-gravitational acceleration in the radial, transverse, and normal directions at 1 au from the Sun. There are two models: the symmetric and asymmetric models. In the symmetric model, only two parameters ($A_1$ and $A_2$) are used to estimate the non-gravitational acceleration, $J$:
\begin{equation}
    J = \sqrt{A_1^2 + A_2^2} \times g[r(t)]
\end{equation}
while, for the asymmetric model, two more parameters, $A_3$ and $DT$, where $DT$ is the time offset of maximum brightness, are involved:
\begin{equation}
    J = \sqrt{A_1^2 + A_2^2 + A_3^2} \times g[r(t-DT)]
\end{equation}
Since the comet's emission looks asymmetric in Fig. \ref{afp_2211} and \ref{afp_1612}, we will use the asymmetric method. Here, we consider a comet of mass $M_N$, which is dominated by water molecules, and $Q(r)$, which is the production rate of water $H_2O$. At 3 au, we are at the limit where water sublimation starts to decay. Since we have observed $Q(OH)$ from TRAPPIST, using the relation mentioned in \citet{Lamy_2009}, i.e., $Q(H_2O) = 1.1 \times Q(OH)$, we get the production rate of water, $Q(H_{2}O)$. We then used equation (3) given in \citet{Sosa_2011},
\begin{equation}
    M_N J = Qmu
\end{equation}
where $J$ is the non-gravitational acceleration, $m$ is the mass of a dominant sublimating molecule, i.e., water ($H_2O$), and $u$ is the effective outflow velocity, which is taken to be $0.27$ km s$^{-1}$ \citep{Sosa_2011}. The mass of the comet is then defined as,
\begin{equation}\label{mass}
    M_N = \frac{Qmu}{J} = \frac{Qmu}{Ag}
\end{equation}
Here, $A$ is $\sqrt{A_1^2 + A_2^2 + A_3^2}$, and, for comet V2, $A_1 = (5.85 \pm 0.01) \times 10^{-8}$ au/day$^2$, $A_2 = (5 \pm 2)\times 10^{-9}$ au/day$^2$, $A_3 = (-6.0 \pm 0.3)\times 10^{-9}$ au/day$^2$ and $DT = (-179 \pm 9) $ days, from NASA's JPL website. After calculating it, we get the median mass to be $(3.4 \pm 0.8) \times 10^{12}$ kg. Here, the error in mass is calculated using the error propagation of all the variables as shown in equation (\ref{mass_error}).
\begin{equation}\label{mass_error}
    \delta M = M \times (\left(\delta Q / Q \right)^2 + \left(\delta J / J \right)^2
\end{equation}
where, $\delta J$ would be calculated as:
\begin{equation}
\delta J = J \times
\sqrt{
\begin{aligned}
&\frac{A_1^2}{A_1^2 + A_2^2 + A_3^2} \times \delta A_1^2 + \frac{A_2^2}{A_1^2 + A_2^2 + A_3^2} \times \delta A_2^2 \\
&+ \frac{A_3^2}{A_1^2 + A_2^2 + A_3^2} \times \delta A_3^2
\end{aligned}
}
\end{equation}
Now, considering the bulk density $\rho$ = $537.8 \pm 0.6$  kgm$^{-3}$ which has been derived \citep{rosetta_2018} using RSI experiment onboard Rosetta mission, the radius of the comet is computed as:
\begin{equation}
    R_N = \left(\frac{3M_N}{4\pi \rho}\right)^{1/3}
\end{equation}
Using the above equation, we get the radius of the cometary nucleus to be $1.1 \pm 0.1$ km.

\section{Conclusions}\label{sec:conclusion}
We have observed comet C/2020 V2 using multiple telescopes in imaging and spectroscopic mode. The observations span 32 months across perihelion. From these observations, we conclude the following:
\begin{enumerate}
    \item The gas production rate ratios, such as C$_2$/CN and C$_3$/CN, indicate that the comet has a typical carbon composition. 
    
    \item The dust-to-gas ratio shows that the comet has a dust-rich composition.

    \item The small variations in production rate ratios and the gas-to-dust ratios over the long-term observations show that the nucleus of the comet is homogeneous. 
    
    \item The mean colours, such as $B-V = 0.77$ $\pm$ $0.04$, $V-R = 0.43$ $\pm$ $0.04$, $R-I = 0.42$ $\pm$ $0.06$, and $B-R = 1.19$ $\pm$ $0.04$, are similar to the median values found for LPCs ($B-V = 0.78$ $\pm$ $0.02$, $V-R = 0.47$ $\pm$ $0.02$, $R-I = 0.42$ $\pm$ $0.03$, $B-R = 1.24$ $\pm$ $0.02$) and median value of DNCs ($B-V = 0.80$ $\pm$ $0.07$, and $V-I = 0.44$ $\pm$ $0.07$).  Departures from the average colours with respect to the heliocentric distance could be due to changes in the activity of the comet. However, the averages of the pre- and post-perihelion colours of comet V2 are quite similar. Moreover, the effect of gas contamination in broadband colours is found to be negligible in this comet.

    \item The mean value of the reflectivity gradient is found to be $10.9 \pm 3.6$ $\%/1000$ \AA\ for $B-V$, $6.2 \pm 3.5$ $\%/1000$ \AA\ for $V-R$, which is similar to the other DNCs. For $R-I$, the value is found to be $4.9 \pm 3.6$ $\%/1000$ \AA.

    \item Spectroscopic calculation of Af$\rho$ shows asymmetry on either side of the photocenter.  LS processing technique on the images of the same dates shows excess emission regions, where one is along the comet's dust tail, while the other is along the anti-tail.
    
    \item The non-gravitational force model provides an estimate of the comet's nuclear radius to be $1.1 \pm 0.1$ km.
    
    \item The compositionally homogenous nature of comet V2 indicates that the formation location of this comet in the early Solar system was homogeneously mixed. However, this is an indirect inference and is open to interpretation.
\end{enumerate}

This work shows that DNCs that exhibit steady behaviour in their early orbital evolution may continue with a similar stable nature throughout their orbit.  This is an important input for the Comet Interceptor mission, 
to select a suitable DNC early enough in its orbital path, allowing for more time to plan the encounter. It also emphasizes the requirement to closely monitor newly discovered comets photometrically alongside the usually carried out astrometric observations to determine their orbital characteristics.  Periodic spectroscopic observations would help confirm the narrow band photometric measurements and also help to fully understand the cometary composition. 

\section*{Conflict of Interest}
The authors declare that there are no conflicts of interest regarding this work.

\section*{Acknowledgements}
We would like to thank the anonymous reviewer for providing insightful comments and suggestions. We acknowledge the local staff at the Mount Abu Observatory for providing the necessary support. We thank the staff of IAO, Hanle and CREST, Hosakote, who made these observations possible. The facilities at IAO and CREST are operated by the Indian Institute of Astrophysics, Bangalore. Work at the Physical Research Laboratory is supported by the Department of Space, Govt. of India.

This publication makes use of data products from the TRAPPIST project, under the scientific direction of Dr. Emmanuel Jehin, Director of Research at the Belgian National Fund for Scientific Research (F.R.S.-FNRS). TRAPPIST-South is funded by F.R.S.-FNRS under grant PDR T.0120.21, and TRAPPIST-North is funded by the University of Liège in collaboration with Cadi Ayyad University of Marrakech. The authors thank NASA, D. Schleicher, and the Lowell Observatory for the loan of a set of HB comet filters.

This work is a result of the bilateral Belgo-Indian projects on Precision Astronomical Spectroscopy for Stellar and solar system bodies, BIPASS, funded by the Belgian Federal Science Policy Office (BELSPO, Government of Belgium; BL$/$33$/$IN22\texttt{\_}BIPASS) and the International Division, Department of Science and Technology, (DST, Government of India; DST/INT/BELG/P-01/2021(G)).

\section*{Data Availability}
The data underlying this article will be shared on reasonable request
to the corresponding author.
 



\bibliographystyle{mnras}
\bibliography{mnras_template} 




\appendix
\onecolumn 
\section{\texorpdfstring{Variation of gas production rates and dust proxy parameter A \textit{\MakeLowercase{f}}$\rho$ in comet C/2020 V2}{Variation of gas production rates and dust proxy parameter Afrho in comet C/2020 V2 (ZTF)}}
\begin{figure*}
\begin{subfigure}{0.49\linewidth}
\centering
\includegraphics[width = 0.95\textwidth]{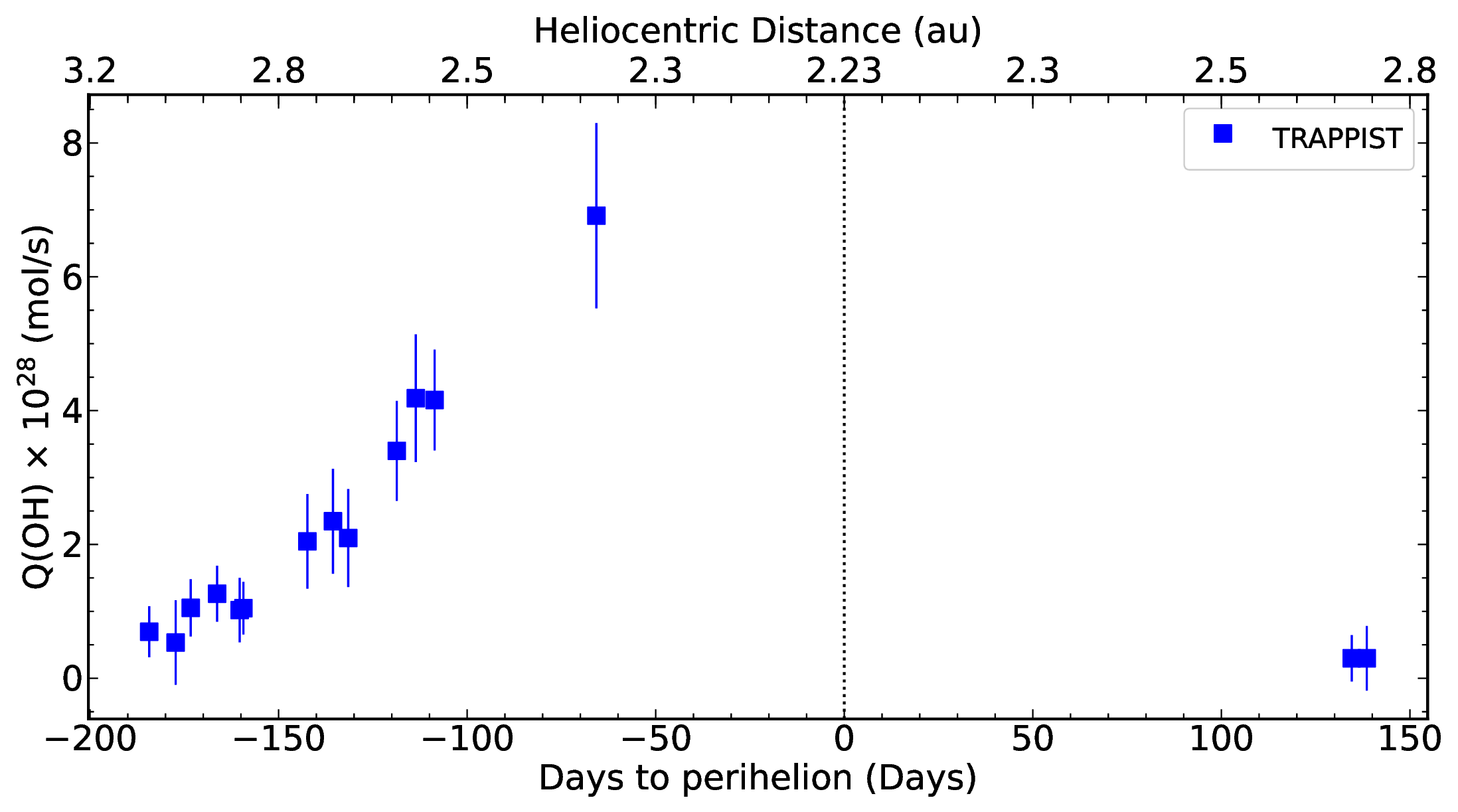}
\subcaption{Activity trend of Q(OH)}
\label{QOH}
\end{subfigure}\hfill
\begin{subfigure}{0.49\linewidth}
\centering
\includegraphics[width = 0.95\textwidth]{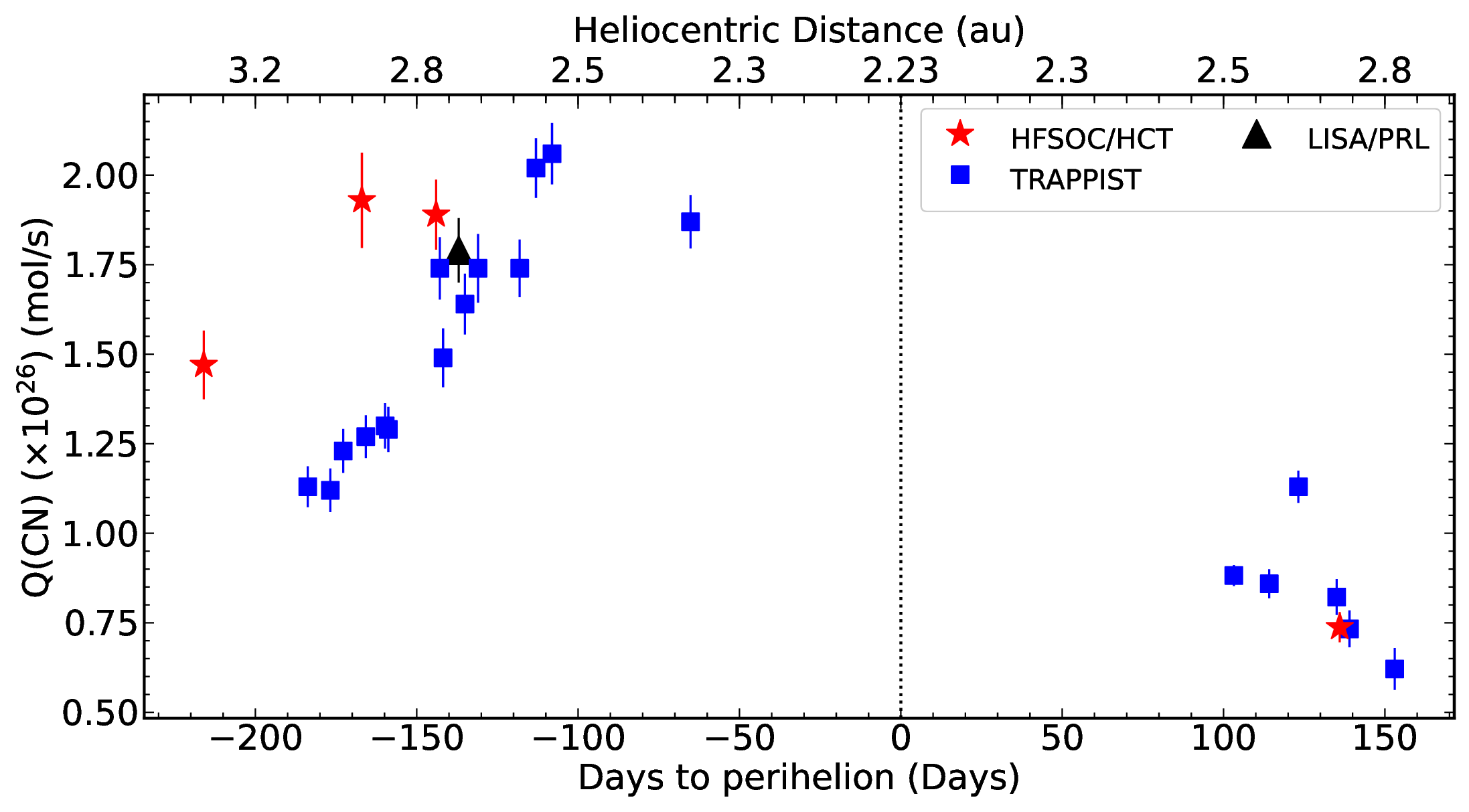}
\subcaption{Activity trend of Q(CN).}
\label{QCN}
\end{subfigure}\hfill
\begin{subfigure}{0.49\linewidth}
\centering
\includegraphics[width = 0.95\textwidth]{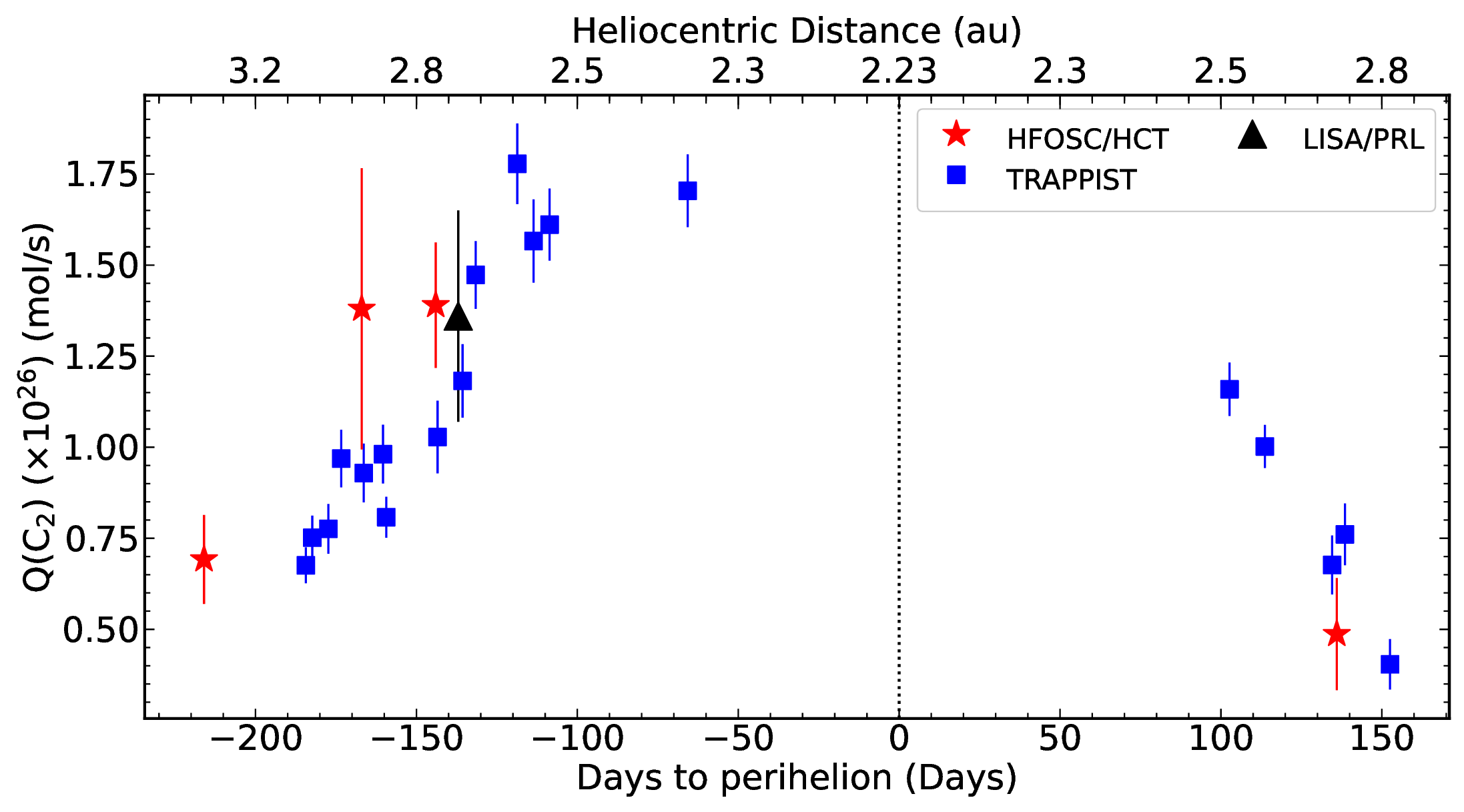}
\subcaption{Activity trend of Q(C$_2$).}
\label{QC2}
\end{subfigure}\hfill
\begin{subfigure}{0.49\linewidth}
\centering
\includegraphics[width = 0.95\textwidth]{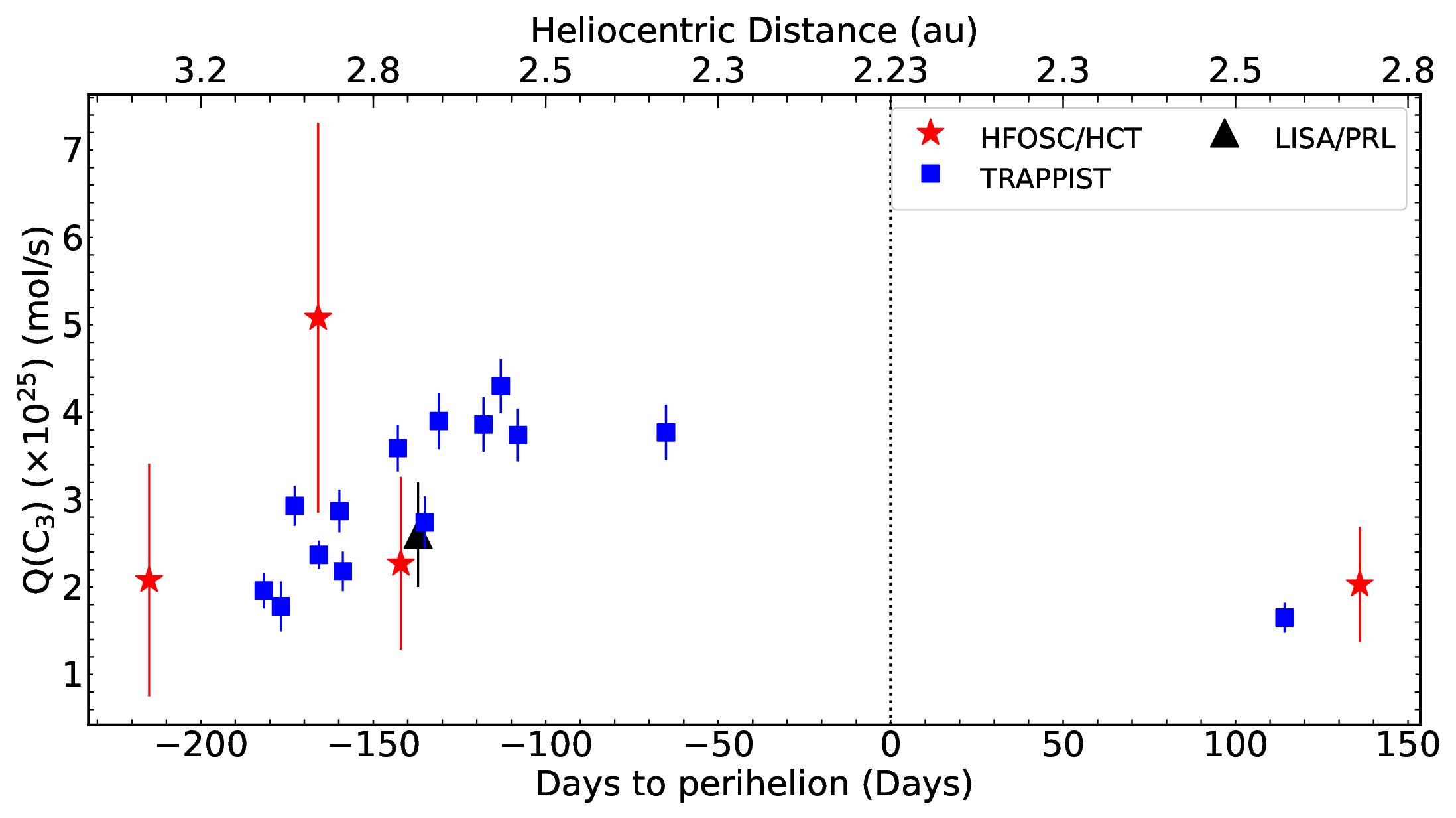}
\subcaption{Activity trend of Q(C$_3$).}
\label{QC3}
\end{subfigure}\hfill
\begin{subfigure}{0.49\linewidth}
\centering
\includegraphics[width = 0.95\textwidth]{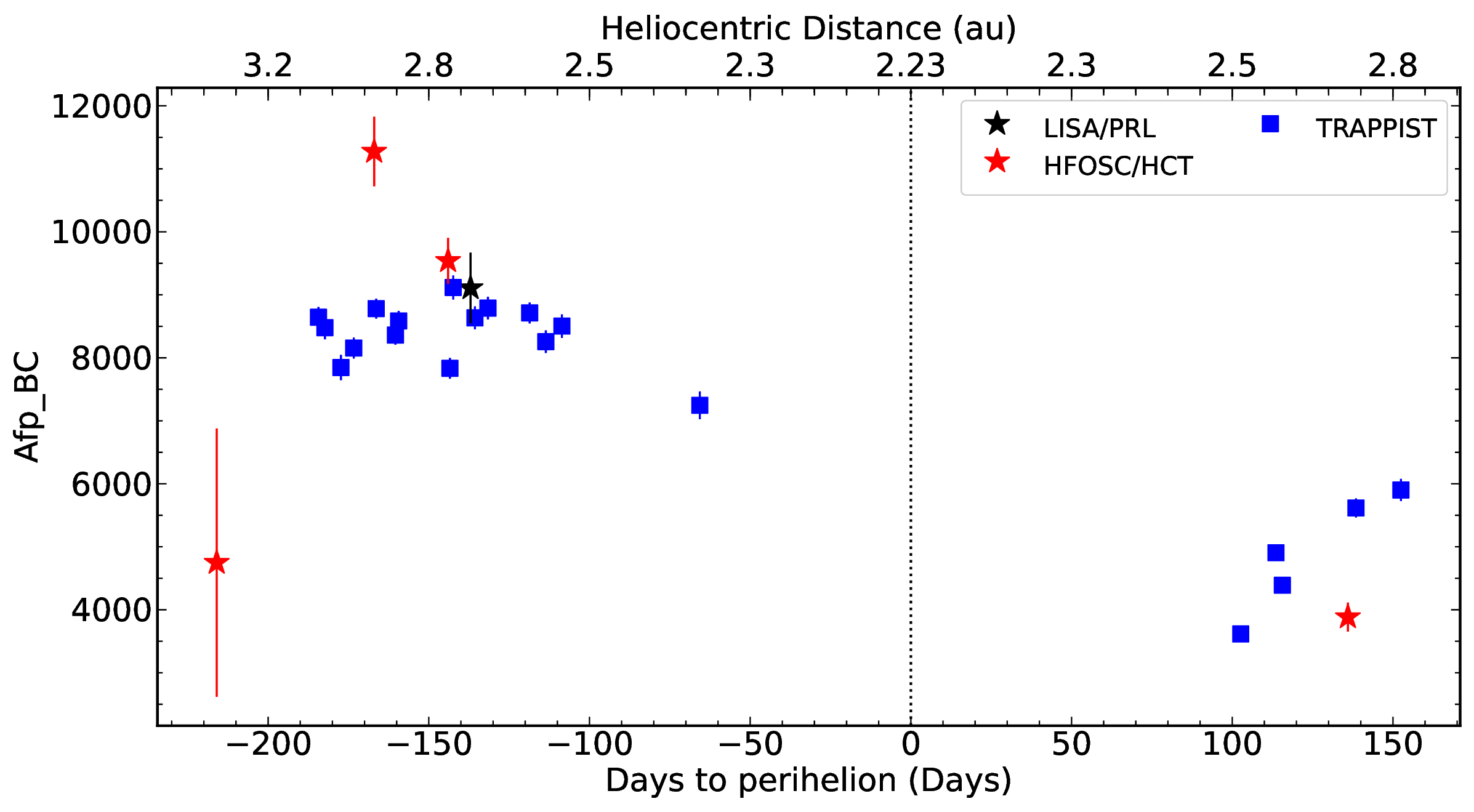}
\subcaption{Activity trend of A$(\theta = 0^{\circ})f\rho$(BC).}
\label{A_BC}
\end{subfigure}\hfill
\begin{subfigure}{0.49\linewidth}
\centering
\includegraphics[width = 0.95\textwidth]{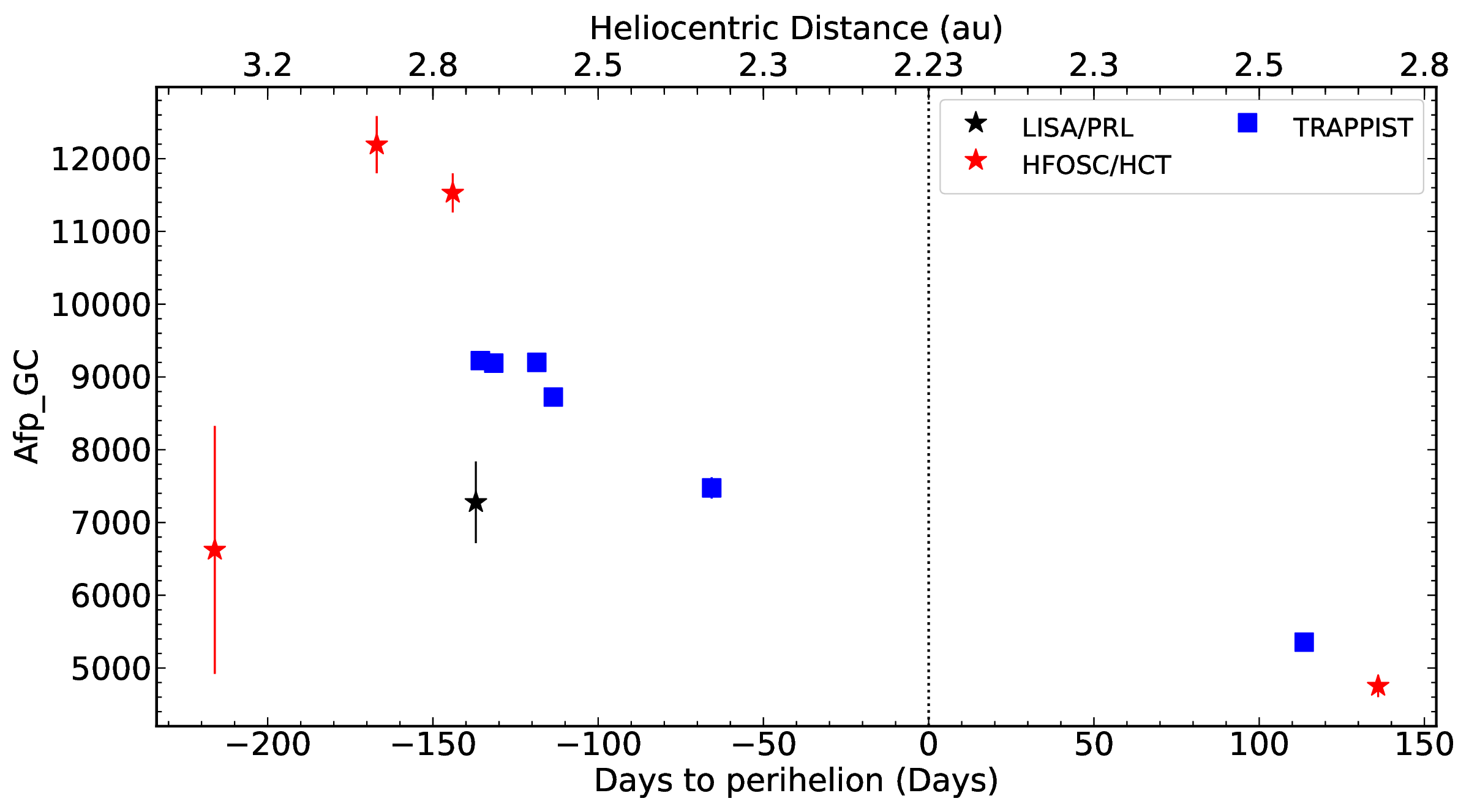}
\subcaption{Activity trend of A$(\theta = 0^{\circ})f\rho$(GC).}
\label{A_GC}
\end{subfigure}
\begin{subfigure}{0.49\linewidth}
\centering
\includegraphics[width = 0.95\textwidth]{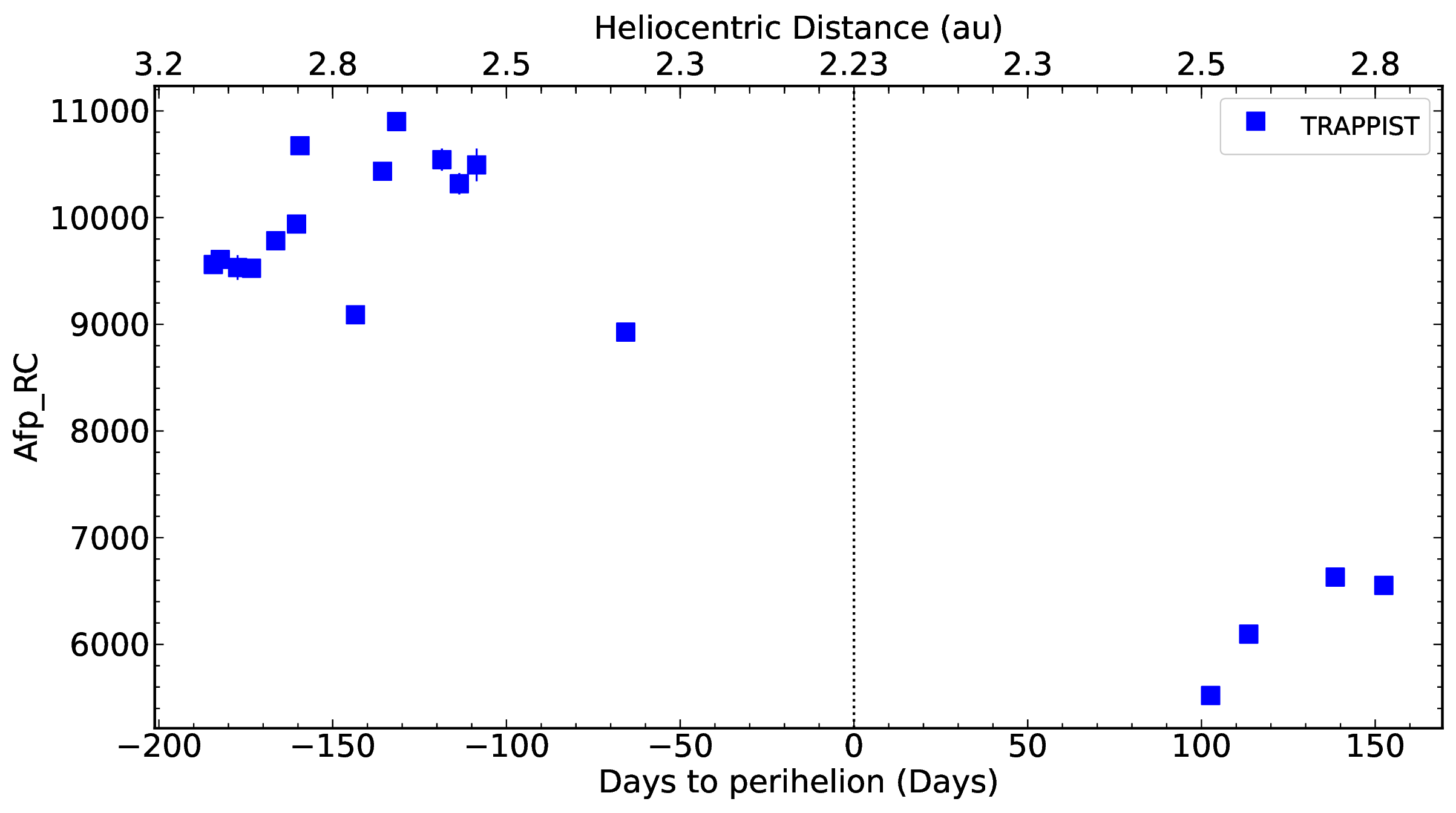}
\subcaption{Activity trend of A$(\theta = 0^{\circ})f\rho$(RC).}
\label{A_RC}
\end{subfigure}
\begin{subfigure}{0.49\linewidth}
\centering
\includegraphics[width = 0.95\textwidth]{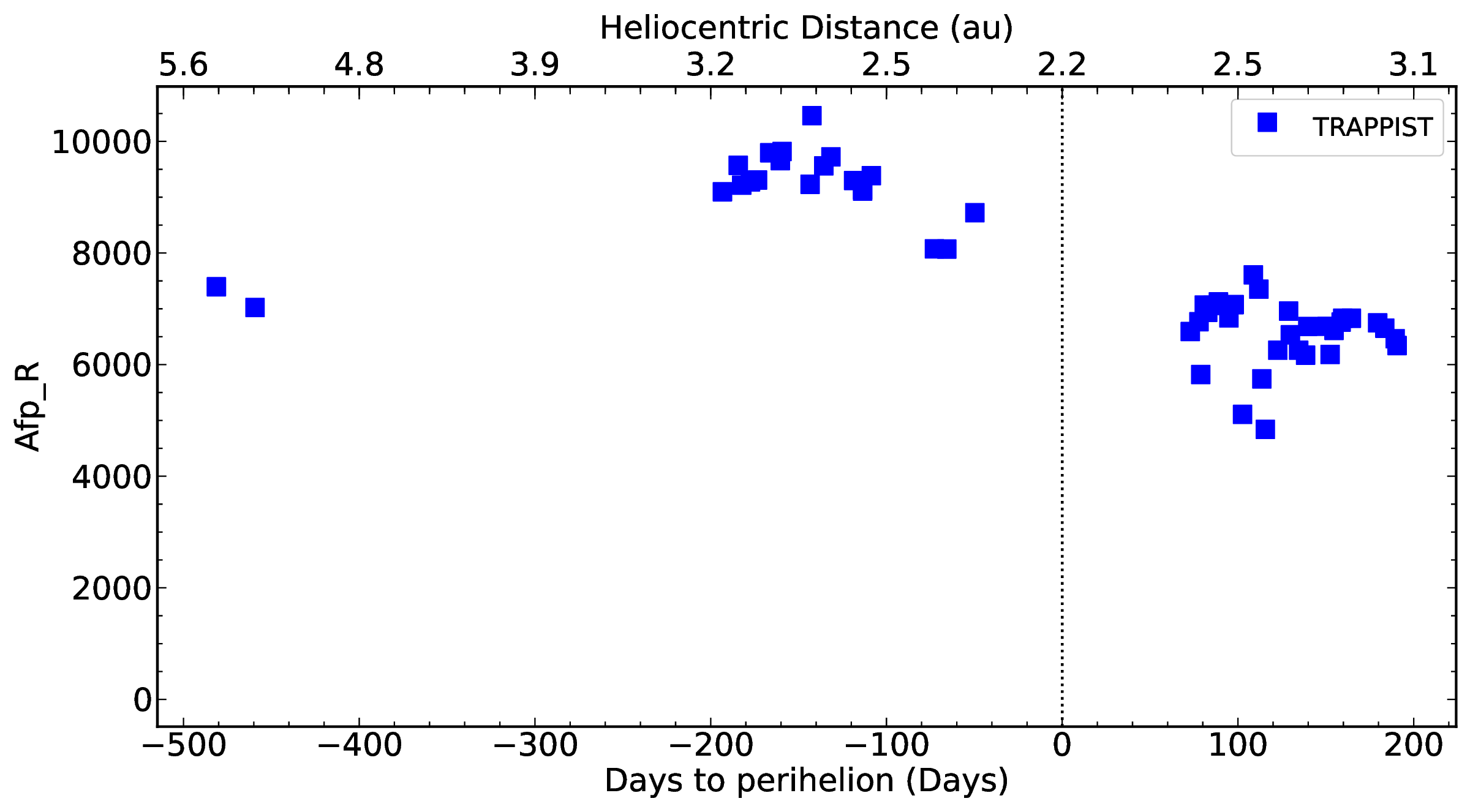}
\subcaption{Activity trend of A$(\theta = 0^{\circ})f\rho$(R).}
\label{A_R}
\end{subfigure}
\caption{Activity trends of comet C/2020 V2 as a function of days to perihelion and heliocentric distance (au).}
\end{figure*}
\newpage
\section{LS Processing on TRAPPIST Narrow-Band Images}
\begin{figure*}
\centering
\subcaptionbox{TRAPPIST CN filter image observed on 2022 November 22 \label{CN_2211}}%
  [0.48\linewidth]{\includegraphics[width=0.8\linewidth]{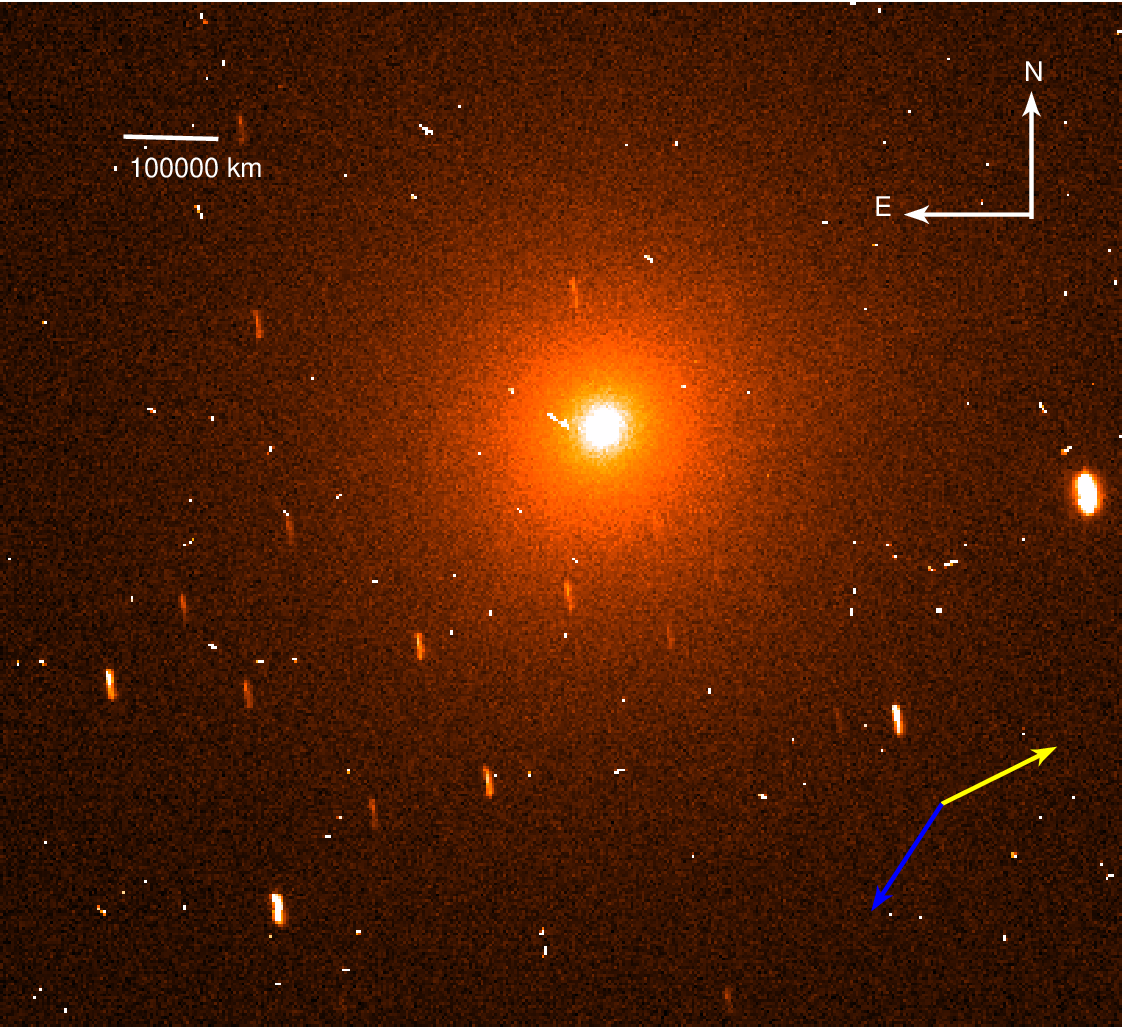}}
\hfill
\subcaptionbox{LS processing on the CN image observed on 2022 November 22 \label{LS_CN_2211}}%
  [0.48\linewidth]{\includegraphics[width=0.8\linewidth]{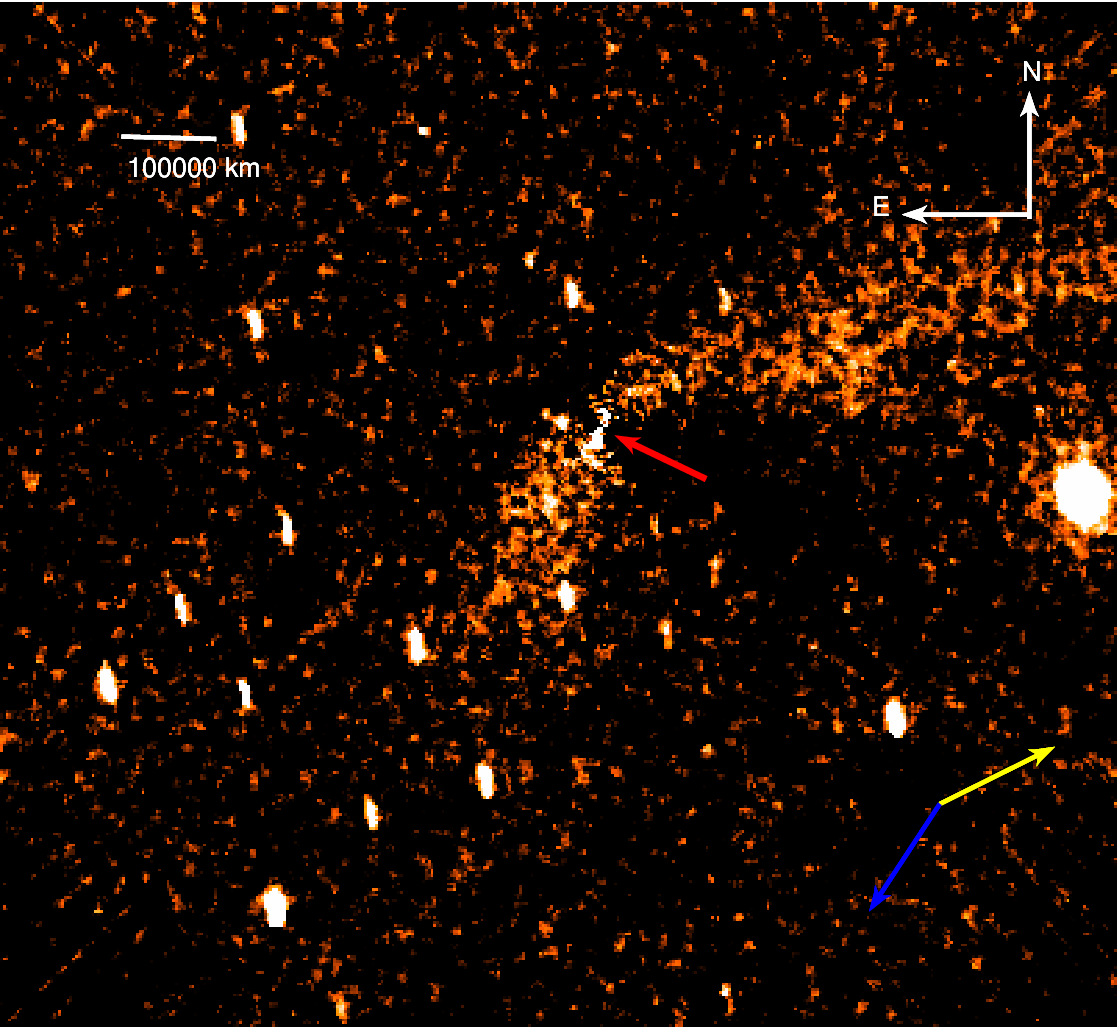}}

\caption{Image of CN filter and its corresponding LS processed. The dust-tail orientation and the Sun's direction are marked in \textit{blue} and \textit{yellow}. The \textit{red} arrow is pointing towards the photocenter.}
\label{fig: CN_combined_2211}
\end{figure*}

\begin{figure*}
\centering
\subcaptionbox{TRAPPIST C$_2$ filter image observed on 2022 November 22 \label{C2_2211}}%
  [0.48\linewidth]{\includegraphics[width=0.8\linewidth]{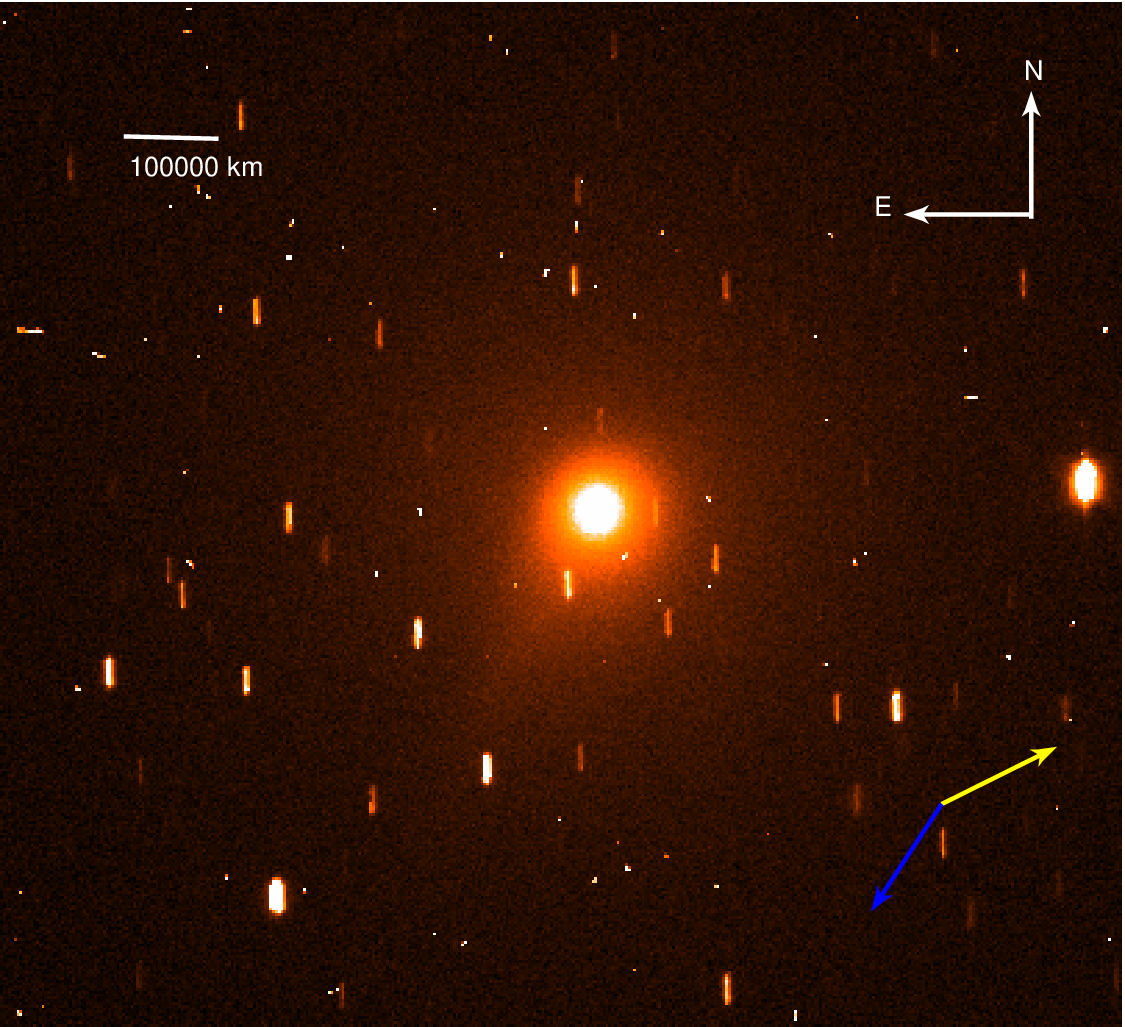}}
\hfill
\subcaptionbox{LS processing on the C$_2$ image observed on 2022 November 22 \label{LS_C2_2211}}%
  [0.48\linewidth]{\includegraphics[width=0.8\linewidth]{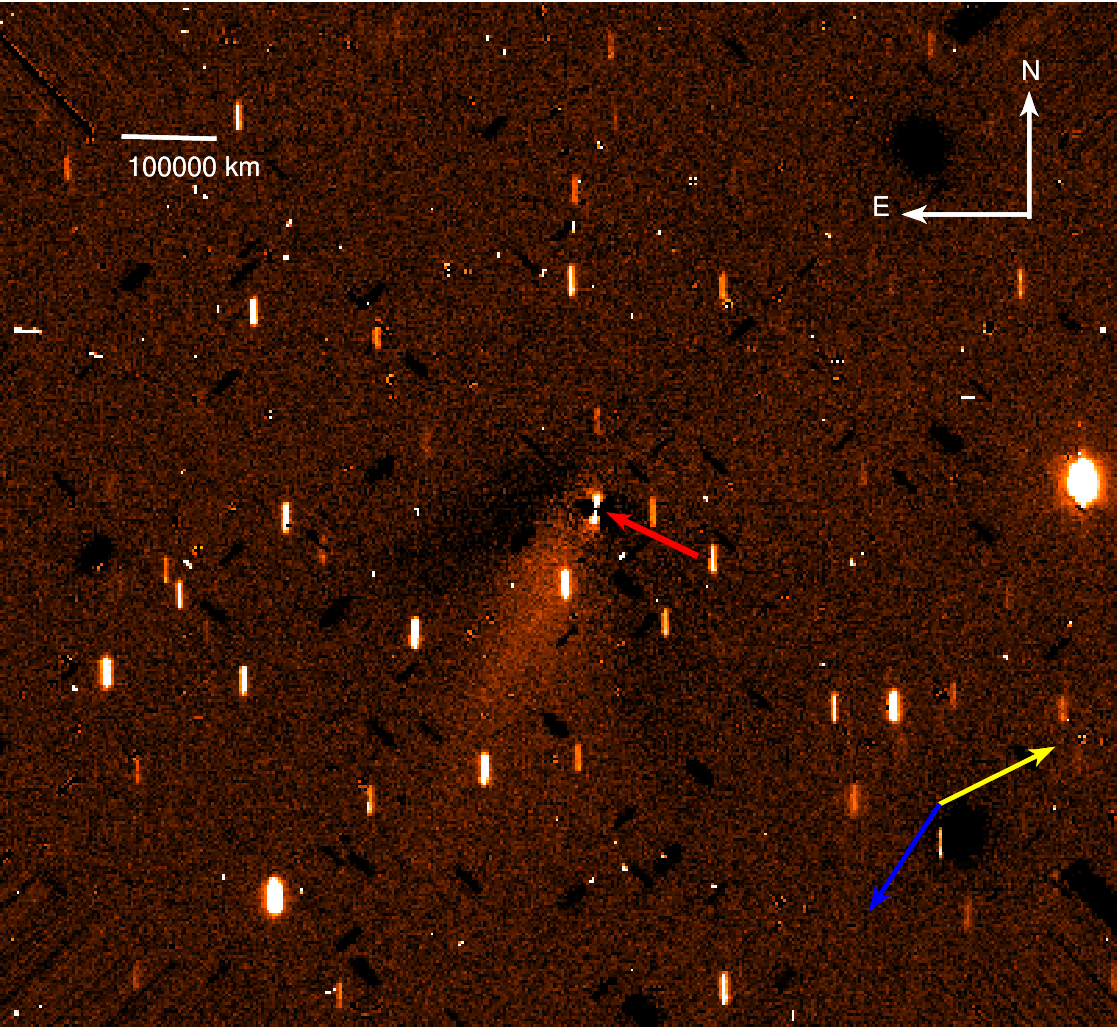}}

\caption{Image of C$_2$ filter and its corresponding LS processed. The dust-tail orientation and the Sun's direction are marked in \textit{blue} and \textit{yellow}. The \textit{red} arrow is pointing towards the photocenter.}
\label{fig: C2_combined_2211}
\end{figure*}

\begin{figure*}
\centering
\subcaptionbox{TRAPPIST C$_3$ filter image observed on 2022 November 22 \label{C3_2211}}%
  [0.48\linewidth]{\includegraphics[width=0.8\linewidth]{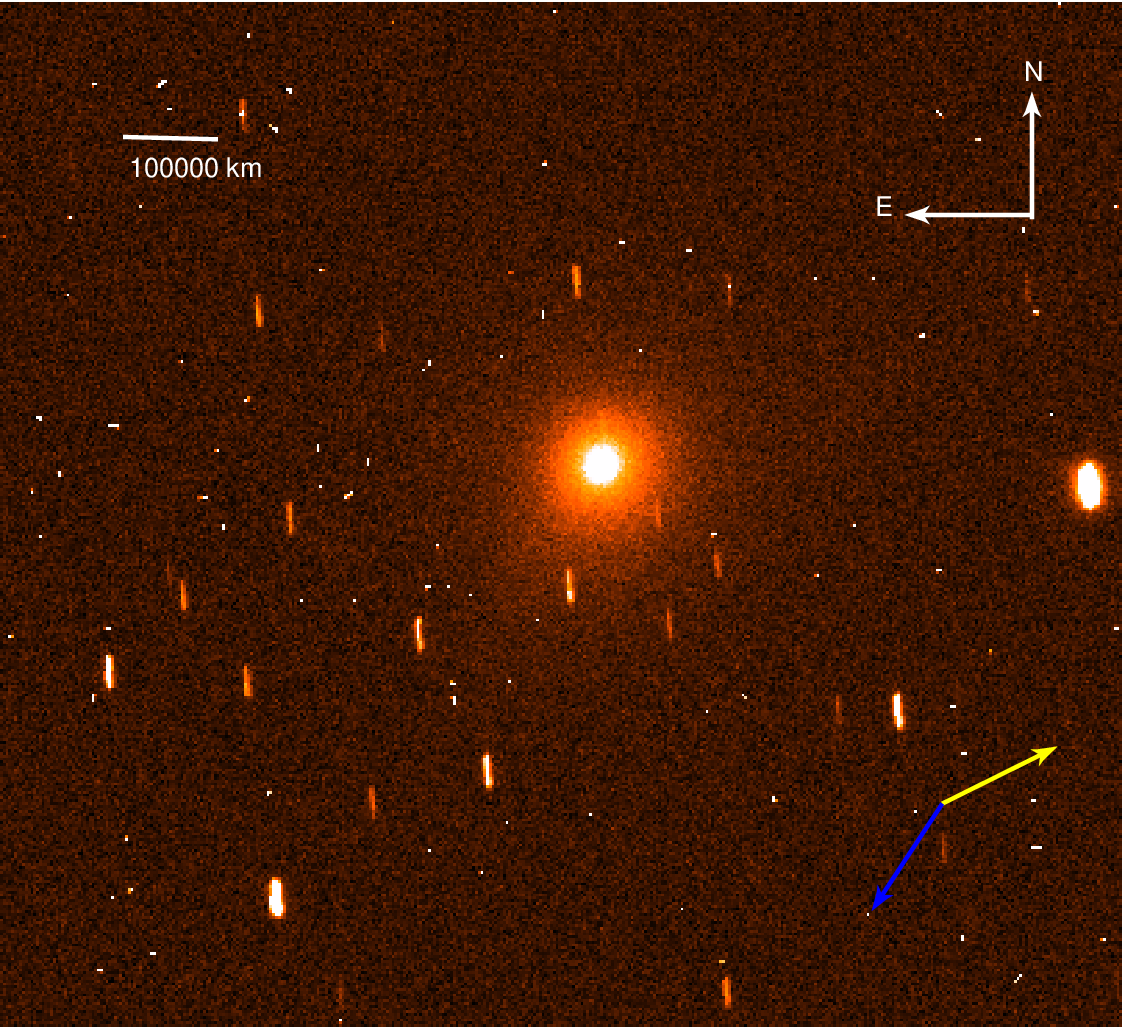}}
\hfill
\subcaptionbox{LS processing on the C$_3$ image observed on 2022 November 22 \label{LS_C3_2211}}%
  [0.48\linewidth]{\includegraphics[width=0.8\linewidth]{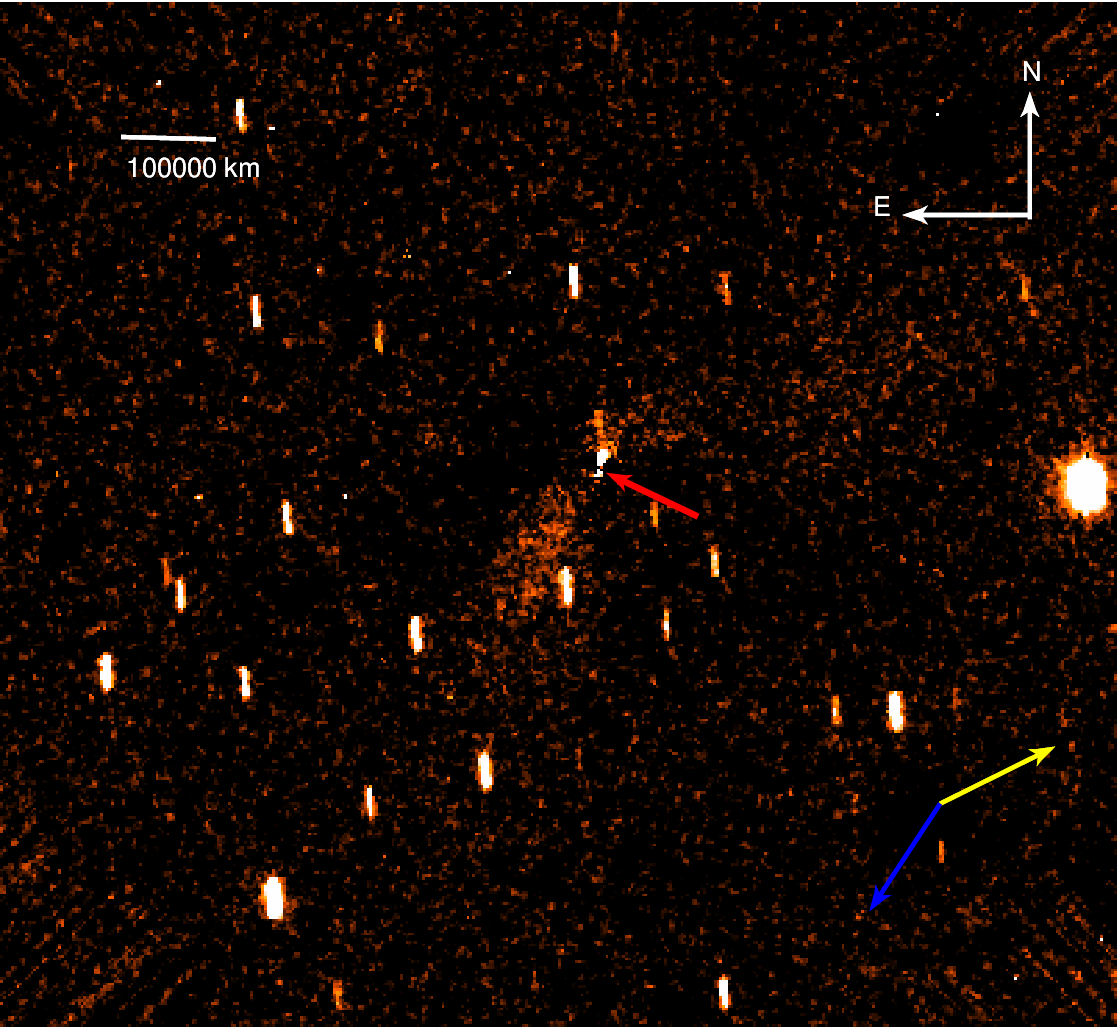}}

\caption{Image of C$_3$ filter and its corresponding LS processed. The dust-tail orientation and the Sun's direction are marked in \textit{blue} and \textit{yellow}. The \textit{red} arrow is pointing towards the photocenter.}
\label{fig: C3_combined_2211}
\end{figure*}

\begin{figure*}
\centering
\subcaptionbox{TRAPPIST CN filter image observed on 2022 December 16 \label{CN_1612}}%
  [0.48\linewidth]{\includegraphics[width=0.8\linewidth]{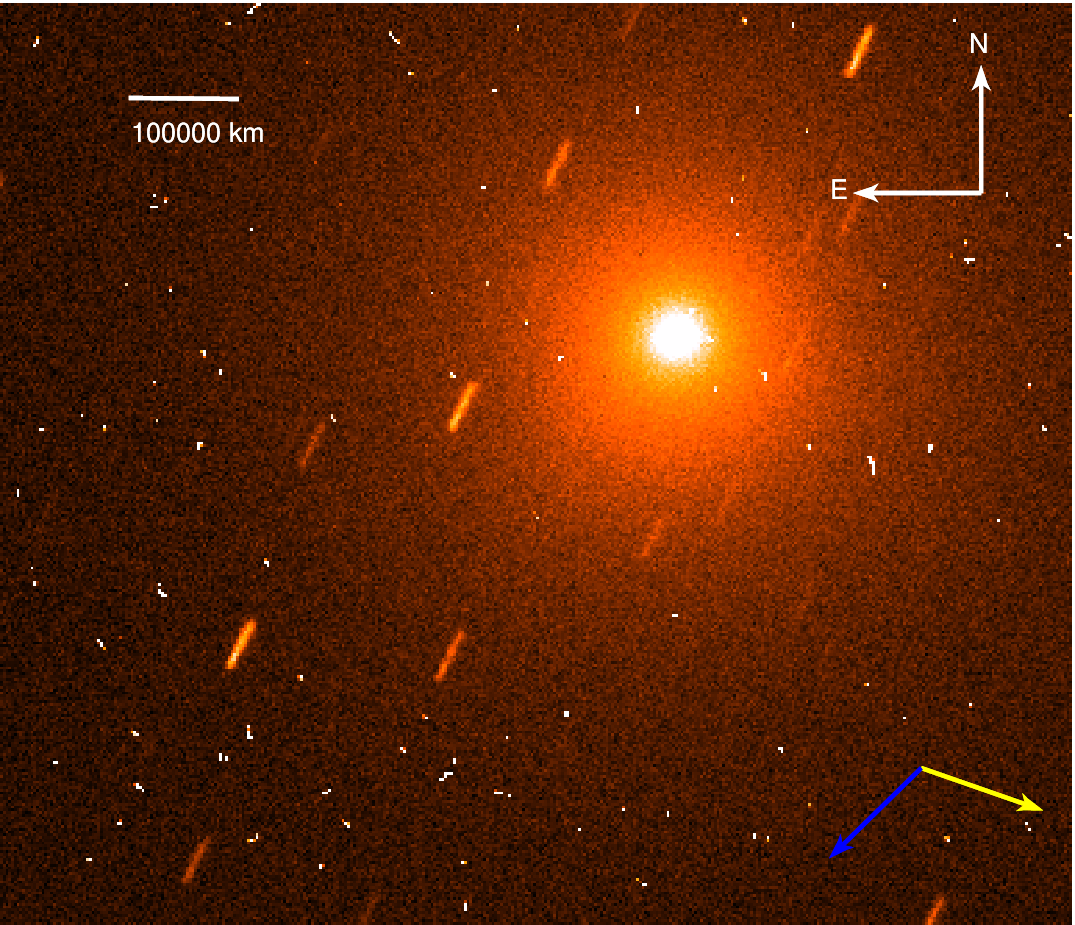}}
\hfill
\subcaptionbox{LS processing on the CN image observed on 2022 December 16 \label{LS_CN_1612}}%
  [0.48\linewidth]{\includegraphics[width=0.8\linewidth]{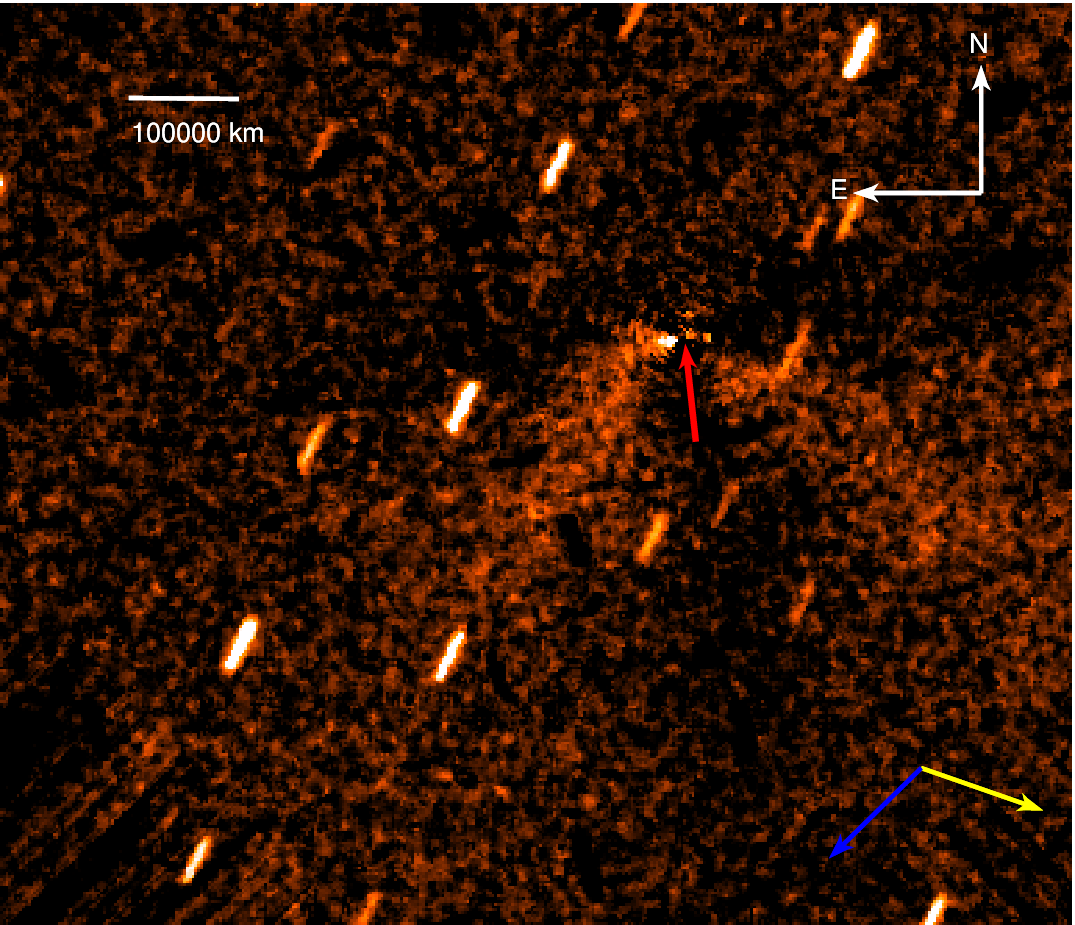}}

\caption{Image of CN filter and its corresponding LS processed. The dust-tail orientation and the Sun's direction are marked in \textit{blue} and \textit{yellow}. The \textit{red} arrow is pointing towards the photocenter.}
\label{fig: CN_combined_1612}
\end{figure*}

\begin{figure*}
\centering
\subcaptionbox{TRAPPIST C$_2$ filter image observed on 2022 December 16 \label{C2_1612}}%
  [0.48\linewidth]{\includegraphics[width=0.8\linewidth]{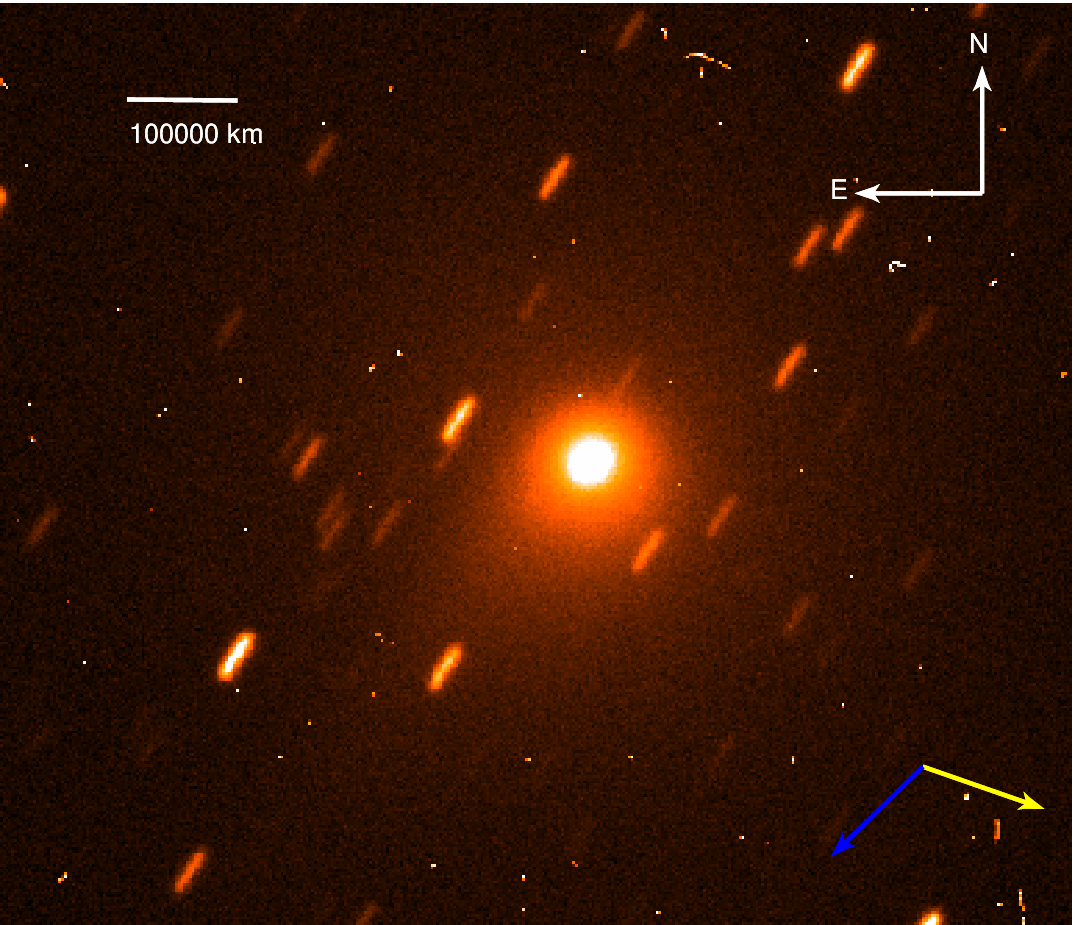}}
\hfill
\subcaptionbox{LS processing on the C$_2$ image observed on 2022 December 16 \label{LS_C2_1612}}%
  [0.48\linewidth]{\includegraphics[width=0.8\linewidth]{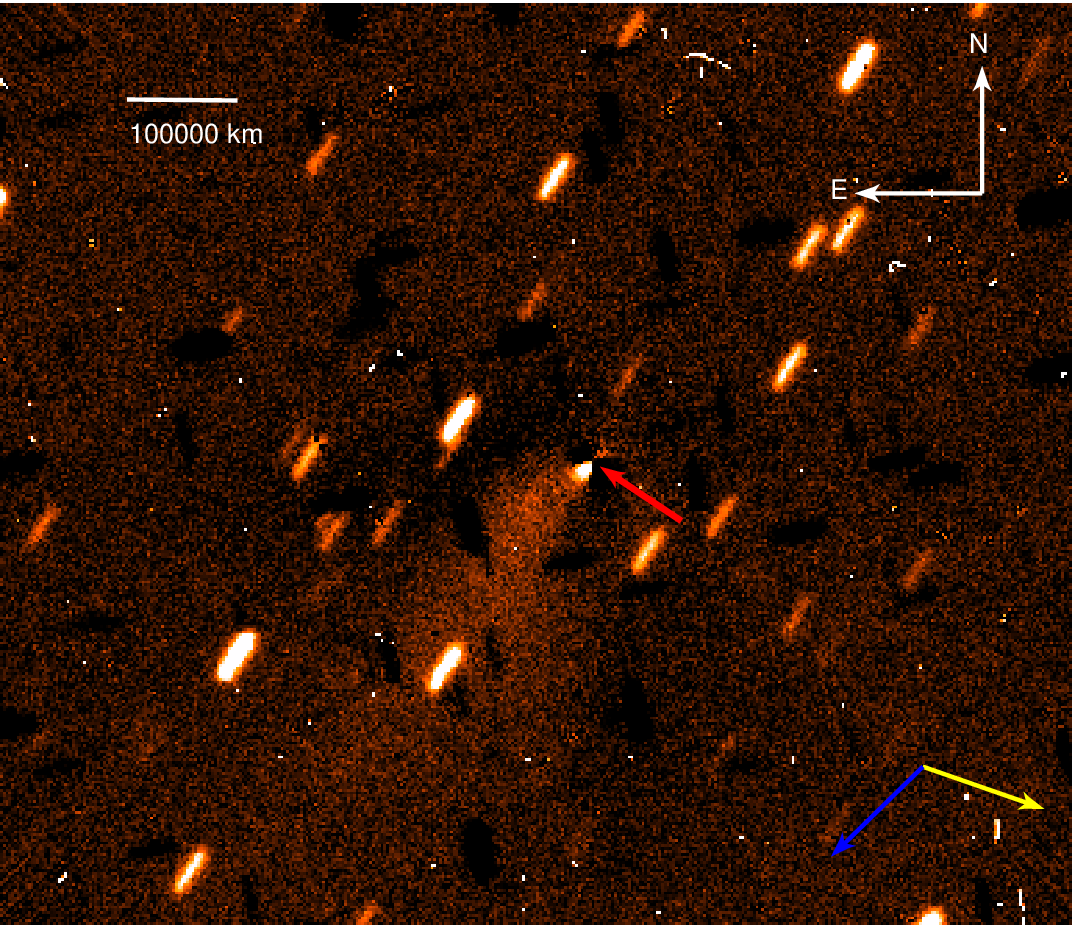}}

\caption{Image of C$_2$ filter and its corresponding LS processed. The dust-tail orientation and the Sun's direction are marked in \textit{blue} and \textit{yellow}. The \textit{red} arrow is pointing towards the photocenter.}
\label{fig: C2_combined_1612}
\end{figure*}

\begin{figure*}
\centering
\subcaptionbox{TRAPPIST C$_3$ filter image observed on 2022 December 16 \label{C3_1612}}%
  [0.48\linewidth]{\includegraphics[width=0.8\linewidth]{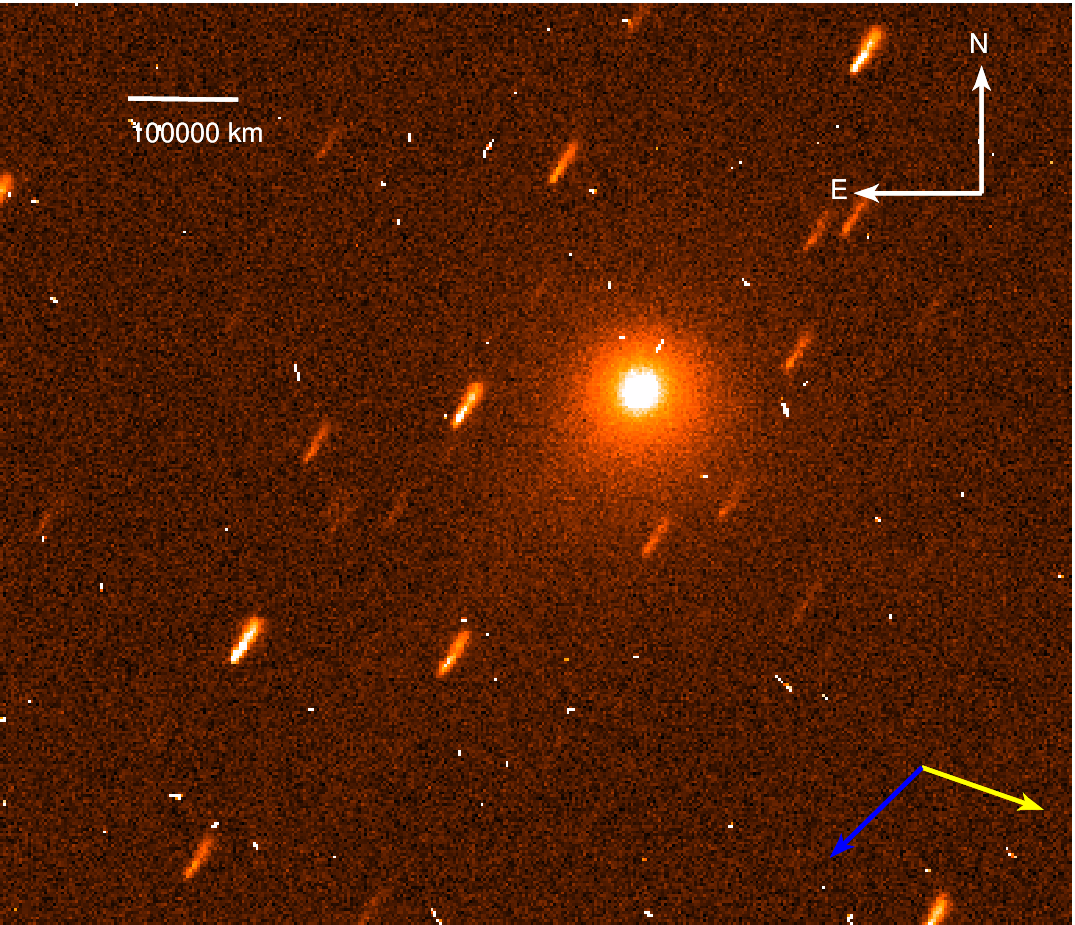}}
\hfill
\subcaptionbox{LS processing on the C$_3$ image observed on 2022 December 16 \label{LS_C3_1612}}%
  [0.48\linewidth]{\includegraphics[width=0.8\linewidth]{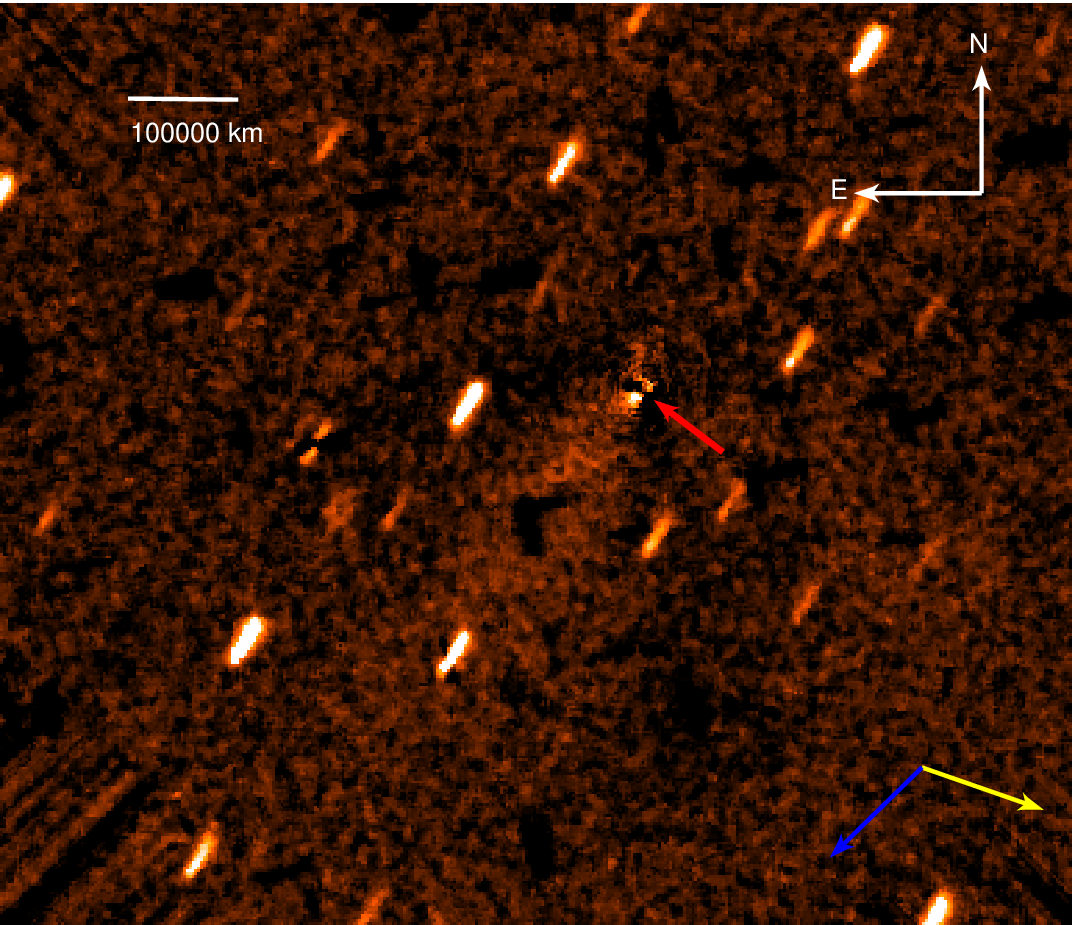}}

\caption{Image of C$_3$ filter and its corresponding LS processed. The dust-tail orientation and the Sun's direction are marked in \textit{blue} and \textit{yellow}. The \textit{red} arrow is pointing towards the photocenter.}
\label{fig: C3_combined_1612}
\end{figure*}

Here, we present the TRAPPIST narrow-band filter images of CN, C$_2$ and C$_3$ molecules observed on 2022 November 22, and 2022 December 16 (Fig. \ref{CN_2211}, Fig. \ref{C2_2211} and Fig. \ref{C3_2211}) and (Fig. \ref{CN_1612}, Fig. \ref{C2_1612} and Fig. \ref{C3_1612}), along with the corresponding LS processing on these images (Fig. \ref{LS_CN_2211}, Fig. \ref{LS_C2_2211} and Fig. \ref{LS_C3_2211}) and (Fig. \ref{LS_CN_1612}, Fig. \ref{LS_C2_1612} and Fig. \ref{LS_C3_1612}), using the details mentioned in section \ref{sec: coma morphology}.

\newpage
\section{\texorpdfstring{TRAPPIST Magnitudes in different filters and A \textit{\MakeLowercase{f}}$\rho$ in Broad-Band R filter}{TRAPPIST Magnitudes in different filters and Afrho in Broad-Band R filter}}

\begin{table*}
\caption{TRAPPIST magnitudes of comet V2 with different Broad-Band filters, B, V, R and I, and the Af$\rho$ in Broad-Band R filter observed from both TRAPPIST-North and South telescopes}
\setlength{\tabcolsep}{4.8pt}
        \resizebox{\textwidth}{!}{%
\begin{tabular}{crrccccccc}
\hline
\multicolumn{1}{|c|}{{Date}} &
  \multicolumn{1}{c|}{{DTP}} &
  \multicolumn{1}{c|}{{r}} &
  \multicolumn{1}{c|}{{$\Delta$}} &
  \multicolumn{1}{c|}{{Telescope}} &
  \multicolumn{4}{c|}{\hspace{0.5cm}{Magnitudes (Dimensionless)}}&
  \multicolumn{1}{c|}{\hspace{0.5cm}{A($\theta=0^{\circ}$)f$\rho$ (cm)}}\\
\multicolumn{1}{|c|}{} &
  \multicolumn{1}{c|}{{(Days)}} &
  \multicolumn{1}{c|}{{(au)}} &
  \multicolumn{1}{c|}{{(au)}} &
  \multicolumn{1}{c|}{TN/TS} &
  \multicolumn{1}{c|}{{B}} &
  \multicolumn{1}{c|}{{V}} &
  \multicolumn{1}{c|}{{R}} &
  \multicolumn{1}{c|}{{I}} &
  \multicolumn{1}{c|}{{R}} \\\hline
11-01-2022 & -481.28 & -5.41 & 5.00 & TN & 16.27 $\pm$ 0.02 & 15.49 $\pm$ 0.01 & 15.05 $\pm$ 0.01 & 14.65 $\pm$ 0.01 & 7395 $\pm$ 42\\
02-02-2022 & -459.33 & -5.23 & 4.62 & TN & 16.08 $\pm$ 0.01 & 15.28 $\pm$ 0.01 & 14.89 $\pm$ 0.01 & 14.56 $\pm$ 0.01 & 7023 $\pm$ 37\\
26-10-2022 & -193.34 & -3.10 & 3.10 & TN & 14.48 $\pm$ 0.02 & 13.70 $\pm$ 0.01 & 13.29 $\pm$ 0.01 & 12.86 $\pm$ 0.01 & 9097 $\pm$ 75\\
04-11-2022 & -184.37 & -3.04 & 2.91 & TN & 14.27 $\pm$ 0.02 & 13.58 $\pm$ 0.01 & 13.13 $\pm$ 0.01 & 12.74 $\pm$ 0.01 & 9572 $\pm$ 71\\
06-11-2022 & -182.36 & -3.02 & 2.86 & TN & 14.34 $\pm$ 0.02 & 13.59 $\pm$ 0.01 & 13.12 $\pm$ 0.01 & 12.70 $\pm$ 0.01 & 9216 $\pm$ 72\\
11-11-2022 & -177.38 & -2.98 & 2.76 & TN & 14.27 $\pm$ 0.02 & 13.51 $\pm$ 0.01 & 13.05 $\pm$ 0.01 & 12.67 $\pm$ 0.01 & 9275 $\pm$ 96\\
15-11-2022 & -173.38 & -2.96 & 2.67 & TN & 14.23 $\pm$ 0.02 & 13.46 $\pm$ 0.01 & 13.01 $\pm$ 0.01 & 12.56 $\pm$ 0.01 & 9307 $\pm$ 70\\
22-11-2022 & -166.39 & -2.91 & 2.53 & TN & 13.95 $\pm$ 0.02 & 13.26 $\pm$ 0.01 & 12.87 $\pm$ 0.01 & 12.48 $\pm$ 0.01 & 9796 $\pm$ 72\\
28-11-2022 & -160.40 & -2.87 & 2.42 & TN & 13.95 $\pm$ 0.02 & 13.22 $\pm$ 0.01 & 12.81 $\pm$ 0.01 & 12.42 $\pm$ 0.01 & 9655 $\pm$ 70\\
29-11-2022 & -159.41 & -2.86 & 2.41 & TN & 13.93 $\pm$ 0.02 & 13.22 $\pm$ 0.01 & 12.76 $\pm$ 0.01 & 12.42 $\pm$ 0.01 & 9819 $\pm$ 71\\
15-12-2022 & -143.45 & -2.76 & 2.18 & TN & 13.79 $\pm$ 0.02 & 13.06 $\pm$ 0.01 & 12.63 $\pm$ 0.01 & 12.23 $\pm$ 0.02 & 9231 $\pm$ 83\\
16-12-2022 & -142.42 & -2.75 & 2.17 & TN & 13.68 $\pm$ 0.02 & 12.97 $\pm$ 0.01 & 12.49 $\pm$ 0.01 & 12.21 $\pm$ 0.03 & 10462 $\pm$ 126\\
23-12-2022 & -135.67 & -2.71 & 2.11 & TN & 13.65 $\pm$ 0.02 & 12.90 $\pm$ 0.01 & 12.48 $\pm$ 0.01 & 12.22 $\pm$ 0.04 & 9561 $\pm$ 94\\
27-12-2022 & -131.61 & -2.68 & 2.08 & TN & 13.65 $\pm$ 0.02 & 12.87 $\pm$ 0.01 & 12.46 $\pm$ 0.01 & -- & 9723 $\pm$ 234\\
09-01-2023 & -118.68 & -2.60 & 2.07 & TN & 13.64 $\pm$ 0.02 & 12.81 $\pm$ 0.01 & 12.47 $\pm$ 0.01 & -- & 9298 $\pm$ 86\\
14-01-2023 & -113.55 & -2.58 & 2.08 & TN & 13.74 $\pm$ 0.02 & -- & 12.48 $\pm$ 0.01 & 12.11 $\pm$ 0.01 & 9110 $\pm$ 123\\
19-01-2023 & -108.56 & -2.55 & 2.11 & TN & 13.64 $\pm$ 0.02 & 12.85 $\pm$ 0.01 & 12.43 $\pm$ 0.01 & 12.04 $\pm$ 0.01 & 9386 $\pm$ 100\\
24-02-2023 & -72.69 & -2.38 & 2.55 & TN & 13.85 $\pm$ 0.02 & 13.03 $\pm$ 0.01 & 12.69 $\pm$ 0.01 & 12.29 $\pm$ 0.01 & 8074 $\pm$ 65\\
03-03-2023 & -65.65 & -2.35 & 2.65 & TN & 13.83 $\pm$ 0.03 & 13.06 $\pm$ 0.02 & 12.73 $\pm$ 0.02 & -- & 8070 $\pm$ 110\\
19-03-2023 & -49.71 & -2.30 & 2.87 & TN & 13.70 $\pm$ 0.03 & 12.98 $\pm$ 0.01 & 12.54 $\pm$ 0.01 & 12.15 $\pm$ 0.01 & 8722 $\pm$ 94\\
26-06-2023 & 49.91 & 2.30 & 2.88 & TS & 14.04 $\pm$ 0.03 & 13.29 $\pm$ 0.01 & 12.88 $\pm$ 0.01 & 12.44 $\pm$ 0.01 & --\\
19-07-2023 & 72.88 & 2.38 & 2.55 & TS & 14.18 $\pm$ 0.02 & -- & 12.95 $\pm$ 0.01 & -- & 6592 $\pm$ 55\\
24-07-2023 & 77.84 & 2.40 & 2.47 & TS & 14.10 $\pm$ 0.02 & -- & 12.98 $\pm$ 0.01 & -- & 6769 $\pm$ 63\\
25-07-2023 & 78.83 & 2.40 & 2.45 & TS & 14.15 $\pm$ 0.02 & -- & 13.05 $\pm$ 0.01 & -- & 5822 $\pm$ 56\\
27-07-2023 & 80.86 & 2.41 & 2.42 & TS & 14.02 $\pm$ 0.02 & 13.28 $\pm$ 0.01 & 12.87 $\pm$ 0.01 & 12.44 $\pm$ 0.01 & 7068 $\pm$ 53\\\
29-07-2023 & 83.00 & 2.42 & 2.40 & TS & -- & -- & 12.89 $\pm$ 0.01 & -- & 6937 $\pm$ 53\\
01-08-2023 & 85.85 & 2.44 & 2.34 & TS & 14.01 $\pm$ 0.02 & -- & 12.87 $\pm$ 0.01 & 12.35 $\pm$ 0.01 & 7061 $\pm$ 59\\
04-08-2023 & 88.89 & 2.45 & 2.29 & TS & 14.01 $\pm$ 0.02 & -- & 12.80 $\pm$ 0.01 & 12.38 $\pm$ 0.01 & 7124 $\pm$ 54\\
10-08-2023 & 94.87 & 2.48 & 2.19 & TS & 14.09 $\pm$ 0.02 & 13.23 $\pm$ 0.01 & 12.82 $\pm$ 0.01 & 12.34 $\pm$ 0.01 & 6837 $\pm$ 42\\
13-08-2023 & 97.85 & 2.49 & 2.15 & TS & 13.94 $\pm$ 0.02 & 12.77 $\pm$ 0.01 & 12.79 $\pm$ 0.01 & 12.33 $\pm$ 0.01 & 7075 $\pm$ 66\\
16-08-2023 & 101.00 & 2.51 & 2.10 & TS & -- & 13.25 $\pm$ 0.01 & -- & 12.38 $\pm$ 0.01 & --\\
18-08-2023 & 102.64 & 2.52 & 2.08 & TN & 14.25 $\pm$ 0.02 & -- & 13.05 $\pm$ 0.01 & 12.77 $\pm$ 0.01 & 5108 $\pm$ 34\\
24-08-2023 & 108.76 & 2.55 & 2.00 & TS & 13.82 $\pm$ 0.02 & 13.06 $\pm$ 0.01 & 12.61 $\pm$ 0.01 & 12.16 $\pm$ 0.01 & 7609 $\pm$ 54\\
27-08-2023 & 111.90 & 2.57 & 1.97 & TS & 13.79 $\pm$ 0.01 & 13.02 $\pm$ 0.01 & 12.62 $\pm$ 0.01 & 12.23 $\pm$ 0.01 & 7352 $\pm$ 42\\
29-08-2023 & 113.60 & 2.58 & 1.95 & TN & 14.05 $\pm$ 0.02 & -- & 12.90 $\pm$ 0.01 & 12.41 $\pm$ 0.01 & 5744 $\pm$ 46\\
31-08-2023 & 116.00 & 2.59 & 1.93 & TS & -- & -- & 13.04 $\pm$ 0.01 & -- & 4840 $\pm$ 60\\
07-09-2023 & 123.00 & 2.63 & 1.88 & TS & -- & -- & 12.53 $\pm$ 0.01 & -- & 6258 $\pm$ 55\\
13-09-2023 & 128.83 & 2.66 & 1.86 & TS & 13.72 $\pm$ 0.01 & 12.96 $\pm$ 0.01 & 12.55 $\pm$ 0.01 & 12.11 $\pm$ 0.01 & 6960 $\pm$ 34\\
23-09-2023 & 138.49 & 2.72 & 1.87 & TS & -- & -- & 12.71 $\pm$ 0.01 & -- & 6169 $\pm$ 72\\
24-09-2023 & 139.72 & 2.73 & 1.87 & TS & 13.73 $\pm$ 0.01 & 12.95 $\pm$ 0.01 & 12.54 $\pm$ 0.01 & 12.14 $\pm$ 0.01 & 6684 $\pm$ 34\\
05-10-2023 & 150.64 & 2.80 & 1.96 & TS & 13.86 $\pm$ 0.01 & 13.12 $\pm$ 0.01 & 12.69 $\pm$ 0.01 & 12.22 $\pm$ 0.01 & 6686 $\pm$ 42\\
07-10-2023 & 152.48 & 2.81 & 1.99 & TN & 14.03 $\pm$ 0.03 & 13.23 $\pm$ 0.02 & 12.75 $\pm$ 0.01 & 12.40 $\pm$ 0.01 & 6182 $\pm$ 34\\
09-10-2023 & 154.72 & 2.83 & 2.01 & TS & 13.97 $\pm$ 0.01 & 13.16 $\pm$ 0.01 & 12.74 $\pm$ 0.01 & 12.32 $\pm$ 0.01 & 6614 $\pm$ 48\\
13-10-2023 & 158.73 & 2.86 & 2.07 & TS & 14.02 $\pm$ 0.01 & 13.28 $\pm$ 0.01 & 12.77 $\pm$ 0.01 & 12.40 $\pm$ 0.01 & 6762 $\pm$ 37\\
14-10-2023 & 159.65 & 2.86 & 2.09 & TS & 14.03 $\pm$ 0.01 & -- & 12.85 $\pm$ 0.01 & 12.43 $\pm$ 0.01 & 6832 $\pm$ 34\\
19-10-2023 & 164.53 & 2.90 & 2.17 & TS & 14.15 $\pm$ 0.02 & 13.40 $\pm$ 0.01 & 12.93 $\pm$ 0.01 & 12.48 $\pm$ 0.01 & 6831 $\pm$ 39\\
03-11-2023 & 179.49 & 3.00 & 2.47 & TS & 14.44 $\pm$ 0.02 & 13.71 $\pm$ 0.01 & 13.24 $\pm$ 0.01 & 12.84 $\pm$ 0.01 & 6749 $\pm$ 40\\
07-11-2023 & 183.57 & 3.03 & 2.56 & TS & 14.59 $\pm$ 0.01 & 13.78 $\pm$ 0.01 & 13.34 $\pm$ 0.01 & 12.90 $\pm$ 0.00 & 6653 $\pm$ 33\\
13-11-2023 & 189.50 & 3.07 & 2.70 & TS & 14.70 $\pm$ 0.01 & 13.88 $\pm$ 0.01 & 13.47 $\pm$ 0.01 & 13.01 $\pm$ 0.00 & 6464 $\pm$ 65\\
14-11-2023 & 190.58 & 3.08 & 2.73 & TS & 14.69 $\pm$ 0.01 & -- & 13.53 $\pm$ 0.01 & -- & 6342 $\pm$ 35\\
05-12-2023 & 212.00 & 3.24 & 3.24 & TS & 15.03 $\pm$ 0.01 & -- & -- & -- & --\\
12-12-2023 & 219.00 & 3.29 & 3.41 & TS & 15.13 $\pm$ 0.01 & -- & 13.90 $\pm$ 0.01 & -- & --\\
13-12-2023 & 220.00 & 3.30 & 3.43 & TS & -- & -- & 13.87 $\pm$ 0.01 & -- & --\\
16-12-2023 & 223.00 & 3.32 & 3.50 & TS & 15.18 $\pm$ 0.01 & 14.41 $\pm$ 0.01 & 13.93 $\pm$ 0.01 & 13.49 $\pm$ 0.01 & --\\
17-12-2023 & 224.00 & 3.33 & 3.52 & TS & 15.18 $\pm$ 0.01 & 14.42 $\pm$ 0.01 & 13.98 $\pm$ 0.01 & 13.54 $\pm$ 0.01 & --\\
09-07-2024 & 429.00 & 4.98 & 4.47 & TS & 16.53 $\pm$ 0.01 & 15.72 $\pm$ 0.01 & 15.29 $\pm$ 0.01 & 14.77 $\pm$ 0.01 & --\\
13-07-2024 & 432.00 & 5.00 & 4.47 & TS & 16.52 $\pm$ 0.01 & 15.67 $\pm$ 0.01 & 15.33 $\pm$ 0.01 & 14.88 $\pm$ 0.01 & --\\
13-08-2024 & 463.00 & 5.26 & 4.47 & TS & -- & -- & 15.37 $\pm$ 0.01 & -- & --\\\hline
\end{tabular}\vspace{-0.5cm}
}
\label{sec: BB Magnitude}
\end{table*}\vspace{-0.5cm} 

\bsp	
\label{lastpage}
\end{document}